\documentclass[epj]{svjour}

\usepackage{graphics}
\usepackage{amsmath}
\usepackage{amssymb}
\usepackage{color}
\usepackage{graphicx}

\usepackage{url}
\usepackage[colorlinks,citecolor=blue,linktoc=all,linkcolor=cyan,urlcolor=black]{hyperref}

\begin{document}
\title{The anomalous magnetic moment of the muon: status of Lattice QCD calculations}
\author{Antoine G\'erardin\inst{1} 
\thanks{\emph{email:} antoine.gerardin@cpt.univ-mrs.fr}
} 
\institute{Aix Marseille Univ, Universit\'{e} de Toulon, CNRS, CPT, Marseille, France.}
\date{Received: date / Revised version: date}

\abstract{In recent years, the anomalous magnetic moment of the muon has triggered a lot of activity in the lattice QCD community because a persistent tension of about $3.5~\sigma$ is observed between the phenomenological estimate and the Brookhaven measurement. The current best phenomenological estimate has an uncertainty comparable to the experimental one and the error is completely dominated by hadronic effects: the leading order hadronic vacuum polarization (HVP) contribution and the hadronic light-by-light (HLbL) scattering contribution. Both are accessible via lattice simulations and a reduction of the error by a factor 4 is required in view of the forthcoming experiments at Fermilab and J-PARC whose results, expected in the next few years, should reduce the experimental precision down to the level of 0.14~ppm. In this article, I review the status of lattice calculations of those quantities, starting with the HVP. This contribution has now reached sub-percent precision and requires a careful understanding of all sources of systematic errors. The HLbL contribution, that is much smaller, still contributes significantly to the error. This contribution is more challenging to compute, but rapid progress has been made on the lattice in the last few years.
\PACS{ \ \ 
     } 
} 

\makeatletter
\g@addto@macro\bfseries{\boldmath}
\makeatother

\newcommand{\dd}{\mathrm{d}}
\newcommand{\ubar}{\overline{u}}
\newcommand{\dbar}{\overline{d}}
\newcommand{\sbar}{\overline{s}}
\newcommand{\cbar}{\overline{c}}
\newcommand{\bbar}{\overline{b}}
\newcommand{\fm}{\mathrm{fm}}
\newcommand{\MeV}{\mathrm{MeV}}
\newcommand{\GeV}{\mathrm{GeV}}
\newcommand{\FF}{{\cal F}_{\pi^0\gamma^*\gamma^*}}
\newcommand{\amu}{a_\mu}
\newcommand{\kernel}{\mathcal{\bar L}}
\newcommand{\PI}{\mathrm{\Pi}}
\newcommand{\ahlbl}{a_\mu^{\rm hlbl}}
\newcommand{\ahvp}{a_{\mu}^{\rm hvp}}
\newcommand{\psib}{\overline{\psi}}
\newcommand{\Fig}[1]{Fig.~\ref{#1}}
\newcommand{\Eq}[1]{Eq.~(\ref{#1})}
\newcommand{\Section}[1]{Section~\ref{#1}}
\newcommand{\Table}[1]{Table~\ref{#1}}
\newcommand{\GammaGG}{\Gamma_{\gamma\gamma}}

\maketitle

\section{Introduction}
\label{intro}

The anomalous magnetic moment of the muon is the deviation of the gyromagnetic ratio $g_{\mu}$ of the muon from 2, the value predicted by the Dirac equation. This deviation is explained by quantum corrections and has been measured with an impressive precision of 0.5 ppm by the Brookhaven experiment~\cite{Bennett:2006fi}. 

Interestingly, a persistent discrepancy of 3.5 standard deviations has been observed in the last few years making this quantity a key observable to search for new physics beyond the Standard Model of particle physics. The main contribution to this observable comes from quantum electrodynamics (QED) and can be accurately computed using a perturbative expansion in the fine-structure constant $\alpha$~\cite{Aoyama:2012wk,Aoyama:2019ryr}. The small electroweak corrections are also under control~\cite{Czarnecki:2002nt,Gnendiger:2013pva}. Finally, although quarks and gluons do not couple directly to the muon, they do interact via loop diagrams. Even if hadronic contributions are relatively small, they completely dominate the error budget and are the limiting factor in view of reducing the theory error. The current status of the Standard Model contributions is summarized in \Table{tab:status}.

Significant efforts have been made on the experimental side and two new experiments (E989 at Fermilab and E34 at J-PARC) plan to reduce the error of the  measurement by a factor four in the next few years~\cite{Venanzoni:2014ixa,Otani:2015lra}. This has sparked a lot of activity in the community to reduce the theory error at the same level of precision and the \textit{Muon $g-2$ Theory Initiative}\footnote{https://muon-gm2-theory.illinois.edu/} was created to facilitate interactions among the theoretical and experimental $g-2$ communities. Recently, this activity has been summarized in a white-paper~\cite{Aoyama:2020ynm}.

\begin{figure}[t]
\begin{center}
\resizebox{0.17\textwidth}{!}{  \includegraphics{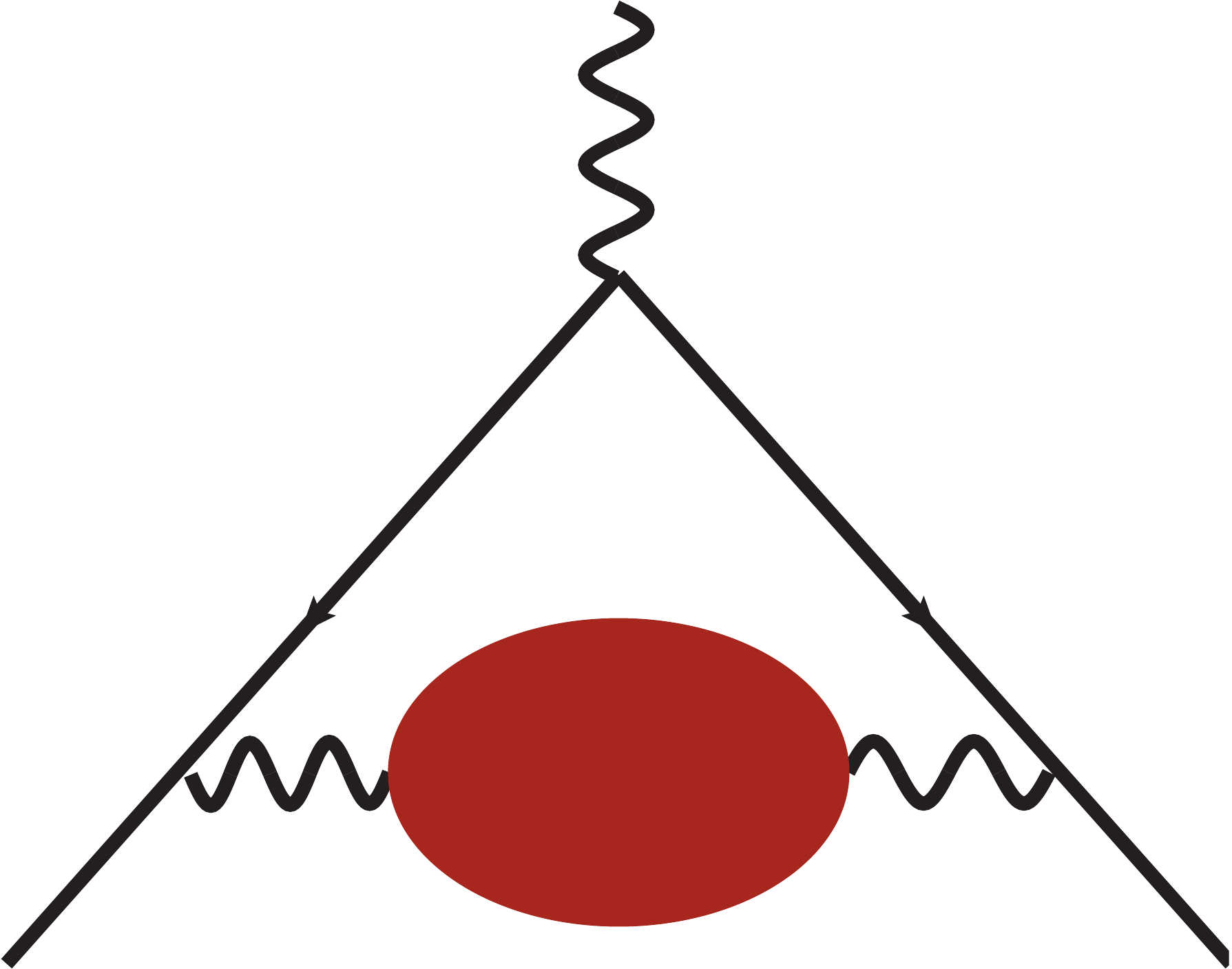}  } \qquad
\resizebox{0.17\textwidth}{!}{  \includegraphics{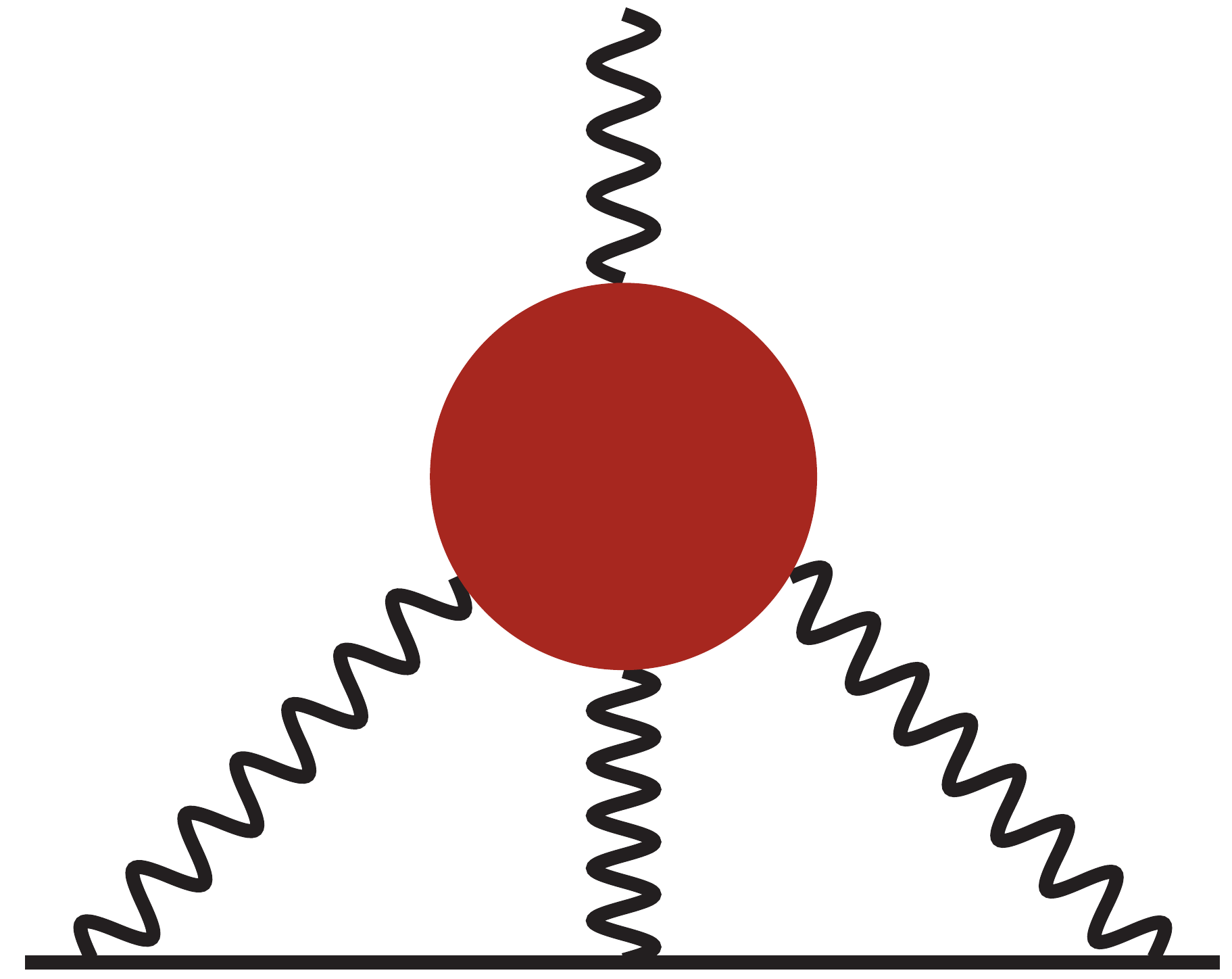}  }
\end{center}
\caption{Diagrams corresponding to the LO-HVP (left) and HLbL (right) contribution to the muon $(g-2)_{\mu}$. The red blobs represents the non-perturbative contribution. Muon and photon are represented by plain and wiggly lines respectively. }
\label{fig:diag}
\end{figure}

More specifically, the theory error is dominated by effects of the strong interaction that can be separated into two distinct contributions depicted in \Fig{fig:diag}: the Hadronic Vacuum Polarization (HVP) that enters at order $\alpha^2$ in the electromagnetic coupling and the Hadronic Light-by-Light scattering (HLbL) contribution, at order $\alpha^3$. For the HVP contribution, the most precise determination based on the dispersive approach has reached a precision of 0.58\%~\cite{Davier:2019can,Keshavarzi:2019abf,Jegerlehner:2017lbd}. This is a data driven method that relies on the analytical properties of the theory: the hadronic contribution is obtained as a convolution integral between a QED weight function $K$, that is known analytically, and the $R-$ratio, obtained from the combination of $e^+ e^- \to$~hadrons cross section data that can be measured experimentally
\begin{equation}
\ahvp = \left( \frac{\alpha m_{\mu}}{3\pi} \right)^2 \, \int_0^{\infty} \mathrm{d}s \, \frac{K(s) R(s)}{s^2} \,,
\label{eq:disp}
\end{equation}
\begin{equation}
R(s) = \frac{\sigma(e^+e^- \to \mathrm{hadrons})}{ 4\pi \alpha^2/(3s)} \,.
\end{equation}
This method is mostly limited by the availably of precise experimental data. The dominant contribution comes from the region $[0.6-0.9]~\GeV$ which contains the rho resonance. Unfortunately, in this energy region, the $\pi^+\pi^-$ channel suffers from  tensions between different data sets~\cite{Davier:2019can,Keshavarzi:2019abf}. Thus, a second, independent determination would be valuable and lattice QCD is the ideal tool to provide a first-principle result. In this case, a precision below $0.25\%$ would be needed to reach the future experimental precision.  Most lattice QCD simulations are less precise, with uncertainties around 2\%, but significant progress has been made in recent years and a very recent determination by the BMW collaboration has sub-percent precision~\cite{Borsanyi:2020mff}.  Several lattice groups are working on this contribution and independent checks of this latest result are expected in the coming months.  

The HLBL contribution appears at order $\alpha^3$ and is suppressed compared to the HVP contribution such that an overall precision of 10\%, or better, is needed. But its determination is also much more challenging, contributing significantly to the theory error. Until recently, this contribution was estimated using model estimates~\cite{Bijnens:1995cc,Bijnens:1995xf} and the Glasgow consensus~\cite{Prades:2009tw} largely relies on model calculations where systematic errors are difficult to quantify.
Recently, two model-independent approaches have been proposed. 
First, the dispersive approach: as compared to the HVP, this data-driven method is much more complex and involves a four-point correlation function~\cite{Colangelo:2014dfa,Colangelo:2014pva,Pauk:2014jza,Colangelo:2015ama}. Moreover, experimental data are often incomplete or missing. However, the lack of experimental data can be partly compensated by lattice QCD, especially for the dominant contributions. Second, it has been shown that lattice QCD can also directly access this hadronic contribution~\cite{Hayakawa:2005eq} and two collaborations have presented results so far. This year, a first lattice estimate, with controlled errors, has been published by the RBC/UKQCD collaboration~\cite{Blum:2019ugy}.

This review is organized as follows: in \Section{sec:lohvp}, I present the current status of the lattice determination of the hadronic vacuum polarisation contribution. Different contribution are discussed with emphasis on the challenges for reaching sub-percent precision. The section ends with a compilation of lattice results. In \Section{sec:bench}, I discuss related quantities that can be used to perform cross-checks between lattice collaborations or with the phenomenological estimate. In \Section{sec:hlbl}, I summarize the status on the hadronic light-by-light contribution. I first briefly explain the methodology before presenting the latest results for the direct lattice calculation. Finally, I present results for the pseudoscalar-pole contributions, that are expected to provide the dominant contribution, and the forward light-by-light scattering amplitudes that can provide valuable information on form factors that are used in phenomenological models to estimate the HLbL contribution.

\begin{table}[h] 
\begin{center} 
\renewcommand{\arraystretch}{1.3}
\begin{tabular}{l@{\hskip 0.05in}cl}
\hline 
Contribution &  	 $a_{\mu} \times 10^{11}$	& Ref.\\ 
\hline 
QED	(order $\mathcal{O}(\alpha^5)$)		& $116\ 584\ 718.93 \pm 0.10$ 	&\cite{Aoyama:2012wk,Aoyama:2019ryr} \\ 
Electroweak			& 	$153.6 \pm 1.0$ 	 		&\cite{Czarnecki:2002nt,Gnendiger:2013pva} \\ 
QCD	&   												\\
\qquad HVP (LO)		&	$6\ 931 \pm 40$			&\cite{Davier:2017zfy,Keshavarzi:2018mgv,Colangelo:2018mtw,Hoferichter:2019mqg,Davier:2019can,Keshavarzi:2019abf} \\ 
\qquad HVP (NLO)		&   	$-98.3 \pm 0.7$				&\cite{Keshavarzi:2019abf} \\  
\qquad HVP (NNLO)		&  	$12.4 \pm 0.1$				&\cite{Kurz:2014wya} \\ 
\qquad HLbL	 		& 	$94 \pm 19$				&\cite{Colangelo:2014qya,Aoyama:2020ynm} \\
 \hline
Total (theory)			& 	$116\ 591\ 810 \pm 43$		&\cite{Aoyama:2020ynm} \\ 
\hline 
Experiment			& 	$116\ 592\ 089 \pm 63$		&\cite{Bennett:2006fi} \\ 
\hline 
\end{tabular} 
\end{center}
\caption{List of the Standard Model contributions to the anomalous magnetic moment of the muon.}
\label{tab:status}
\end{table} 

\section{Hadronic vacuum polarization \label{sec:lohvp}}

The leading-order hadronic vacuum polarization is the dominant hadronic contribution to the muon anomalous magnetic moment. The corresponding diagram is shown on the left panel of \Fig{fig:diag}. The first lattice calculation started with~\cite{Blum:2002ii} where it was realized that this quantity is accessible via lattice simulations. In recent years, several collaborations, with different lattice discretizations, have published results and, this year, the first publication with sub-percent precision has been presented~\cite{Borsanyi:2020mff}. A further increase in precision is still needed in view of the forthcoming experimental results and, in the next sections, I summarize the current status of lattice results.

\subsection{Lattice determination}

The LO-HVP contribution can be obtained through a convolution integral of the subtracted polarization function $\hat{\PI}$ with a QED weight function over spacelike momenta
\begin{equation}
a_{\mu}^{\rm hvp} = \left( \frac{\alpha}{\pi} \right)^2 \int_0^{\infty} \dd Q^2 K(Q^2) \hat{\PI}(Q^2) \,,
\label{eq:masterhvp}
\end{equation}
\begin{equation}
\hat{\PI}(Q^2) = 4\pi \left[ \PI(Q^2) - \PI(0) \right] \,.
\end{equation}
Here $\alpha$ is the electromagnetic coupling and $K$ is a known analytic QED weight function~\cite{Blum:2002ii} that depends on the lepton mass. 
The polarization function is a scalar function that parametrizes the hadronic vacuum polarization tensor, defined in the continuum through the two-point correlation function of electromagnetic currents
\begin{eqnarray}
\PI_{\mu\nu}(Q^2) &=& \int \dd^4x \, \langle J_{\mu}(x) J_{\nu}(0) \rangle \, e^{i Q\cdot x}  \\
&=& (Q_{\mu} Q_{\nu} - \delta_{\mu\nu} Q^2 ) \PI(Q^2)
\label{eq:vactensor}
\end{eqnarray}
where $J_{\mu} = \frac{2}{3} \ubar \gamma_{\mu} u - \frac{1}{3} \dbar \gamma_{\mu} d - \frac{1}{3} \sbar \gamma_{\mu} s + \frac{2}{3} \cbar \gamma_{\mu} c - \frac{1}{3} \bbar \gamma_{\mu} b$ is the hadronic part of the electromagnetic curent and $Q$ denotes the space-like momentum. The top quark is not shown here: its contribution is known can be estimated using perturbation theory.  The b-quark contribution has been determined by the HPQCD collaboration~\cite{Colquhoun:2014ica} who quote $a_{\mu}^{\rm b} = 0.271(37)\times 10^{-10}$.

In recent lattice calculations, the vacuum polarisation function is obtained from the vector 2-point correlation function at vanishing three-momentum
\begin{equation}
\PI(Q^2) = 4\pi^2 \int \dd t \, G(t) \left[ t^2 - \frac{4}{Q^2} \sin^2\left(\frac{Q t}{2} \right) \right]
\label{eq:tmrPi}
\end{equation}
with
\begin{equation}
G(t) = \frac{1}{3} \sum_{k=1}^3 \int \dd^3 x \, \langle J_{k}(\mathbf{x},t) J_{k}(0) \rangle \,.
\end{equation}
The time momentum representation (TMR)~\cite{Bernecker:2011gh} is then obtained by inverting the integration over imaginary time and momentum in Eqs.~(\ref{eq:masterhvp}) and (\ref{eq:tmrPi})
\begin{equation}
a_{\mu}^{\rm hvp} = \left( \frac{\alpha}{\pi} \right)^2 \int_0^{\infty} \dd x_0 \, \tilde{K}(t) G(t) \,.
\label{eq:lohvp}
\end{equation}
Equivalently, one can start with the definition of the time-moments~\cite{Chakraborty:2014mfa,DellaMorte:2017dyu}
\begin{align}
\Pi_k &= (-1)^{k+1} \frac{G_{2k+2}}{(2k+2)!} \,, \ G_{2k} = 2 \ \int_0^{\infty} \mathrm{d}t \, t^{2k} \, G(t) \,, \label{eq:timemoment}
\end{align}
which are the Taylor coefficients of the vacuum polarization function at vanishing four-momentum transfer squared $Q^2=0$~\cite{DellaMorte:2017dyu}, 
\begin{equation}
\Pi(Q^2) = \Pi_0 + \sum_{k=1}^{\infty} \Pi_k \, Q^{2k} \,.
\end{equation}
From the first few time moments, Pad\'e approximants can be constructed to estimate $\ahvp$~\cite{Chakraborty:2014mwa}. 

\begin{figure}[t]
\begin{center}
\resizebox{0.2\textwidth}{!}{  \includegraphics{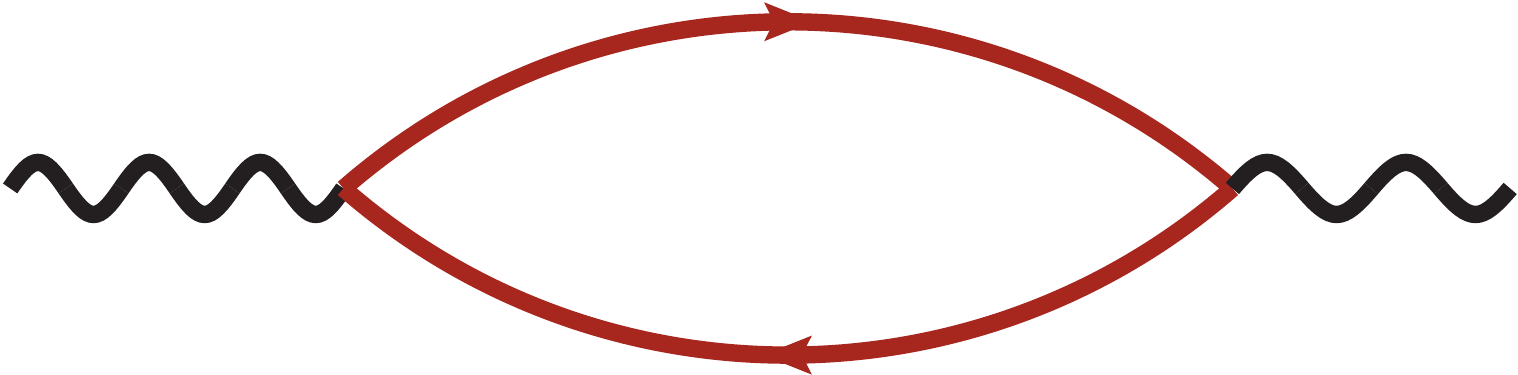}  } \quad
\resizebox{0.2\textwidth}{!}{  \includegraphics{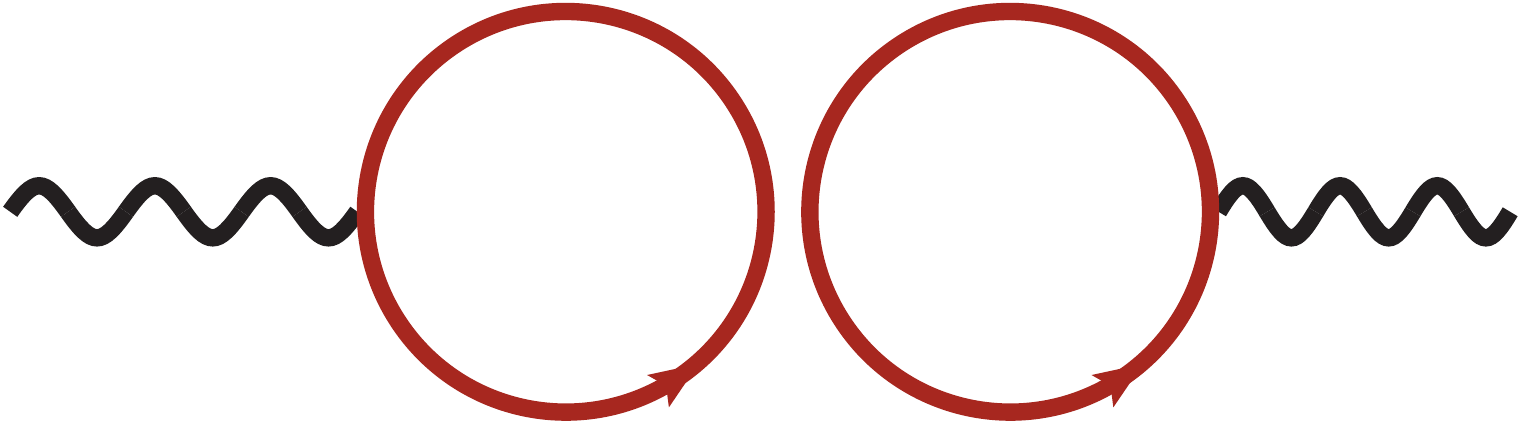}  }
\end{center}
\caption{The quark-connected (left) and quark-disconnected (right) Wick contractions for the LO-HVP in the isospin limit.}
\label{fig:hvp_iso} 
\end{figure}

\begin{figure}[h!]
\resizebox{0.45\textwidth}{!}{  \includegraphics{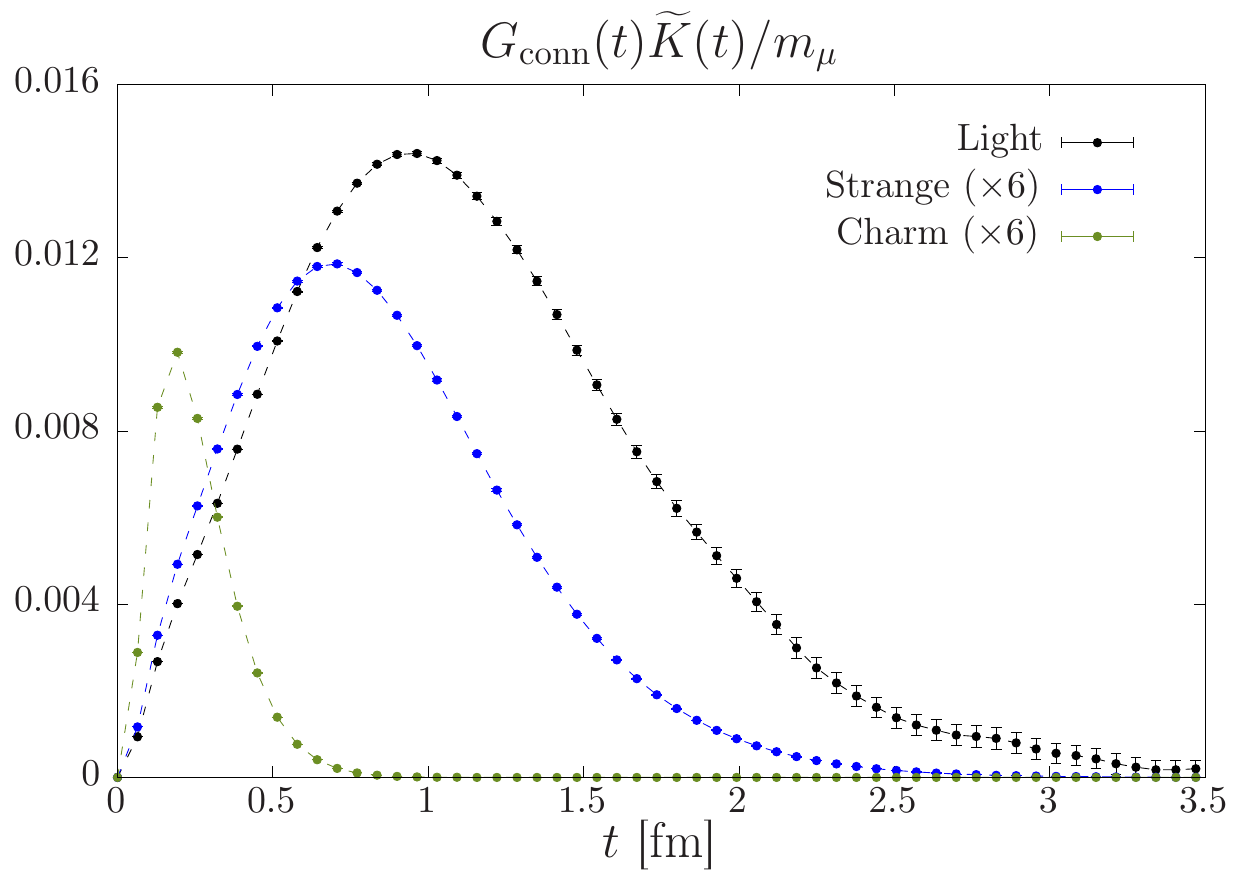}  } 
\caption{Integrand for the light, strange and charm quark contributions as a function of the Euclidean time. Extracted from~\cite{Gerardin:2019rua}. }
\label{fig:hvp_int}
\end{figure}

We first consider the iso-symmetric theory, where the up and down quark masses are degenerate and the electromagnetic effects are neglected. The definition of iso-symmetric QCD depends on the choice made to set the scale and the quark masses in lattice simulations. It will be discussed in \Section{sec:IB} along with isospin breaking corrections. In this approximation, the two-point correlation function can be written as a sum of connected contributions for the light, strange, charm and bottom quarks and a quark disconnected contribution  
\begin{equation}
G(t) = \frac{5}{9} G_l(t) + \frac{1}{9} G_s(t) + \frac{4}{9} G_c(t) + \frac{1}{9} G_b(t) + G_{\rm disc}(t) \,.
\label{eq:dec1}
\end{equation}
The corresponding diagrams are depicted in~\Fig{fig:hvp_iso} and the shape of the integrand is shown on~\Fig{fig:hvp_int} for light, strange and charm quarks. It is also interesting to consider the isospin decomposition 
\begin{equation}
G(t) = G^{I=0}(t) + G^{I=1}(t) 
\label{eq:dec2}
\end{equation}
with 
\begin{multline}
G^{I=1}(t) = \frac{1}{2} G_l(t) \,, \quad G^{I=0}(t) = \frac{1}{18} G_l(t) + \frac{1}{9} G_s(t) \\ + \frac{4}{9} G_c(t) + \frac{1}{9} G_b(t) +  G_{\rm disc}(t) 
\end{multline}
where the iso-vector part contains $9/10$ of the light quark contribution while other $1/10$ belongs to the iso-scalar part. 

A lattice calculation with a sub-percent precision requires to go beyond the iso-symmetric limit of QCD by including QED and strong isospin-breaking corrections due to the mass difference of the up and down quarks. If it is often convenient to separate the calculation into an iso-symmetric part plus isospin breaking corrections, this separation is scheme dependent and a direct comparison between lattice collaborations becomes less trivial for intermediate quantities. This will be discussed in~\Section{sec:IB}. 

\begin{figure}
\begin{center}
\resizebox{0.4\textwidth}{!}{  \includegraphics{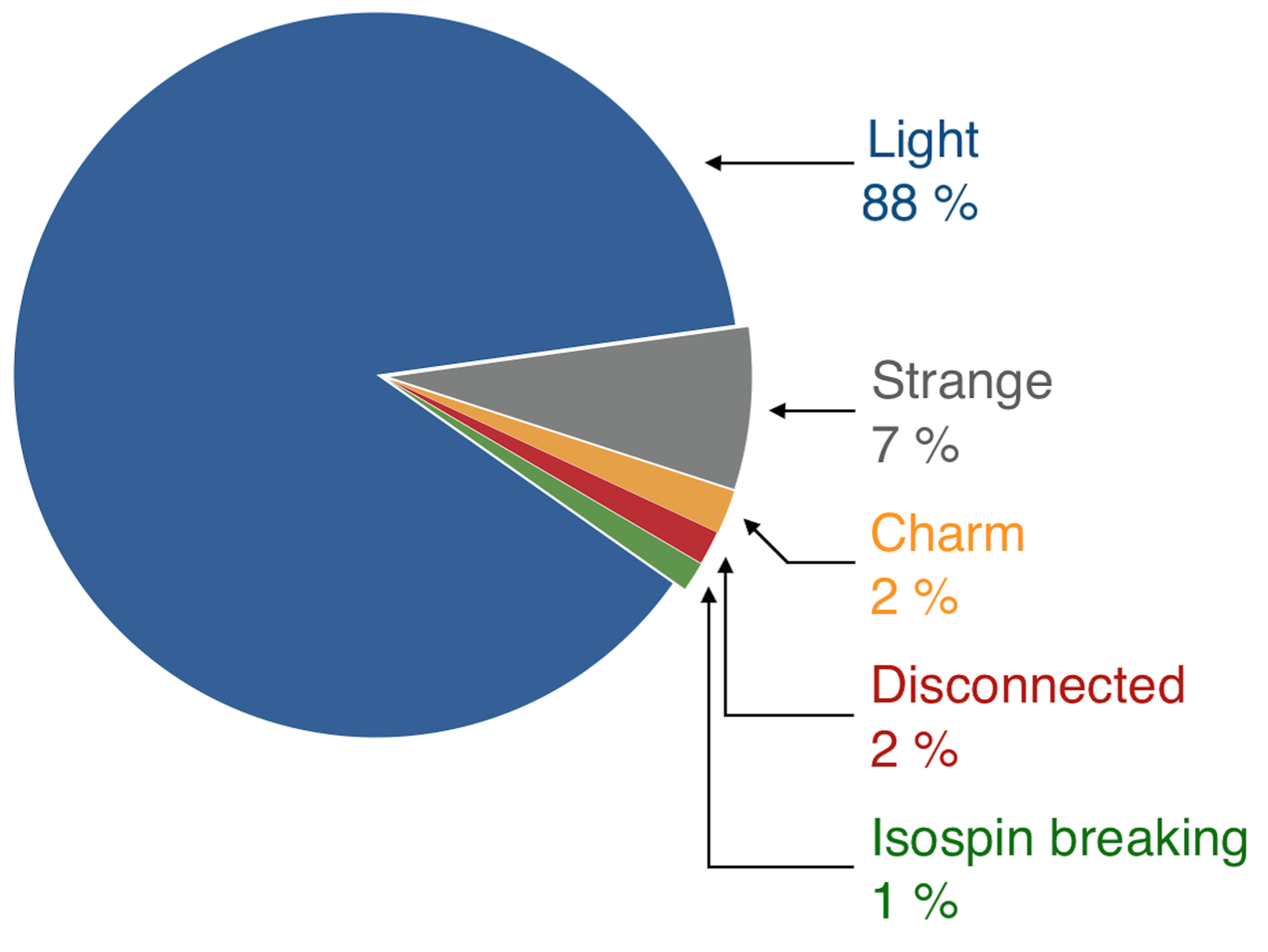}  }
\end{center}
\caption{Different contributions to the leading order HVP in the muon $(g-2)$ as defined in \Eq{eq:dec1}.}
\label{fig:sizes}
\end{figure}

In \Fig{fig:sizes}, I show the relative size of each contribution to the LO-HVP and, in the following sub-sections, I will present the status and the challenges associated with each of them. I first discuss the four contributions in \Eq{eq:dec1} in the isospin limit. Then, I present the status of the isospin breaking corrections before concluding with a comparison between lattice results for the total contribution.

\subsection{The light quark contribution}
\label{sec:light}

The connected light-quark contribution is the dominant contribution to the leading-order HVP and accounts for almost 90\% of the total. It is obtained by inserting $G_l$ from \Eq{eq:dec1} into the master formula~(Eq.~\ref{eq:lohvp}). By definition, we consider the isospin-symmetric  limit and the isospin-breaking corrections will be discussed later in \Section{sec:IB}. Among the main challenges to reach the desired precision are the statistical precision, the control of finite-size effects and a precise scale-setting. Most collaborations are now working with at least one ensemble at (or very close to) the physical pion mass, eliminating the error from the chiral extrapolation. 

\subsubsection{Statistical precision}
\label{sec:statprec}

\begin{figure}[t]
\begin{center}
\resizebox{0.4\textwidth}{!}{  \includegraphics{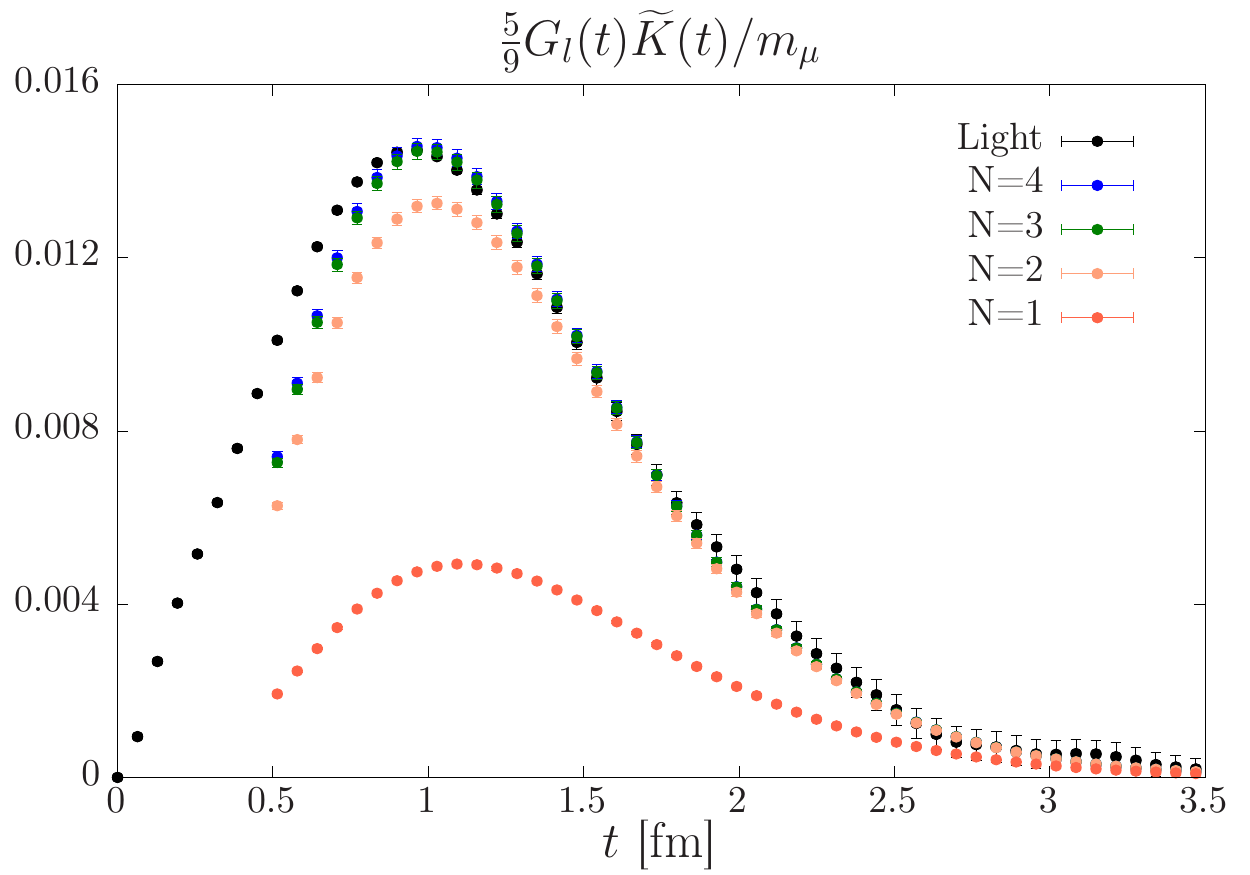}   } \\[3mm]
\resizebox{0.4\textwidth}{!}{  \includegraphics{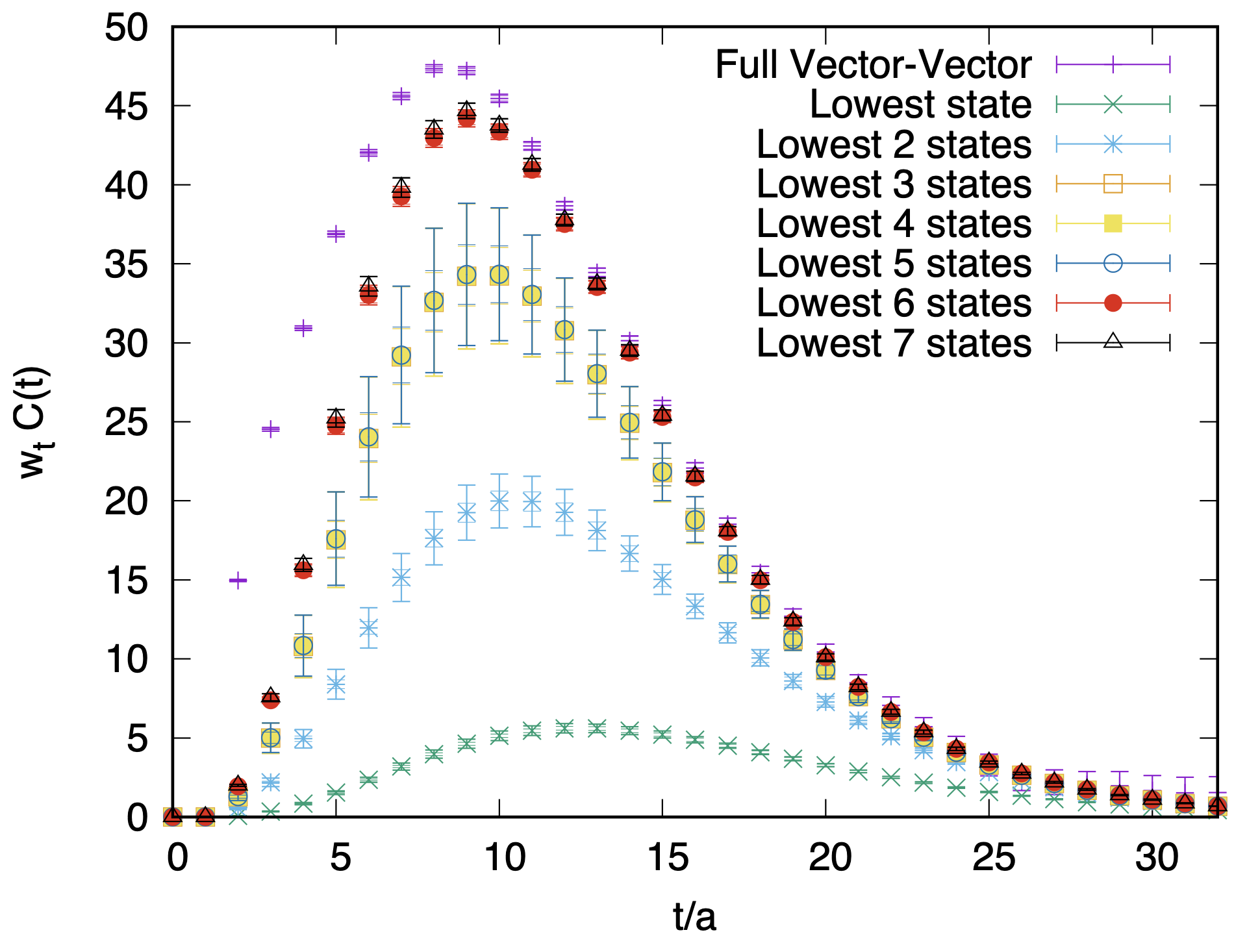}   } 
\end{center}
\vspace{-0.3cm}
\caption{Reconstruction of the vector correlator using the method described in the main text. Top : Mainz~\cite{Gerardin:2019rua} with a pion mass of $200~\MeV$ and $a\approx0.064~\fm$. Bottom : RBC/UKQCD at the physical pion mass~\cite{RBC:seattle} and $a\approx0.114~\fm$. The different curves are obtained by adding more and more states in the spectral decomposition. One can see that more sates are required as le pion mass is lowered.}
\label{fig:tailhvp}      
\end{figure}

The tail of the integrand for the light quark contribution suffers from a noise-to-signal problem. For the vector two-point correlation function, one expects the noise over signal ratio to increases exponentially as $\exp( (m_ V- m_{\pi})t)$ where $m_V$ is the mass of the reseonance and the signal is eventually lost at large Euclidean times. At the physical pion mass, the integrand probes distances above $3~\fm$ (see \Fig{fig:tailhvp}). Methods to overcome the noise problem can be separated into two categories:
\begin{itemize}
\item algorithmic improvements,
\item description of the tail based on theoretical grounds.
\end{itemize}

\begin{figure}[h!]
\resizebox{0.45\textwidth}{!}{  \includegraphics{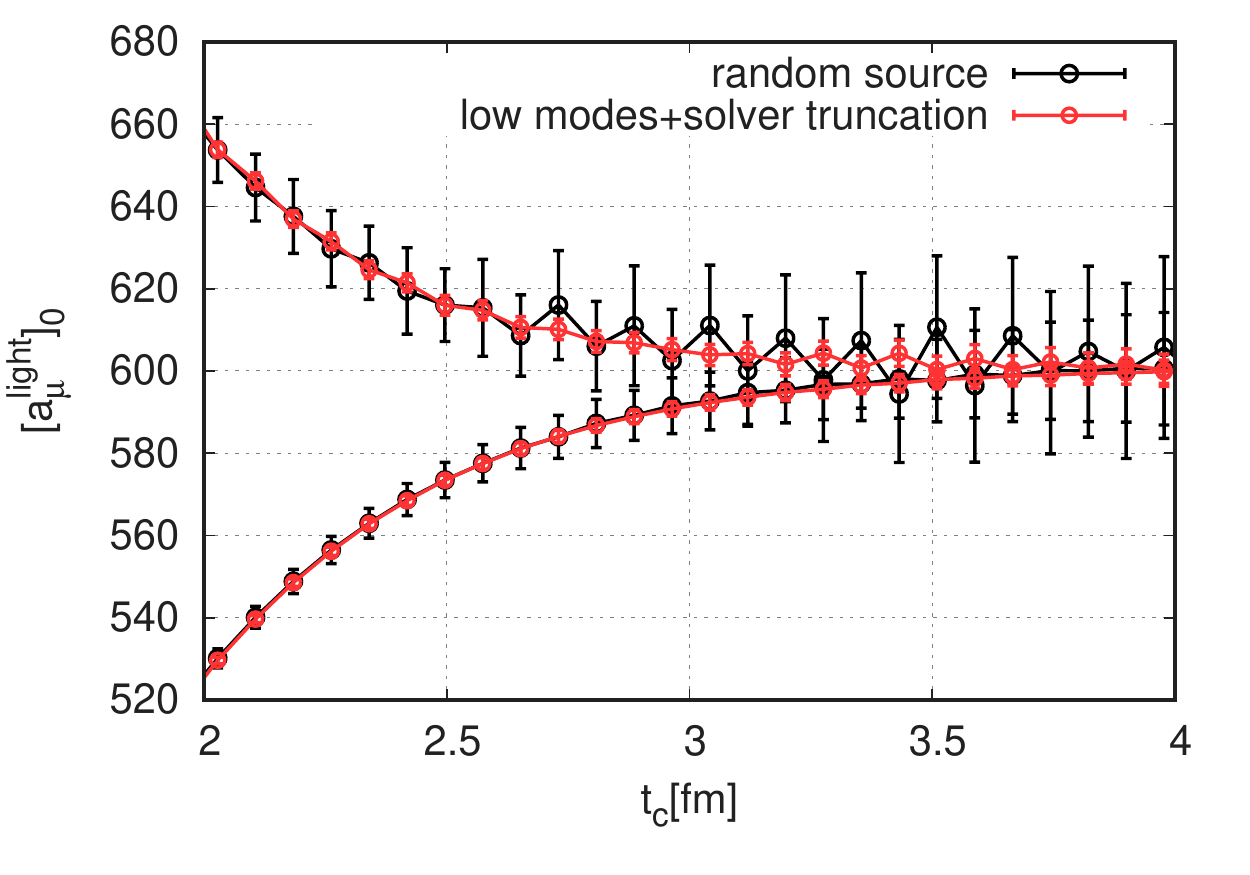}  } 
\caption{Comparison of the low-mode averaging technique combined with all-mode-averaging with the traditional method with stochastic sources. The two sets of curves correspond to the lower and upper bound of the integrand as described in \Eq{eq:bndmeth}. The lattice data are obtained at the physical pion mass. The plot is taken from~\cite{Borsanyi:2020mff}.}
\label{fig:lma}      
\end{figure}

Several algorithmic improvements are used to reduce the cost of the simulation and to increase the statistical precision.  
The low-mode averaging technique~\cite{DeGrand:2004wh,Giusti:2004yp} is used by the BMW~\cite{Borsanyi:2020mff} and the RBC/UKQCD~\cite{Blum:2018mom} collaboration and in Ref.~\cite{Aubin:2019usy}. It consists in calculating exactly the low-modes of the Dirac operator and it provides an exact estimate of the low-part of the all-to-all propagator. A comparison with stochastic sources is shown in \Fig{fig:lma}.
The part orthogonal to these modes is generally evaluated using standard methods such as stochastic sources. This can be combined with the all-mode-averaging (or truncated solver) technique~\cite{Bali:2009hu,Blum:2012uh,Shintani:2014vja}  to reduce further the numerical cost. The latter is also used in~\cite{Lehner:2020crt,Shintani:2019wai} and by the Fermilab/HPQCD/MILC (FHM) collaboration where a reduction of the cost by a factor of 2 has been observed compared to standard stochastic sources~\cite{Davies:2019efs}. 
Recently, a multi-level Monte Carlo integration approach has been proposed to reduce exponentially the variance of the  correlator~\cite{DallaBrida:2020cik,Ce:2016idq,Ce:2016ajy}. 
The application of this method to the LO-HVP has been presented in~\cite{DallaBrida:2020cik} in the case of Wilson-Clover fermions and with a pion mass of 270~MeV where a significant reduction of the error is observed at large Euclidean times as compared to standard HMC simulations.

In addition to algorithmic improvements, a specific treatment of the tail of the integrand is usually needed and the long-distance behavior is treated in a different manner by splitting the integration range at some Euclidean time $t^*$. This separation relies on the observation that, at large Euclidean times, the contribution to the signal comes from a relatively small number of states.

A first strategy is to perform a (multi)-exponential fit to extend the correlator above some Euclidean time $t^*$. Below the cut, the lattice data are used to evaluate \Eq{eq:lohvp} and, above the cut, the approximate correlator, reconstructed from the fit, is used. This strategy is the one followed by the ETM collaboration with the value $t^*$ in the range $[1.6-1.8~\fm]$ and where a single state is included above the cut.
The Fermilab/HPQCD/MILC collaboration also uses this method with the value $t^* \approx 2~\fm$~\cite{Davies:2019efs}. In this case, more than one state is included and additional smeared interpolating operators are used to stabilize the fit. 
This method is expected to work well as long as the resonance state in the vector channel is above the threshold, a valid assumption for large unphysical quark masses or at sufficiently coarse lattice spacing in the staggered formalism. Otherwise, the two-pion like state might be difficult to resolve on the lattice. Since the later is lighter, it might lead to an underestimate of the light quark contribution. In~\cite{Davies:2019efs}, the FHM collaboration has provided evidence that the bias introduced by this method is below their statistical uncertainty.

Another strategy, followed by Mainz~\cite{Gerardin:2019rua,Andersen:2018mau} and more recently by the RBC/UKQCD collaboration~\cite{RBC:seattle}, is to perform a dedicated spectroscopy analysis in the vector channel using a large basis of interpolating operators that couple to single- and multi-particle state and using the L\"uscher formalism~\cite{Luscher:1991cf}. In this way, one can extract the masses and overlaps of the low-lying eigenstates and the correlator for $t>t^*$ can be reconstructed in a systematic way by introducing more and more eigenstates in the lattice spectral decomposition
\begin{equation}
G(t) = \sum_{n=0}^{N-1} \frac{Z_n}{2E_n} e^{-E_n t} \, +\, O(e^{-E_{N} t})\,,
\end{equation}
where $Z_n$ is an overlap factor. The important point is that only a few states are required to saturate the integrand. The overlaps and energies can be determined with relatively high precision and the error on the reconstructed integrand grows linearly with time, solving the noise/signal problem. A reconstruction of the integrand is shown in \Fig{fig:tailhvp} for both collaborations. Recently, the FHM collaboration has presented preliminary result using the staggered formalism~\cite{OnlineWorkshop_FHM_GEVP}. In this case, an additional difficulty comes from the contribution of the 15 copies of the meson with different tastes.

Finally, the bounding method introduced in~\cite{Lehner:bnd,Borsanyi:2017zdw,Blum:2018mom}, or its further improvements, provide a systematic way to cut the integration range in time. The idea is to find a compromise between the statistical precision that increases with $t^*$ and the systematic error introduced by the specific treatment of the tail for $t>t^*$. In practice, one can derive rigorous lower and an upper bounds for the integrand and $t^*$ is chosen such that both  agree within statistical error. In its original formulation, the bounds are
 \begin{equation}
0 \leq G(t) \leq G(t_c) e^{-E_0 (t-t_c)}  \,, \quad t \geq t_c \,.
\label{eq:bndmeth}
\end{equation}
The lower bound is a consequence of the positivity of the correlator. For the upper bound, one notices that the correlator decreases faster than the lowest lying state. Here $E_0$ is the ground state of the spectrum in finite volume and in the isovector channel. More stringent lower bounds have been proposed in the literature~\cite{Blum:2018mom,Gerardin:2019rua} and these methods has been employed  in~\cite{Blum:2018mom,Gerardin:2019rua,Aubin:2019usy,Borsanyi:2020mff,Lehner:2020crt}

\subsubsection{Finite-size effects}

Lattice simulations are inevitably performed in a finite volume and need to be corrected for this. 
In the case of the HVP, one expects the isovector channel to be mostly affected and one usually assumes finite-size effects (FSE) to be negligible in the isoscalar channel, dominated by three-pions states at long distances.  
If FSE are exponentially suppressed with the physical size of the lattice, they are not negligible for typical lattice sizes that are currently used to compute $\ahvp$.
At the physical pion mass and with a physical volume of about 6~fm ($m_{\pi} L \approx 4$), this correction turns out to be at the level of 3\% of the total contribution. Thus, for a few permil precision, this correction needs not only to be included, but also to be known with a relative precision better than 10\%. This is currently one of the most important systematic corrections that need to be applied to lattice data and a careful treatment is thus mandatory for the target precision. Several theory-based approaches have been proposed, with major improvements in the recent years, and the results have been confirmed by direct, large-volume, lattice simulations.

\paragraph{Chiral Perturbation Theory.} 
The natural framework to study finite-size corrections to $\amu$ is Chiral Perturbation Theory (ChiPT). In early calculations, FSE were estimated using NLO ChiPT~\cite{Borsanyi:2017zdw,Blum:2018mom}. 
The FHM collaboration~\cite{Davies:2019efs} uses an extended version of ChPT that also includes the $\rho$ meson and photons~\cite{Gasser:1983yg,Ecker:1988te}. 
More recently, the NNLO correction in ChiPT have been determined in~\cite{Bijnens:2017esv,Aubin:2019usy} and, in a recent paper~\cite{Aubin:2020scy}, the authors have shown that ChiPT can, in principle, be used to compute FSEs at any order in the effective field theory.
The NNLO correction is found to be quite large and of the order of 0.4–0.45 times the NLO correction at physical pion mass and for spatial volumes in the range $L=5.4-5.8~\fm$. In~\cite{Aubin:2019usy}, the systematic error is obtained by assuming a similar relative correction for the next order. It corresponds to a 15\% uncertainty. It is interesting to note that it translates into a 0.5\% error on the total contribution, underlining the importance of this correction for a few permil precision.

\begin{figure}[h!]
\begin{center}
\resizebox{0.41\textwidth}{!}{  \includegraphics{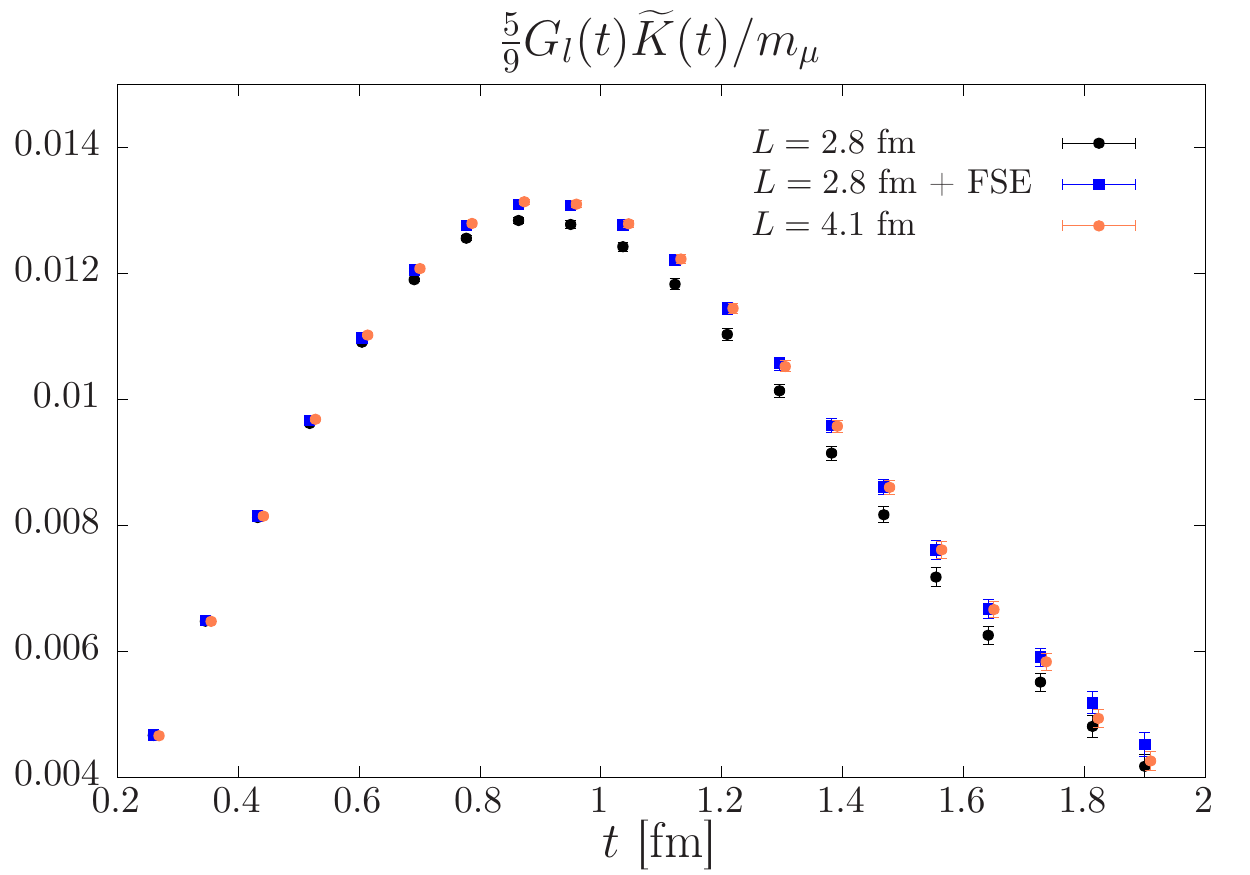}  } 
\end{center}
\caption{Test of the finite-size correction procedure using the time-like pion form factor and the Luescher formalism. The pion mass is 280 MeV and two volumes with $L=2.8~\fm$ and $L=4.1~\fm$ are used (it corresponds to $m_{\pi} L \approx 4$ and 6 respectively). Figure extracted from~\cite{Gerardin:2019rua}.}
\label{fig:fse_light}   
\vspace{-1cm}   
\end{figure}

\paragraph{Time-like pion form factor and the L\"uscher formalism.} 
A second approach is based on the L\"uscher formalism and the knowledge of the time-like pion form factor. In this method, the isovector part of the correlation function is evaluated in both finite and infinite volumes using a spectral decomposition. The time-like pion form factor enters in the determination of the finite-volume matrix elements. The theoretical framework has been derived in~\cite{Luscher:1991cf,Lellouch:2000pv,Bernecker:2011gh} and has been used for the first time in~\cite{DellaMorte:2017dyu}. This method is used by the Mainz collaboration~\cite{Gerardin:2019rua}, where the pion form factor is obtained from a dedicated lattice calculation~\cite{Andersen:2018mau}. The procedure has been explicitly tested on ensembles with a pion mass of $280~\MeV$ as can be seen on \Fig{fig:fse_light}. In~\cite{Giusti:2018mdh}, the ETM collaboration uses a similar strategy to estimate FSE: they assume a Gounaris-Sakurai parametrization of the pion form factor and the associated parameters $M_{\rho}$ and $\Gamma_{\rho}$ are obtained from a fit to the lattice data. This strategy is embedded in the so-called \textit{analytical representation} of the correlator that also includes a description of the correlator at short distances. In their recent publication~\cite{Borsanyi:2020mff}, the BMW collaboration has also studied this method assuming a Gounaris-Sakurai parametrization of the pion form factor  with parameters estimated from phenomenology. They obtained results perfectly compatible with their direct lattice calculation using large volumes. 

\paragraph{Hansen-Patella method.}  
More recently, a new method has been proposed where finite-size effects are expressed in terms of the forward Compton amplitude of the pion~\cite{Hansen:2019rbh} in an expansion in $\exp[-|\vec{n}| m_{\pi}L]$. The first publication was restricted to the dominant $\exp[-M_{\pi} L]$ contribution and the sub-dominant contributions were neglected. However, the latter are numerically relevant and this limitation has been overcome in a more recent publication~\cite{Hansen:2020whp,APLAT20:Patella} where sub-leading terms $\exp[- \sqrt{2} M_{\pi} L]$  and $\exp[- \sqrt{3} M_{\pi} L]$ have been included. Numerically, one observes a nice convergence of this expansion. This methods has been employed on lattice data for the first time by the BMW collaboration~\cite{Borsanyi:2020mff}. The results are found to be numerically compatible with NNLO ChPT and the previous method based on the time-like pion form factor. Interestingly, this method also provides the leading correction for the finite time-extent of the lattice.

\paragraph{Direct lattice calculation.}  

\begin{figure}[b]
\begin{center}
\resizebox{0.41\textwidth}{!}{  \includegraphics{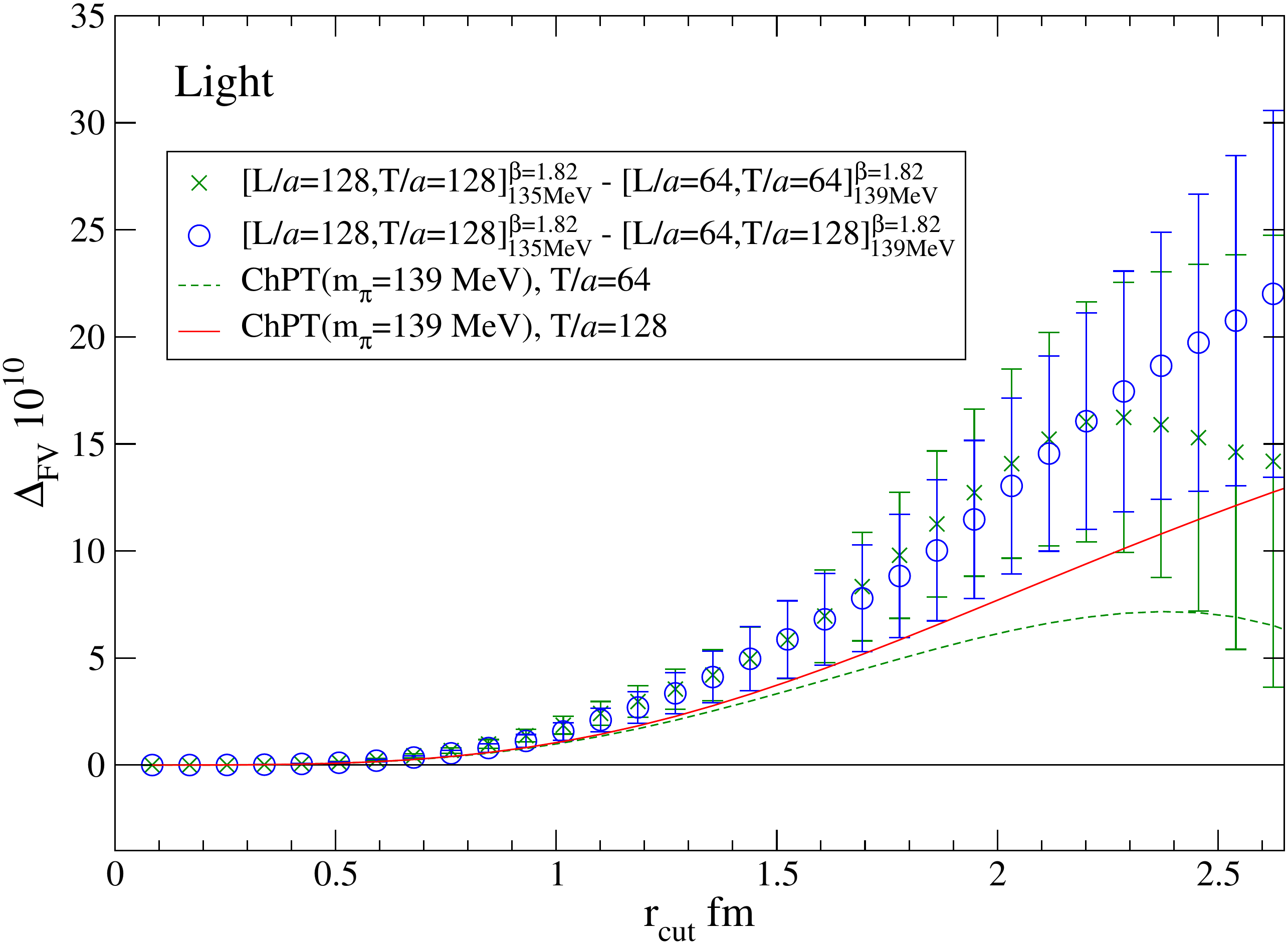}  } 
\end{center}
\caption{Difference of the light quark contribution $\ahvp(r_{\rm cut})$ on three lattices with $64^4$, $128^4$ and $64^3 \times 128$ (the spatial extents in physical units are $L=5.4~\fm$ and $L=10.8~\fm$ and the pion mass is $m_{\pi} \approx 135~\MeV$. The plot is taken from~\cite{Shintani:2019wai}.}
\label{fig:fselat}      
\end{figure}

An other approach is to perform dedicated lattice calculations using small and large physical volumes at the same bare lattice parameters. The main challenge of such calculations is to get a signal at large Euclidean times where FSEs are most important. Due to the large volumes required by this method, it is numerically expensive even if a relatively coarse lattice spacings can be used. 
The PACS collaboration have presented results using two volumes of $L=5.4~\fm$ and $L=10.8~\fm$~\cite{Shintani:2019wai}. On the larger volume, ChiPT can be used to estimate the small persistant FSE. As can be seen on Fig.~\ref{fig:fselat}, their result differs by about 1 standard deviation with the NLO ChiPT estimate, confirming the prediction obtained from other approaches, while with relatively big uncertainties of about 30\%. 
A similar work has been presented by RBC/UKQCD \cite{Lehner_hvp_fse} where they found a correction of $21.6(6.3) \times 10^{-10}$ at the physical pion mass between two volumes $L=6.22~\fm$ and  $L=4.66~\fm$. This correction is almost twice as big at the correction obtained using NLO ChiPT. 
This method is also the one used by the BMW collaboration where a lattice with $L\approx10.7~\fm$ was used~\cite{Borsanyi:2020mff}. They found a correction of $18.7(2.5) \times 10^{-10}$ at the physical pion mass as compared to $L=6.3~\fm$. This observation is in line with previous studies but with higher precision.

Most lattice collaborations correct lattice data for FSE on each ensemble prior to the continuum (and chiral) extrapolation. When using staggered quarks, finite-size effects are entangled with the taste-breaking corrections and therefore with the continuum extrapolation. 
This motivates the BMW collaboration to first compute $\ahvp$ in the continuum limit, at a reference volume with $L\approx 6.3~\fm$ and, in a second step, to  correct this continuum extrapolation for FSEs.

To conclude, three theory-based approaches are known to estimate FSEs. Remarkably, they provide very similar estimates suggesting that this correction is under control at the level of $10\%$. This has been further confirmed by the recent large-volume calculation performed by the BMW collaboration. FSEs are also important for the quark disconnected contribution, as discussed in Section~\ref{sec:dischvp}, and for the isospin-breaking correction discussed in Section~\ref{sec:IB}. %

\subsubsection{Chiral extrapolation/interpolation}

The chiral extrapolation is still a significant source of error for lattice collaborations that are not working directly at the physical pion mass. In Ref.~\cite{Gerardin:2019rua}, the Mainz group observes a significant enhancement of the signal for the light quark contribution between the ensemble at the physical pion mass and another ensemble with a 220~MeV pion mass. The ETM collaboration does not yet include ensemble directly at the physical pion mass~\cite{Giusti:2018mdh}. However, such an ensemble is now available and has been used in a recent paper on the ratios of the hadronic contributions to the lepton $g-2$~\cite{Giusti:2020efo}. For collaborations working close to the physical pion mass~\cite{Borsanyi:2017zdw,Blum:2018mom,Shintani:2019wai,Davies:2019efs,Aubin:2019usy}, a smooth interpolation is required and does not introduce a significant error. Fortunately, it concerns most collaborations and this issue has become less relevant in recent analysis.
 
\subsubsection{Lattice spacing and continuum extrapolation}
\label{sec:scaleset}

Although the anomalous magnetic moment is a dimensionless quantity, the weight function $\widetilde{K}$ in \Eq{eq:lohvp} depends on the muon mass that eventually needs to be converted into lattice units. To study the impact of the scale setting uncertainty, the authors of~\cite{DellaMorte:2017dyu} have estimated the relative error on $\ahvp$ from error propagation
\begin{equation}
\mathrm{\Delta} \ahvp = \left| a \frac{ \dd \ahvp }{ \dd a } \right| \cdot \frac{ \mathrm{\Delta} a }{ a } = \left| m_\mu \frac{ \dd \ahvp }{ \dd m_\mu } \right| \cdot \frac{ \mathrm{\Delta} a }{ a } \,,
\end{equation}
and obtained $\mathrm{\Delta} \ahvp / \ahvp \approx 1.8\%$. Therefore, a two-permil precision for $\ahvp$ translates into a permil determination of the scale setting. This is probably one of the biggest challenge for lattice QCD and requires the inclusion of isospin-breaking effects. It should be noted that this relation has been derived in the time-momentum representation given by \Eq{eq:lohvp} and might differ if one uses a different estimators. A review of scale setting can be found in~\cite{Sommer:2014mea}.

A continuum extrapolation is also required and this step depends significantly on the action and the choice of operators that are used in lattice simulations. 
All results presented in \Fig{fig:cmp_light} are obtained using lattice formulations where $O(a)$ lattice artifacts have been removed and the leading discretization errors are therefore expected to be $O(a^2)$. 
Recently, the relevance of log-corrections for high precision calculations as been emphasized in~\cite{Husung:2019ytz}. 
In simulations performed with staggered quarks, the leading source of discretization effects are due to taste-breaking effects. Taste splitting is generally corrected for using staggered chiral perturbation theory (S$\chi$PT)~\cite{Davies:2019efs,Borsanyi:2020mff,Aubin:2019usy}. 
The continuum extrapolation turns out to be the main source of systematic error in the recent publication by the BMW collaboration~\cite{Borsanyi:2020mff}. Obviously, at least three lattice spacings should be used to check the expected scaling. 

\subsubsection{Results and conclusion}

\begin{figure}[h!]
\begin{center}
\vspace{-0.2cm}
\resizebox{0.4\textwidth}{!}{  \includegraphics{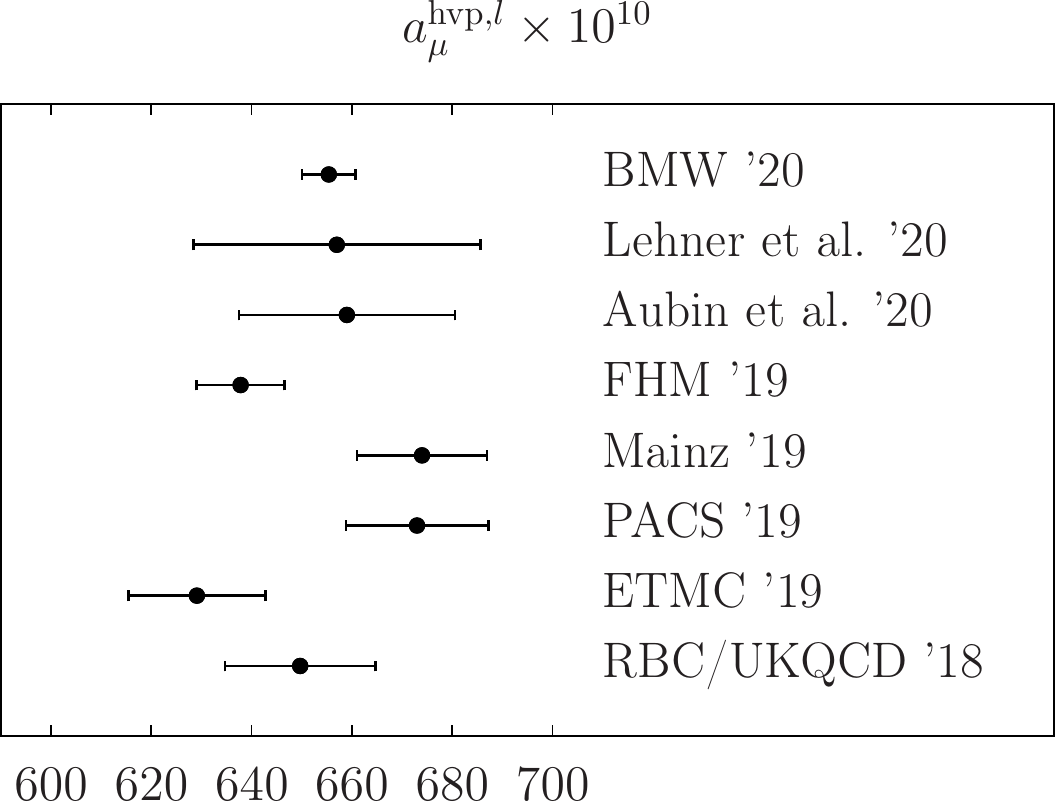}  } 
\end{center}
\caption{Comparison of lattice results for the connected light quark contribution to the leading-order HVP. The results are extracted from~\cite{Borsanyi:2020mff,Lehner:2020crt,Aubin:2019usy,Davies:2019efs,Gerardin:2019rua,Shintani:2019wai,Giusti:2019hkz,Giusti:2019xct,Blum:2018mom}. The BMW point includes the FSE correction for the light quark contribution quoted in~\cite{Borsanyi:2020mff}.}
\label{fig:cmp_light}      
\end{figure}

A summary of the results for the light-quark contribution, at the physical point, in the continuum limit and in infinite volume, but in the isospin-symmetric limit, is depicted in \Fig{fig:cmp_light}. We observe some tension between different collaborations. Since the systematic error is not negligible and is correlated between different collaborations (especially for the FSE correction), further work is needed here. Possible cross-checks among lattice groups but also with phenomenology will be further discussed in \Section{sec:bench}.

\subsection{The strange and charm quark contributions}

\begin{figure}[h!]
\begin{center}
\vspace{-0.5cm}
\resizebox{0.4\textwidth}{!}{  \includegraphics{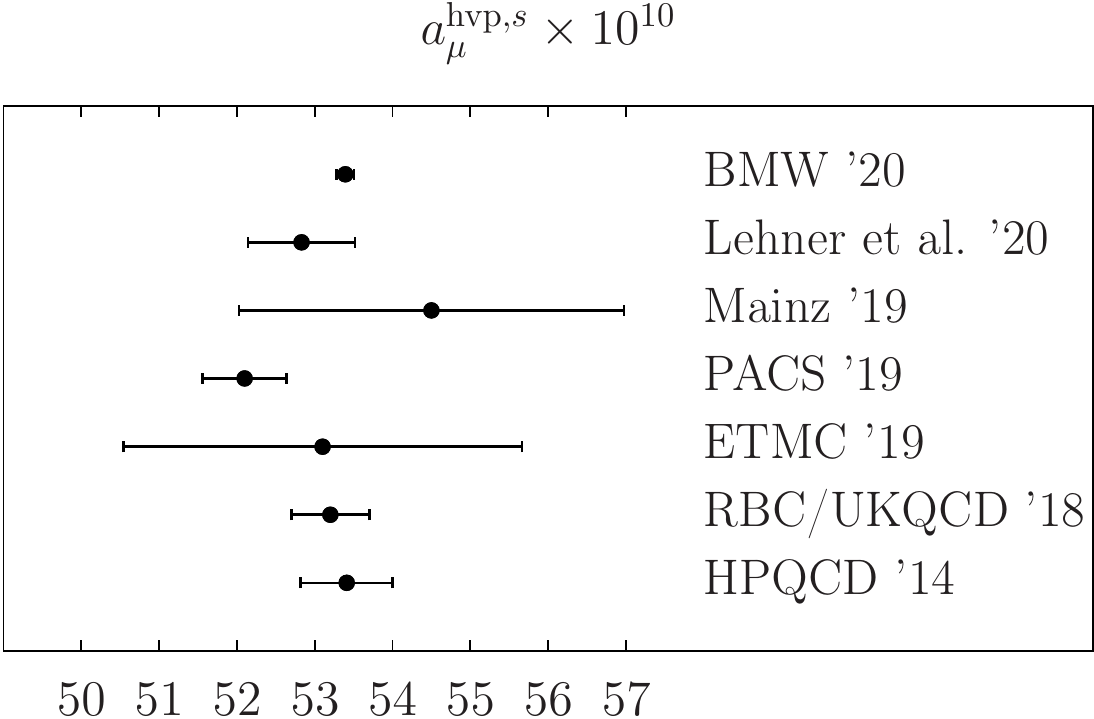}  } 
\end{center}
\caption{Comparison of lattice results for the connected strange quark contribution to the leading-order HVP. The results are extracted from~\cite{Borsanyi:2020mff,Gerardin:2019rua,Shintani:2019wai,Giusti:2019xct,Blum:2018mom,Chakraborty:2014mfa}.}
\label{fig:amu_strange}      
\end{figure}

The strange quark contribution is much simpler to evaluate on the lattice. As can be seen in \Fig{fig:hvp_int}, the integrand decreases faster than the light quark contribution and there is no noise problem at large Euclidean times. No sophisticated treatment of the tail is needed and a high statistical precision is easily achieved by integrating the lattice data only. Finite-size effects are much smaller than for the light-quark contribution and can be safely neglected~\cite{Chakraborty:2014mwa,Giusti:2017jof}. The results for different lattice collaborations are depicted in \Fig{fig:amu_strange} and we observe a nice agreement between all of them. 
A significant source of uncertainty still lies in the determination of the scale setting, as discussed in \Section{sec:scaleset}. It mostly explains the spread of the errors and can be improved independently. An other source of systematic uncertainty originates from the tuning the of the strange quark mass at the physical point. 
The strange-quark contribution amount to about 8\% of the total LO-HVP and a relative precision of 2.5\% is required for a two permil precision. This is already achieved by many of the collaborations and this is not the main challenge for the future. 

\begin{figure}[b]
\begin{center}
\resizebox{0.4\textwidth}{!}{  \includegraphics{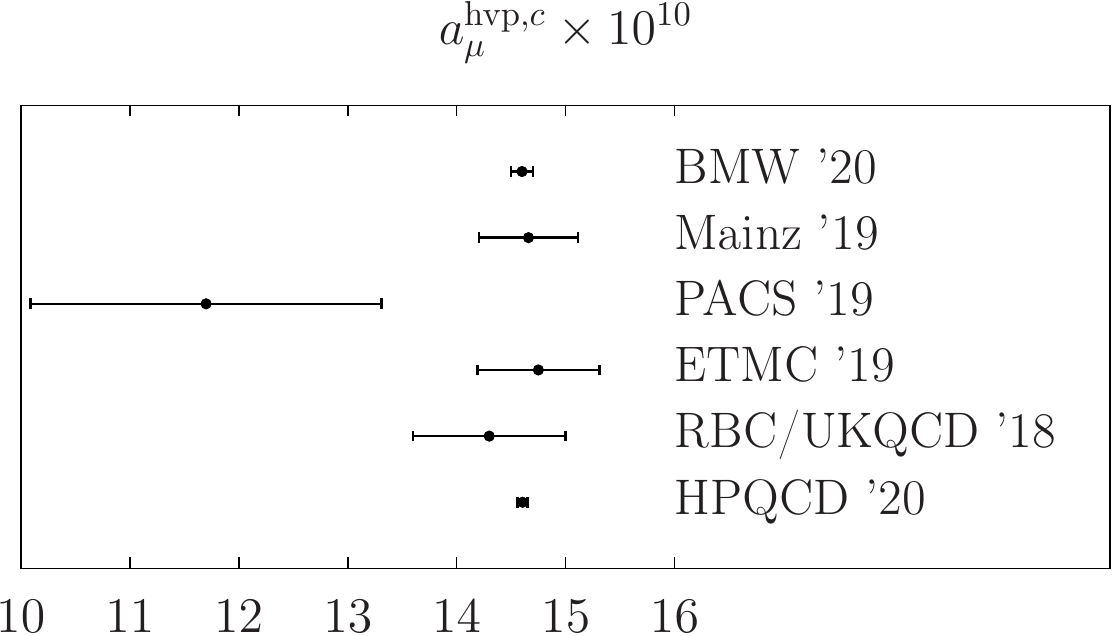}  } 
\end{center}
\caption{Comparison of lattice results for the connected charm quark contribution to the leading-order HVP. The results are extracted from~\cite{Borsanyi:2020mff,Gerardin:2019rua,Shintani:2019wai,Giusti:2019xct,Blum:2018mom,Hatton:2020qhk}.}
\label{fig:amu_charm}      
\end{figure}

The integrand of the charm quark contribution is peaked at even smaller Euclidean times. As for the strange quark contribution, a high statistical precision is easily achieved. In that case, the main difficulty lies in the continuum extrapolation and the control of discretization effects. As can be seen in \Fig{fig:hvp_int}, it becomes difficult to sample the integrand correctly and small lattice spacings are required. The results for various collaborations are shown in \Fig{fig:amu_charm} and, as for the strange quark contribution, we observe a very good agreement between groups. This contribution is about 2\% of the total contribution. For most collaborations, the current uncertainty is already below the required precision for a two permil determination and does not represent a challenge for the future target precision.

\subsection{The quark-disconnected contribution}
\label{sec:dischvp}

Thanks to the structure of the electromagnetic current, the quark-disconnected contribution to the LO-HVP contribution can be factorized as~\cite{Francis:2014hoa}
\begin{equation}
G_{\rm disc}(t) = - \frac{1}{9} \frac{1}{3} \sum_{k=1}^3 \big\langle \left( L^l_k(t) - L^s_k(t) \right)  \left(L^l_k(0) - L^s_k(0)\right)  \big\rangle \\ 
\label{eq:disc}
\end{equation}
where the loop on a given timeslice and with a quark of flavor $f$ is defined by
\begin{equation}
L^{f}_{k}(t) =  \frac{1}{\sqrt{V}} \sum_{\vec{x}} \mathrm{Tr}  \left[ S_f(x,x) \gamma_{k} \right]  \,.
\label{eq:loopf}
\end{equation}
with $S_f$ the  quark propagator. Strictly speaking, Eq.~(\ref{eq:loopf}) holds only for Wilson fermions with a local current and needs to be adapted to other discretizations. 
Here, we assume $m_u=m_d$ and isospin-breaking corrections will be discussed later in Section~\ref{sec:IB}. We also neglect the valence charm quark contribution that has been shown to be much smaller than the current statistical precision~\cite{Borsanyi:2017zdw}. 
From \Eq{eq:disc}, it is clear that this contribution vanishes in the SU(3)$_f$ limit where $m_u=m_d=m_s$. 

\begin{figure}[t!]
\begin{center}
\resizebox{0.45\textwidth}{!}{  \includegraphics{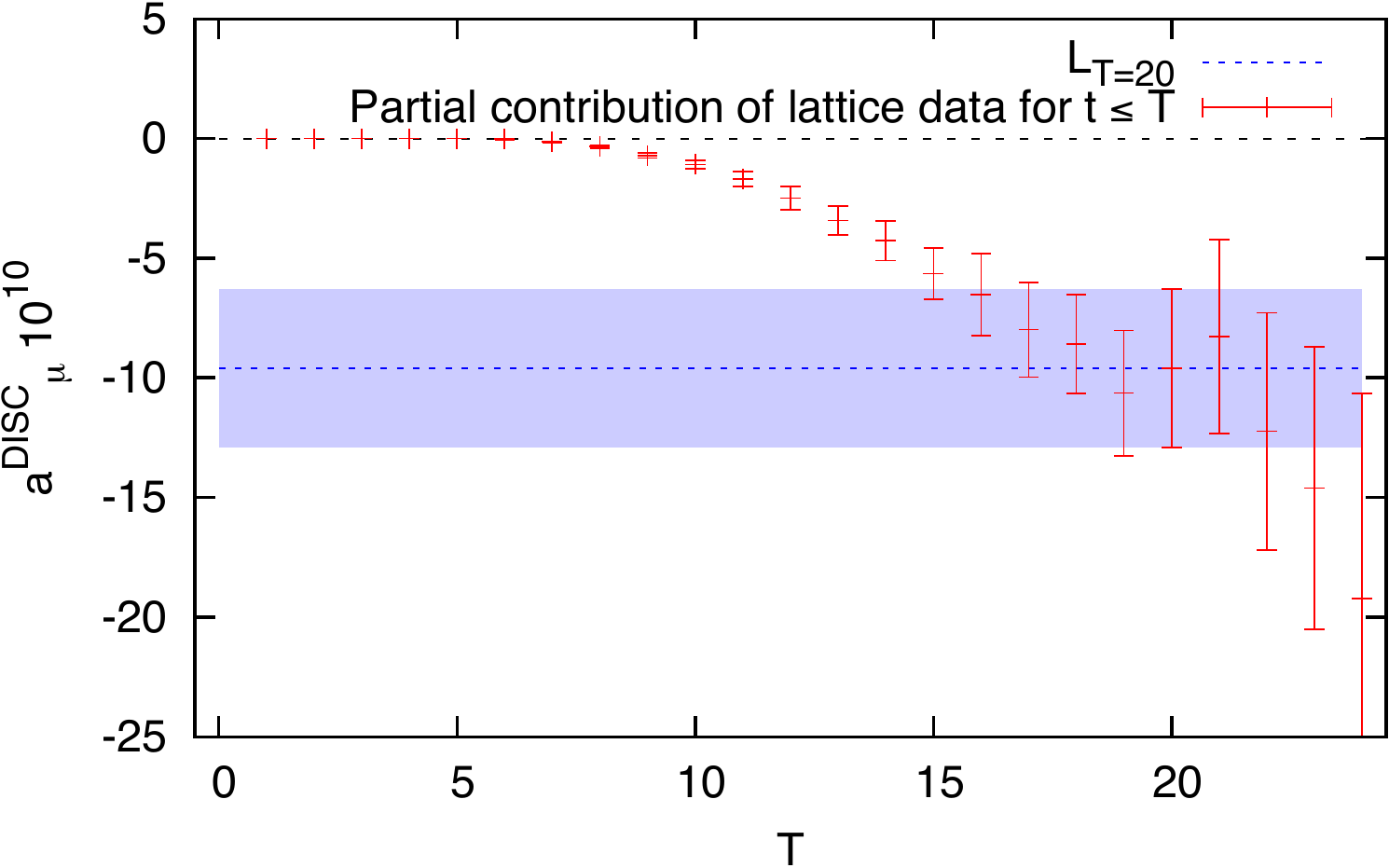}  } 
\end{center}
\caption{Integrated value of $a_{\mu}^{\rm hvp, disc}$ obtaining using \Eq{eq:lohvp} and integrating the lattice data up to $t\leq T$. Extracted from~\cite{Blum:2015you}.}
\label{fig:int_disc}      
\end{figure}

The calculation of the loop functions requires the estimation of all-to-all propagators $S_f$ on each time-slice. The latter are notoriously difficult to estimate in lattice QCD and require the use of sophisticated noise-reduction techniques.
Since this contribution can be expressed as a difference between light and strange loops, we expect the signal to be dominated by low energy physics. 
In particular, it is important that noise reduction techniques maintain the light- minus strange structure to benefit from the cancellation of noise between the light and strange-quark contributions. This is the case of the low-mode averaging technique used by the RBC/UKQCD and BMW collaborations where the low-modes are computed for eigenvalues up to (or close to) the strange quark mass. In practice, the low-mode averaging described in \Section{sec:statprec} is often combined with the all-mode-averaging (or truncated solver) technique~\cite{Bali:2009hu,Blum:2012uh,Shintani:2014vja} to estimate the stochastic part of the estimator. 
In~Ref.~\cite{Gerardin:2019vio}, the Mainz group uses hierarchical probing~\cite{Stathopoulos:2013aci},  which replaces the sequence of noise vectors by one noise vector multiplied with a sequence of Hadamard vectors. The same noise vectors are used for both the light and strange quark inversions. 
In any case, the signal is eventually lost at large Euclidean times $t\approx 2 - 2.5~\fm$ and, as for the light quark contribution, the bounding method can be used to cut the integration range (Fig.~\ref{fig:int_disc}). 

Finite-size effects are also relevant for this contribution. If one assumes FSEs to be negligible in the iso-scalar channel, the quark-disconnected contribution, taken separately, receives the same FSE correction as the light quark contribution but with a factor $-1/10$. At the physical pion mass and with a volume of 5.5~fm, they are about $15\%$ for the RBC/UKQCD collaboration. The systematic error associated to this correction is, however, negligible and under control.

\begin{figure}[h!]
\vspace{-0.2cm}
\begin{center}
\resizebox{0.4\textwidth}{!}{  \includegraphics{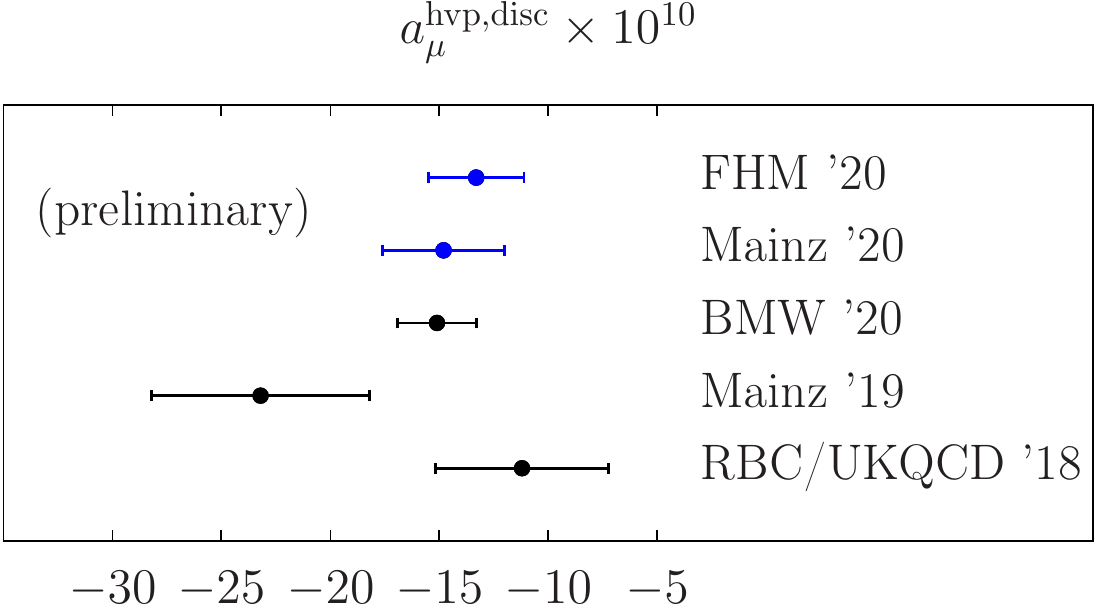}  } 
\end{center}
\caption{Comparison of lattice results for the quark-disconnected contribution, to the leading-order HVP, in the isospin limit. The results are extracted from~\cite{Borsanyi:2020mff,Gerardin:2019rua,Blum:2018mom}. The BMW point includes the FSE correction for the light quark contribution quoted in~\cite{Borsanyi:2020mff}. New preliminary results, in blue, have been presented during the workshop \textit{The hadronic vacuum polarization from lattice QCD at high precision}, 16-20 November 2020.}
\label{fig:amu_disc}      
\vspace{-0.6cm}
\end{figure}

The quark-disconnected contribution has been computed by fewer collaborations, including RBC/UKQCD~\cite{Blum:2015you,Blum:2018mom}, BMW~\cite{Borsanyi:2017zdw,Borsanyi:2020mff} and the Mainz group~\cite{DellaMorte:2017dyu,Gerardin:2019rua}. An early calculation has been presented by the the HPQCD collaboration in~\cite{Chakraborty:2015ugp}, but on a single ensemble with a heavy pion of $391~$MeV. The published results read
\begin{eqnarray*}
\mathrm{BMW} &: \, &a_{\mu}^{\rm hvp,disc} =-13.2(1.3)_{\rm stat}(1.3)_{\rm syst}  \times 10^{-10} \\ 
\mathrm{Mainz}  &: \, &a_{\mu}^{\rm hvp,disc} = -23.2(2.2)_{\rm stat}(4.5)_{\rm syst}  \times 10^{-10} \\
\mathrm{RBC} &: \, &a_{\mu}^{\rm hvp,disc} = -11.2(3.3)_{\rm stat}(2.3)_{\rm syst} \times 10^{-10} 
\end{eqnarray*}
and are summarized in \Fig{fig:amu_disc}. This contribution is negative and, in magnitude, amounts to about $2\%$ of the total contribution. A precision below 10\% is therefore needed to reach the two permil precision. 

\begin{figure}[t!]
\begin{center}
\resizebox{0.45\textwidth}{!}{  \includegraphics{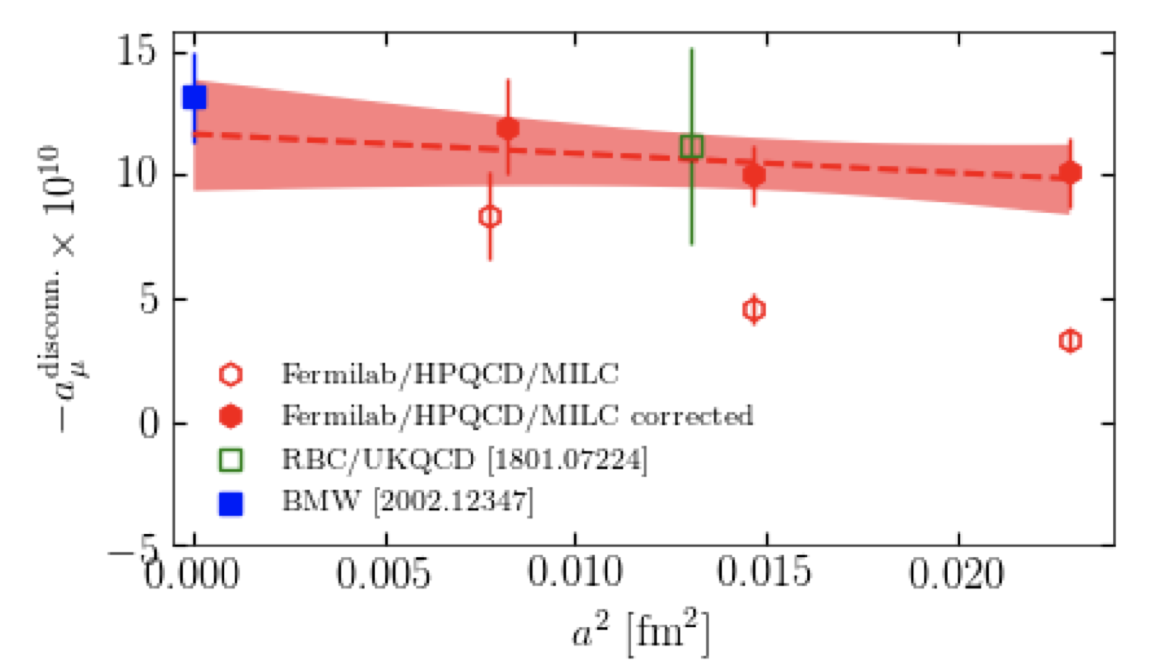}  } 
\end{center}
\caption{Preliminary summary of the disconnected results for the FHM collaboration. Slides presented during the workshop \textit{The hadronic vacuum polarization from lattice QCD at high precision} by the Muon g-2 Theory Initiative~\cite{OnlineWorkshop_FHM}.}    
\label{fig:discFHM}
\end{figure}

Recently, the FHM collaboration~\cite{OnlineWorkshop_FHM} has presented preliminary results shown in \Fig{fig:discFHM}. They are obtained using three lattice spacings at the physical pion mass and the extrapolation is in good agreement with previous results.

We observe a slight tension of about $2\sigma$ between the published Mainz results and the other collaborations. The significance of this tension should however be tempered by noticing that most results are dominated by systematic error and, for the RBC collaboration, a single lattice spacing with $a = 0.11~\fm$ was used. The results of the Mainz group was obtained on a reduced set of ensembles that includes 3 lattice spacings but no ensemble directly at the physical pion mass such that a chiral extrapolation was required and turned out to be the main source of uncertainty (since the isoscalar contribution is not singular in the chiral limit, the disconnected piece has to compensate the singularity present in the light quark contribution~\cite{Gerardin:2019vio}). The Mainz group has recently presented a preliminary update~\cite{OnlineWorkshop_Mainz} that includes more ensembles, including one at the physical pion mass and the new noise reduction technique proposed in~\cite{Giusti:2019kff}. Their preliminary results $a_{\mu}^{\rm hvp,disc} =-14.8 \pm 2.8  \times 10^{-10}$ reduces the tension. The BMW collaboration works directly as the physical pion mass but did not include their finest lattice spacing. 

The precision of the BMW collaboration has reached 10\% and recent new calculations confirm the overall size of this contribution. Further cross-checks on this important contribution would be highly desirable and a first step might be to compare the results obtained from the window method described in \Section{sec:window}. 

\subsection{Isospin-breaking corrections}
\label{sec:IB}

Most lattice simulations are performed in the iso-symmetric limit where $m_u = m_d$. However, the current level of precision of the HVP contribution requires the inclusion of isospin breaking effects. There are two sources of isospin-breaking corrections in nature: first the mass of the up and down quarks are measured to be different, second the electromagnetic corrections due to the different electric charges of the quarks. These effects are expected to be of order $O(\alpha) \approx 1\%$ and $O((m_d - m_u)/\Lambda_{\rm QCD}) \approx 1\%$, respectively. Recent calculations have confirmed that this contribution is below 1\% (see Table~\ref{tab:ib}). 

Those calculations are numerically challenging. Numerous diagrams need to be evaluated and a good statistical precision, at large Euclidean times, is difficult to achieve. Similar methods as the one discussed for the light quark contribution can be used. Naively, finite-size effects can also be large: corrections are expected to fall off as powers of $1/L$, where $L$ is the spatial extent of the lattice. However, in~\cite{Bijnens:2019ejw}, the authors derived analytic expressions for the electromagnetic finite-volume correction to the two-pion contribution to the LO-HVP where it was shown that the leading term is $1/L^3$. For typical large volumes with $m_{\pi}L >4$, this correction is shown to be at the level of a few percent and would thus be negligible for a few per-mil accuracy in the LO-HVP. 
The scale-setting, which is one of the limiting factor in term of precision, as discussed in \Section{sec:scaleset}, needs to include IB corrections as well. Furthermore, if the vector current needs to be renormalized, isospin corrections need to be included there as well.

After a brief overview of the methods used to include iso-spin breaking effects, I present recent results. Until recently, only a sub-part of the diagrams, expected to be dominant, have been estimated. In particular, the ETM and the RBC/UKQCD collaboration have worked within the electro-quenched approximation. Recently, the BMW collaboration has published a first estimate of all diagrams including electromagnetic sea-quark effect~\cite{Borsanyi:2020mff}.

\subsubsection{Methods}

To include QED on the lattice, we are interested in evaluating the path integral in presence of both QCD and QED
\begin{equation}
\left<O\right> = \frac{1}{Z}\! \int \mathcal{D}[U]\,\mathcal{D}[A]\,\mathcal{D}[\psib,\psi] \, O \, e^ { -S[U,A,\psib,\psi]} \,,
\end{equation}
with the total action 
\begin{equation}
S[U,A,\psib,\psi] = S_g[U]+S_{\gamma}[A] + \sum_{f} \psib_f D[U,A;e,q_f,m_f] \psi_f
\end{equation}
where $S_g$ and $S_{\gamma}$ are the gluon and photon actions and $D$ is the Dirac operator associated to  the quark of flavor $f$. The latter includes coupling to photons and it depends on the quark mass $m_f$, the quark electric charge $e_f$ and the gluonic coupling $g$. The first step is to regularize the photon action on the lattice and the associated difficulties are discussed in detail in~\cite{Borsanyi:2014jba,Patella:2017fgk}. All results for the HVP are obtained using the QED$_L$ prescription~\cite{Hayakawa:2008an} where the spatial zero modes of the photon field in finite volume are explicitly removed. 

Two methods are used to include QED effects in $\ahvp$ using lattice simulations.

The first methods treats QED non-perturbatively. It consists in generating gauge configurations for the full QED+QCD theory~\cite{Duncan:1996xy}. Most simulations are performed in the electro-quenched approximation where the the sea-quarks are considered as electrically neutral. This approximation allows to generate the U(1) photon fields independently of the SU(3) gauge fields and then reduces the computational cost. For the strong-isospin-breaking effects, different values for the quark masses can be used in the action. Here, the results are obtained to all orders in $\alpha$ (or within the electro-quenched approximation). For the HVP, this method has been studied by the RBC collaboration~\cite{Boyle:2016lbc} and the QCDSF collaboration~\cite{APLAT20:QCDSF}.

\begin{figure}[t!]
\begin{center}
\resizebox{0.12\textwidth}{!}{  \includegraphics{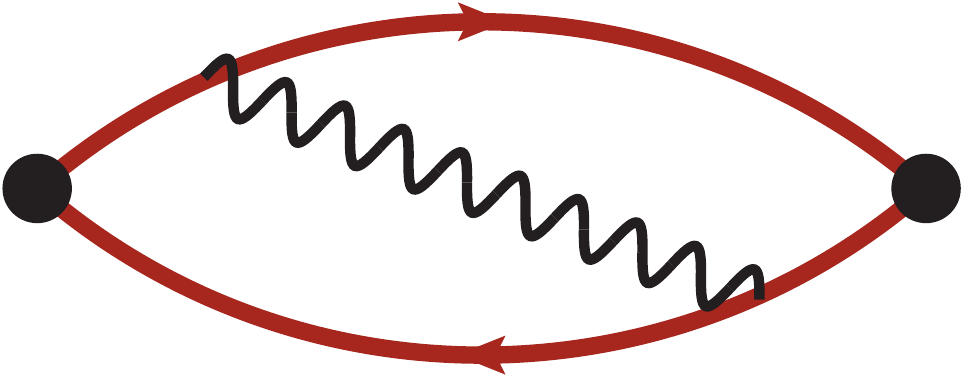}  }  \qquad
\resizebox{0.12\textwidth}{!}{  \includegraphics{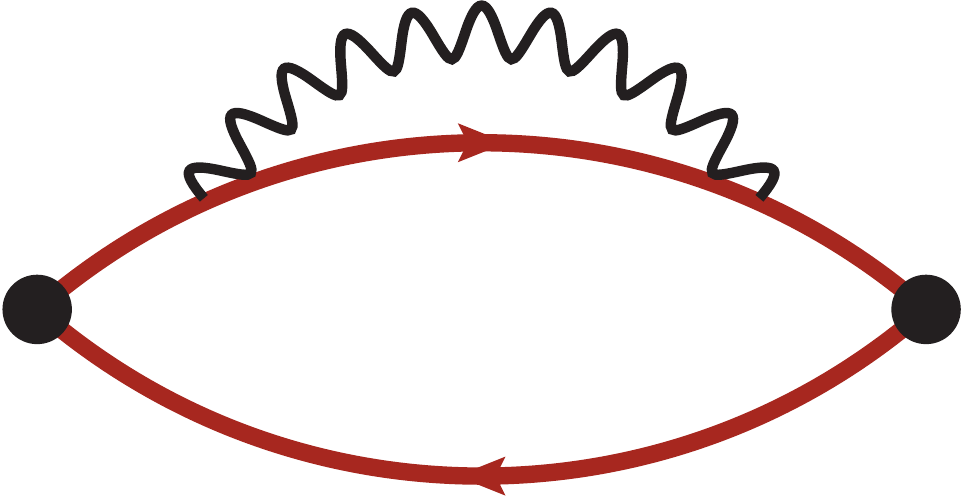}  } \qquad 
\resizebox{0.12\textwidth}{!}{  \includegraphics{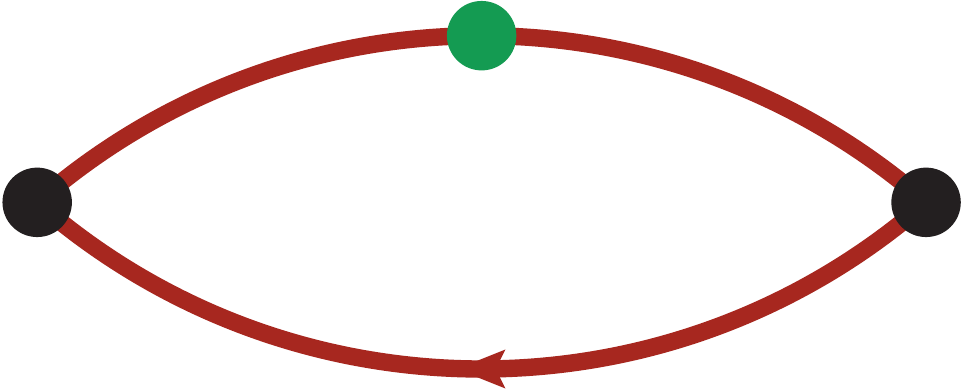}  } \\[7mm]

\resizebox{0.12\textwidth}{!}{  \includegraphics{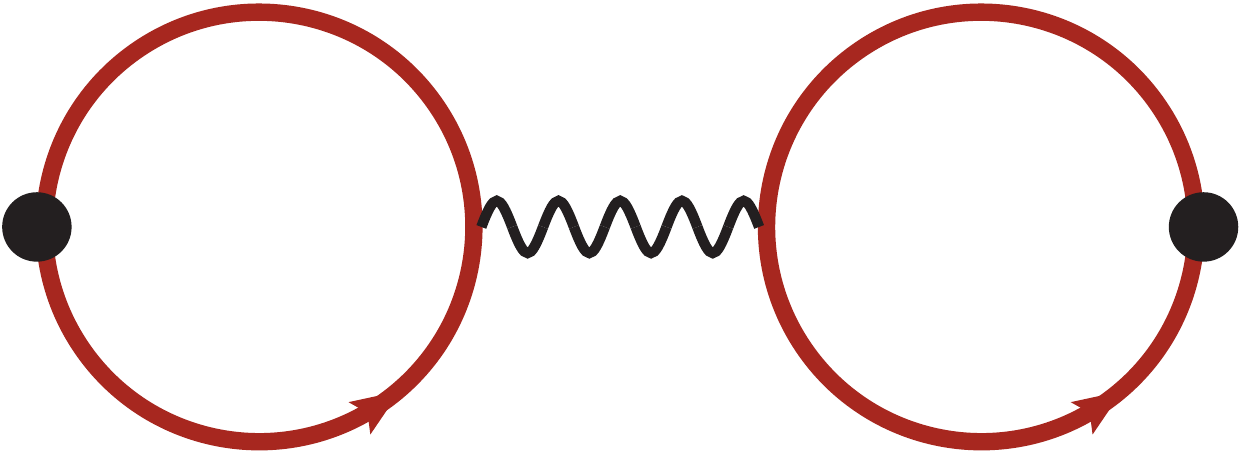}  } \qquad
\resizebox{0.12\textwidth}{!}{  \includegraphics{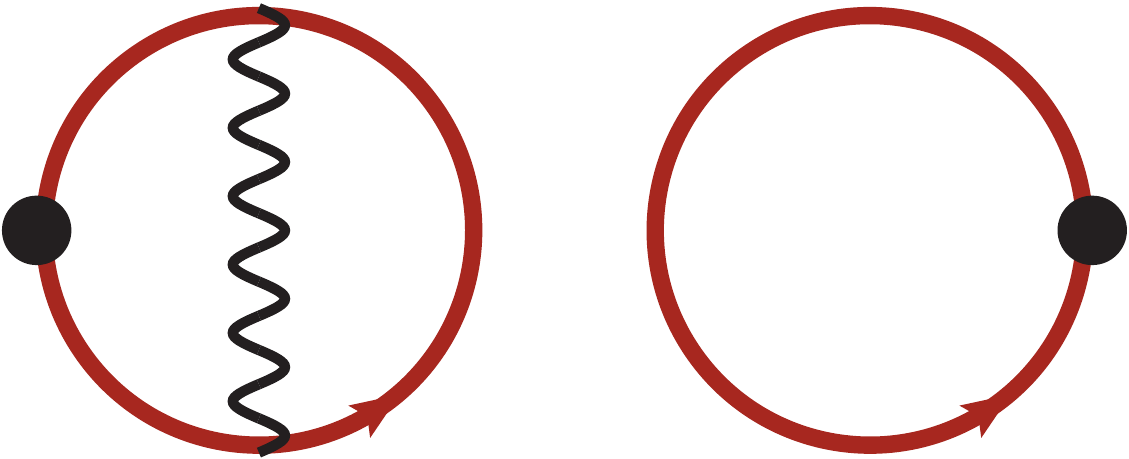}  } \qquad 
\resizebox{0.12\textwidth}{!}{  \includegraphics{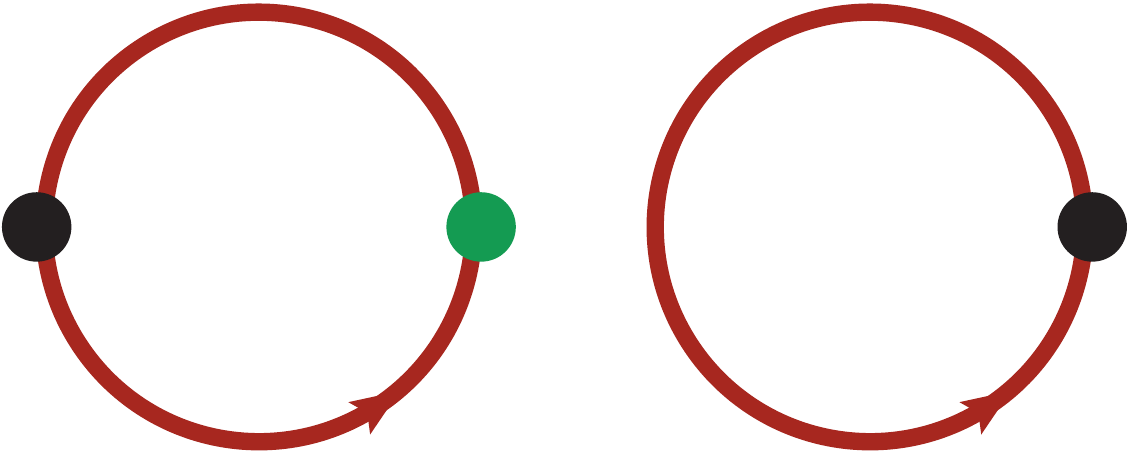}  } 
\end{center}
\caption{Top: Dominant QED and strong isospin-breaking diagrams for the quark-connected (top panel) and the quark-disconnected (bottom panel) contributions.}
\label{fig:ib1}      
%
\begin{center}
\resizebox{0.12\textwidth}{!}{  \includegraphics{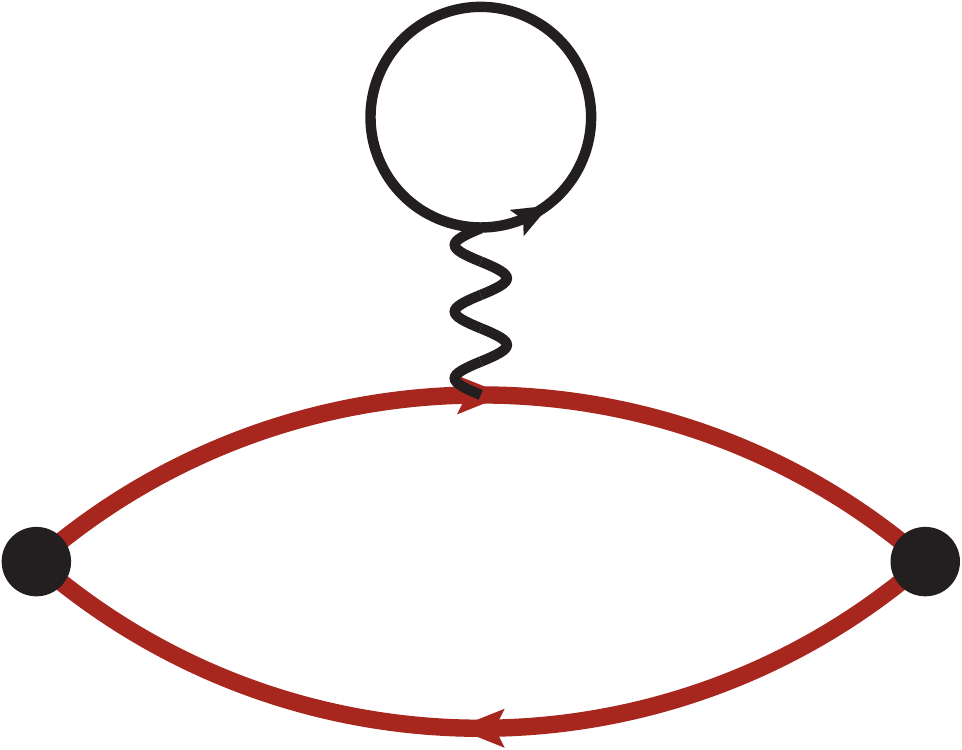}  } \qquad
\resizebox{0.12\textwidth}{!}{  \includegraphics{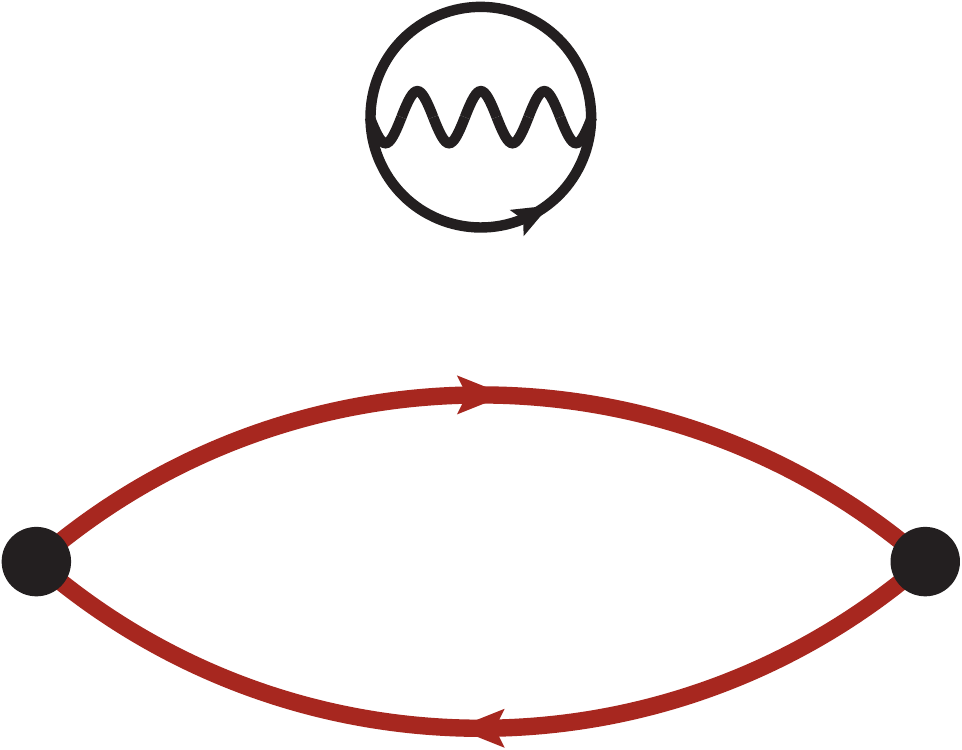}  } \qquad
\resizebox{0.12\textwidth}{!}{  \includegraphics{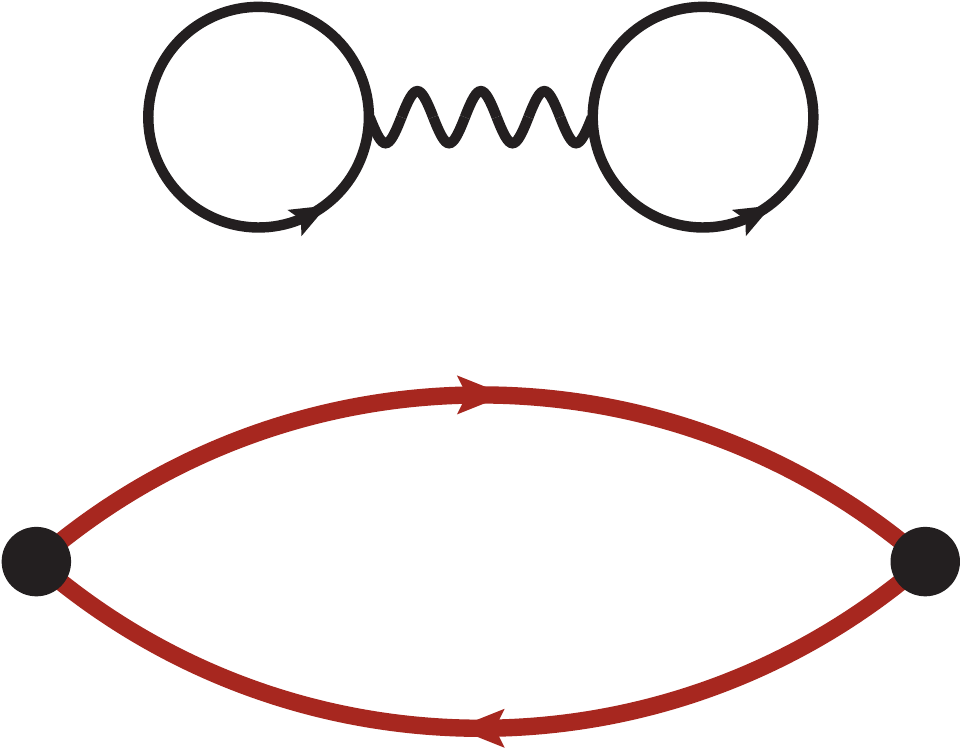}  } \\[7mm]

\resizebox{0.12\textwidth}{!}{  \includegraphics{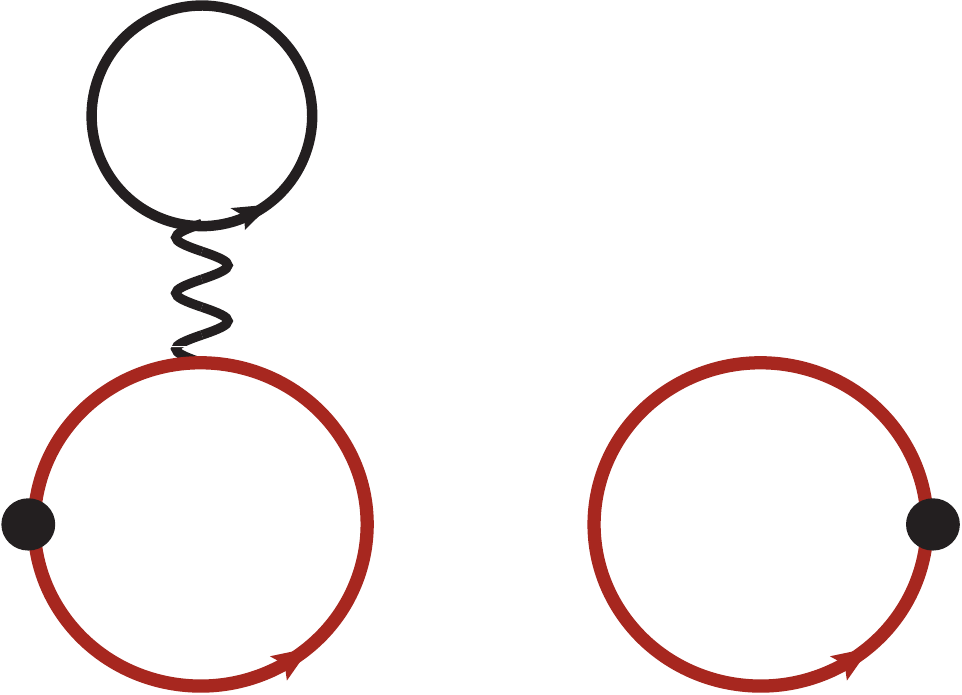}  } \qquad
\resizebox{0.12\textwidth}{!}{  \includegraphics{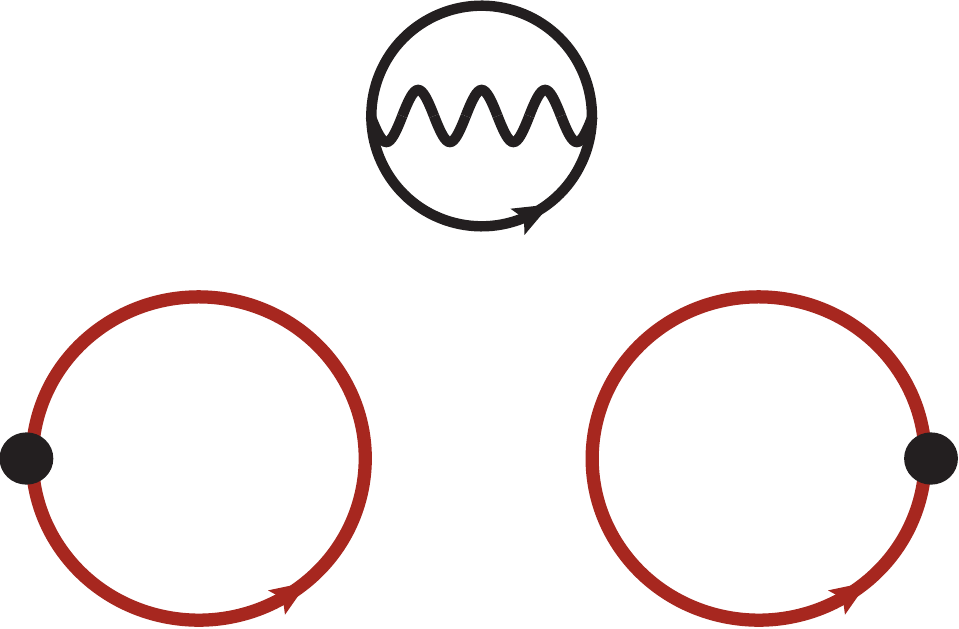}  } \qquad
\resizebox{0.12\textwidth}{!}{  \includegraphics{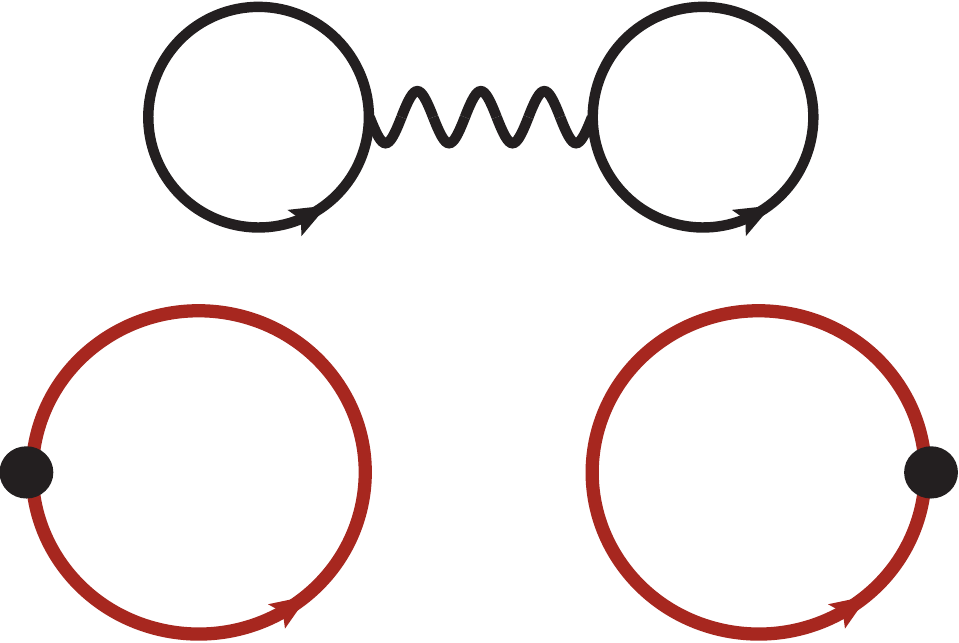}  } 
\end{center}
\caption{Diagrams beyond the electro-quenched approximation for the quark-connected (top panel) and the quark-disconnected (bottom panel) contributions (diagrams are $1/N_c$ suppressed).} 
\label{fig:ib2}      
\end{figure}

The second approach to compute isospin breaking effects is based on the perturbative expansion of the path integral in powers of the fine structure constant, $\alpha$, and the mass difference of the quark compared to the mass in the iso-symmetric theory $\Delta m_f = (m_f - \overline{m})$~\cite{deDivitiis:2011eh,deDivitiis:2013xla}. Since it is a perturbative expansion around the iso-symmetric theory, the same SU(3) gauge ensembles can be used.
The relevant diagrams for the HVP are shown in Figs.~(\ref{fig:ib1}) and (\ref{fig:ib2}). Additional diagrams, that are specific to the regularization used on the lattice, are not shown. Corrections to the connected and disconnected parts have been separated. The second set of plots include diagrams beyond the electro-quenched approximation. This methods is used by the ETM~\cite{Giusti:2017jof,Giusti:2019xct}, RBC/UKQCD~\cite{Blum:2018mom} and BMW~\cite{Borsanyi:2020mff} collaborations
  
A comparison between the two methods has been presented by the RBC/UKQCD collaboration in \cite{Boyle:2016lbc}.

\subsubsection{Separation prescription}

\begin{table*}
\renewcommand{\arraystretch}{1.3}
\begin{center} 
\begin{tabular}{l@{\hskip 0.2in}c@{\hskip 0.2in}c@{\hskip 0.2in}c@{\hskip 0.2in}c@{\hskip 0.2in}c}
\hline 
Figure		&	BMW~\cite{Borsanyi:2020mff}		&	ETMC~\cite{Giusti:2017jof,Giusti:2019xct} 		&	FHM~\cite{Chakraborty:2017tqp}			&	RBC/UKQCD~\cite{Blum:2018mom}	&	QCDSF~\cite{Westin:2019tgc,APLAT20:QCDSF}		\\ 
\hline 
(\ref{fig:ib1}) - top - {\scriptsize  QED}	& 	$-1.27(40)(33)$		&	1.2(1.0)		&	$\times$		&	 5.9(5.7)			\\
(\ref{fig:ib1}) - top - {\scriptsize  SIB}		&	6.59(63)(53) 		&	6.0(2.3)		&	$9.4(4.5)$			&	10.6(4.3)(6.6)		\\
(\ref{fig:ib1}) - bottom - {\scriptsize  QED}		&	$-0.55(15)(11)$		&	$\times$		&	$\times$		&	$-6.9(2.1)(1.4)$	\\
(\ref{fig:ib1}) - bottom - {\scriptsize  SIB}		&	$-4.63(54)(69)$		&	$\times$		&	$\times$		&	$\times$	\\
(\ref{fig:ib2})	&	0.37(20)(19)		&	$\times$		&	$\times$		&	$\times$		 \\
\hline
Charm				&	$\times$	&	$-0.0344(21)$	&	$\times$	&	$\times$	&	$\times$	\\
\hline 
Total			&		0.46(91)(98)			&	7.1(2.9)		&	$9.4(4.5)$		&	9.5(10.2)	&	$<1\%$	\\
\hline
\end{tabular} 
\end{center}
\caption{Summary of the results for different lattice collaborations. We quote separately the connected (\Fig{fig:ib1} - top) and disconnected connected (\Fig{fig:ib1} - bottom) contributions in the electro-quenched approximation from the other sub-leading diagrams of \Fig{fig:ib2}. As explained in the text, those numbers are scheme dependent and a direct comparison between lattice collaborations is not trivial.} 
\label{tab:ib}
\end{table*} 

The separation between strong isospin breaking and QED effects is prescription dependent. Furthermore, the definition of the physical point in the iso-symmetric theory is also scheme dependent. Only the full contribution, in QCD + QED simulations can be unambiguously defined. In that sense, some care needs to be taken when comparing the results from different lattice calculations, as in \Table{tab:ib}. However, the ambiguity between different schemes is expected to be small ($\mathcal{O}(\alpha m_f)$ with $m_f$ the quark mass) compared to the current statistical precision. Further details on the prescriptions used by each group can be found in~\cite{Aoyama:2020ynm} and references therein.

\subsubsection{Results}

\begin{figure}[b]
\vspace{-0.4cm}
\begin{center}
\resizebox{0.45\textwidth}{!}{  \includegraphics{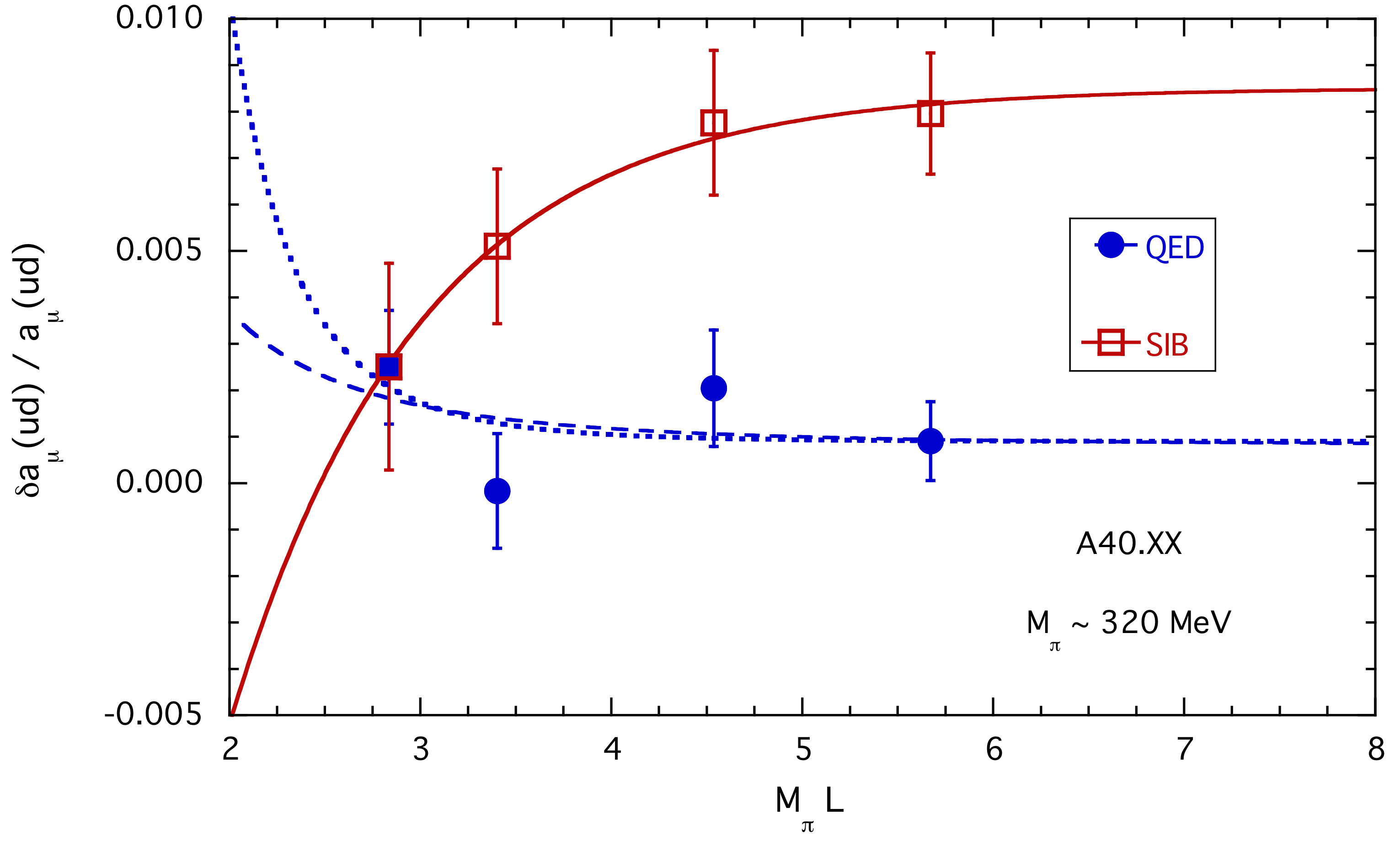}  } 
\end{center}
\vspace{-0.4cm}
\caption{QED and strong-isospin corrections to $\ahvp$ for five ensembles with the same bare lattice parameters but different volumes. The red line corresponds to a fit of the form $A+Be^{-M_{\pi} L}$ and the blue line to a fit of the form  $C+D/L^n$ with $n=3$ (dashed line) or $n=6$ (dotted line). Figure extracted from~\cite{Giusti:2019hkz}}
\label{fig:fse_IB}      
\end{figure}

The first set of diagrams, depicted in the top panel of \Fig{fig:ib1}, represents the correction to the quark-connected contribution in the electro-quenched approximation, where QED corrections for the sea-quarks are not taken into account. The right most diagram is the strong isospin correction while the other two diagrams are QED corrections. Those diagrams have been computed by the ETM~\cite{Giusti:2019xct}, the RBC/UKQCD~\cite{Blum:2018mom} and the BMW~\cite{Borsanyi:2020mff} collaborations.

The second set of diagrams in \Fig{fig:ib1} (bottom panel) has been computed by both the RBC/UKQCD~\cite{Blum:2018mom} and the BMW~\cite{Borsanyi:2020mff} collaborations. They correspond to corrections to the quark-disconnected contribution, in the same electro-quenched approximation. The RBC/UKQCD collaboration has evaluated only the first diagram of \Fig{fig:ib2}, that is expected to be dominant.

Finally,  beyond the electro-quenched approximation, one has to compute the diagrams in Figs.~(\ref{fig:ib2}). Only the BMW collaboration~\cite{Borsanyi:2020mff} has presented results so far. Those diagrams are expected to be at least either SU(3)$_f$ or $1/N_c$ suppressed and are more challenging to compute on the lattice. Fortunately, from \Table{tab:ib}, they appear to be negligible even for a precision of a few permil on $\ahvp$. 

A direct check of finite-size effect corrections has been done in~\cite{Giusti:2019xct} where different volumes with the same bare lattice parameters are used. For strong IB corrections, an exponential suppression with the lattice size is observed. However, the precision of the data is not yet sufficient to test the coefficient of the power-law for the QED corrections. See \Fig{fig:fse_IB}.

The isospin-breaking effects for the charm quark have been estimated by the ETM collaboration and found to be negligible~\cite{Giusti:2017jof,Giusti:2019xct}, see \Table{tab:ib}.

Further results are expected in the near future. The Mainz group have presented preliminary results on the inclusion of isospin-breaking effects using the the ROME123 method~\cite{Risch:2017xxe,Risch:2018ozp}. A first step toward IB correction to $\ahvp$ has been presented in~\cite{Risch:2019xio}. This year, the QCDSF collaboration also presented preliminary results during the APLAT 2020 conference~\cite{APLAT20:QCDSF}. They use full lattice QCD+ QED simulations at a single lattice spacing. They observed a positive increase of about $0.2\%$ on $\ahvp$ due to QED effects. When comparing this value with previous determinations, it should be noted that an unphysical value of the fine structure constant $\alpha \approx 0.1$ has been used.

\subsection{Summary of the HVP contribution}

\begin{figure}[h!]
\begin{center}
\resizebox{0.4\textwidth}{!}{  \includegraphics{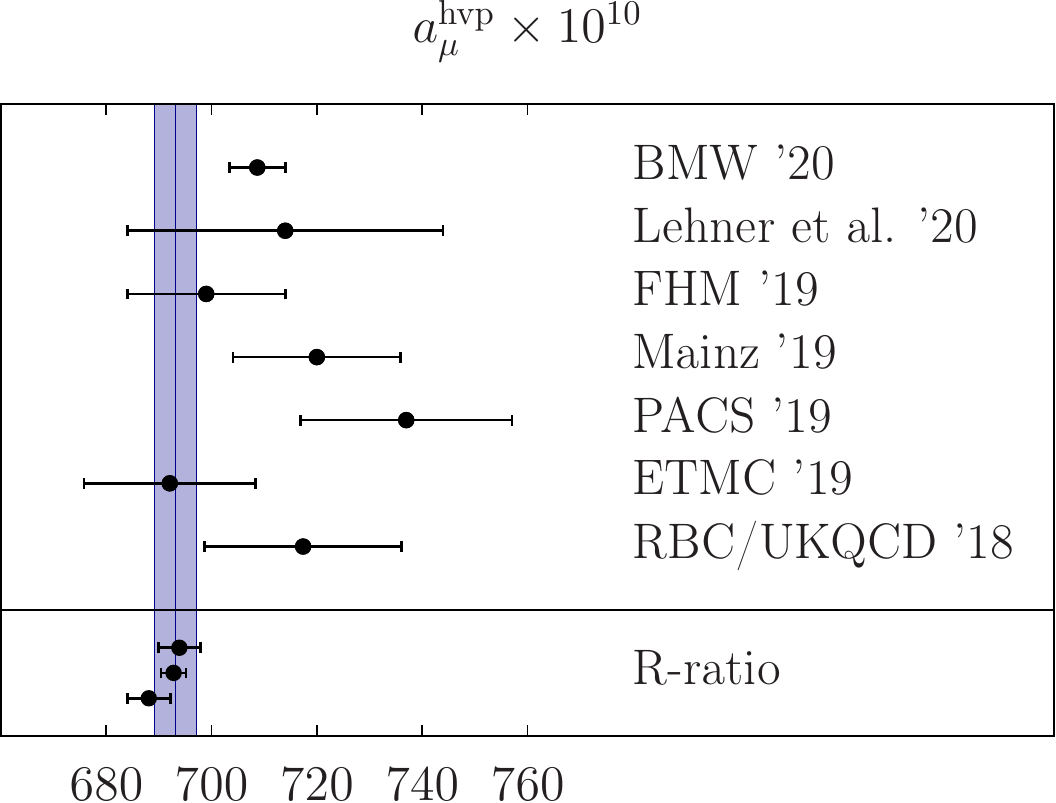}  } 
\end{center}
\caption{Comparison of lattice results for full contribution to the leading-order HVP. The lattice results are extracted from~\cite{Borsanyi:2020mff,Lehner:2020crt,Davies:2019efs,Gerardin:2019rua,Shintani:2019wai,Giusti:2019hkz,Giusti:2019xct,Blum:2018mom}. The R-ratio results are extracted from~\cite{Davier:2019can,Keshavarzi:2019abf,Jegerlehner:2017gek}. The vertical bands correspond to the value quoted in \Table{tab:status}. The later was obtained in~\cite{Aoyama:2020ynm} after merging all R-ratio results and does not include any lattice estimate.}
\label{fig:cmp_tot}      
\end{figure}

A comparison of the most recent lattice results is shown in \Fig{fig:cmp_tot}, along with the more precise results based the data-driven evaluations of the HVP. 
It is reassuring that results obtained with different lattice discretizations, and therefore affected by different systematic errors are in relative good agreement with each other. However, some tensions, especially for the dominant light quark contribution, are still present and need to be understood. Some methods will be discussed in the next section. 

Finally, the first sub-percent calculation has been published this year by the BMW collaboration~\cite{Borsanyi:2020mff}. This result is in tension with the R-ratio estimates and further investigations, among lattice collaborations and with phenomenology are required. Importantly, an even stronger tension is observed for the window method discussed in \Section{sec:window}.

\section{Benchmark quantities for the HVP contribution}
\label{sec:bench}

To match the precision of future experimental measurements, the target precision for the LO HVP contribution is 0.2\% and it represents an enormous challenge for lattice simulations. In previous sections, we have presented an overview of the dominant sources of error common to all lattice simulations. Since current estimates for this observable are usually dominated by systematic errors, it is of major importance to perform further cross-checks between collaborations to provide evidence that lattice simulations are under control. 
In the next subsections, we discuss several strategies that can be used, either as a check among lattice calculations (by designing observables less sensible to some specific source of error), or directly with phenomenology.

\subsection{The "window method"}
\label{sec:window}

The window method has been presented for the first time in Ref.~\cite{Blum:2018mom} as a tool to improve the accuracy of the HVP by supplanting the dispersive results based on R-ratio measurements~\cite{Davier:2019can,Keshavarzi:2019abf,Jegerlehner:2017lbd} by lattice inputs in a time-region where the lattice data turn out to be more precise. The idea is that the lattice calculation is much easier if one discards very short Euclidean times ($<0.3~\fm$), which are affected by large discretization effects, and the long distance contribution ($>1~\fm$), which is noisy and requires significant finite-size corrections. An additional feature of this method is that the chiral extrapolation is much smoother due to the suppression of the tail of the integrand. 

In this method, the integrand for $\ahvp$ in the TMR representation is convoluted with a smooth window function $W(t;t_0,t_1)$ in Euclidean time
\begin{equation}
a_{\mu}^{\rm win} = \left( \frac{\alpha}{\pi} \right)^2 \sum_{t} G(t)\, \widetilde{K}(t) \, W(t;t_0,t_1) \,.
\label{eq:win}
\end{equation}
A convenient definition is given by 
\begin{equation}
W(t;t_0,t_1) = \Theta(t,t_0,\Delta) - \Theta(t,t_1,\Delta) \,,
\end{equation}
where $\Theta$ is a smooth step function defined by $\Theta(t,t^{\prime},\Delta) = \left[ 1 + \tanh[ (t-t^{\prime}) / \Delta ] \right] / 2$. In Ref.~\cite{Blum:2018mom}, the authors have chosen the parameters $t_0 = 0.4~$fm, $t_1 = 1~$fm and $\Delta = 0.15~$fm. This choice was done by minimizing the total error on $\ahvp$ when using lattice data to compute $a_{\mu}^{\rm win}$ and the R-ratio data in the two complementary time regions. 

However, such a combination (between lattice data and R-ratio data) might appear premature. Indeed, the recent result from the BMW collaboration~\cite{Borsanyi:2020mff} presents a significant discrepancy with the one obtained in the dispersive approach. 
In this section, rather than using the window method to provide a more accurate result, we underline the power of this method to compare different lattice calculations and to spot the region where the disagreement might appear.

In \Section{sec:lohvp}, we listed the different challenges inherent to a sub-percent calculation of the leading order HVP. Interestingly, they do not affect the same time ranges. If discretization effects are mostly important at short distances, FSEs and the specific treatment of the tail of the integrand become more relevant at large Euclidean times. Thus, the window method is a useful tool to compare different lattice calculations. The choice made in \cite{Blum:2018mom} has several advantages. First, by removing the short distance contribution, discretization effects are suppressed and the continuum extrapolation might be smoother. Second, the suppression of the tail not only reduces significantly the noise at large Euclidean times but also flattens the chiral behavior. Finally, finite-size effects are much smaller on this quantity. However, some difficulties remain: the uncertainty associated to the scale setting, discussed in \Section{sec:scaleset}, is still present. The situation is even a bit worse since the definition of the window itself depends on the scale setting determination. 

\begin{figure}[h!]
\vspace{-0.5cm}
\begin{center}
\resizebox{0.4\textwidth}{!}{  \includegraphics{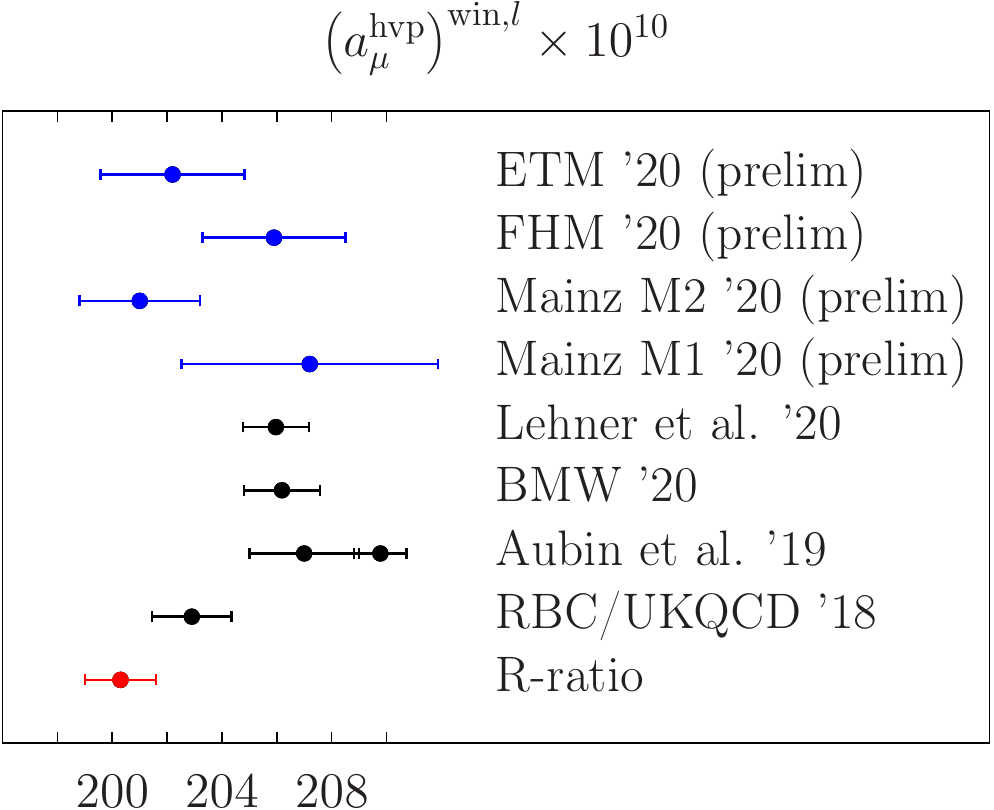}  } 
\end{center}
\caption{Comparison of lattice results for the window observable in the case of the connected contribution in the isospin symmetric limit. The window is defined by $(\Delta,t_0,t_1) = (0.15, 0.4, 1.0)$. The results are extracted from~\cite{Lehner:2020crt,Borsanyi:2020mff,Aubin:2019usy,Blum:2018mom}. For Aubin et al. the two points differ by the procedure used to perform the continuum extrapolation.}
\label{fig:amu_win}      
\end{figure}

Ideally, results for this intermediate window, as well as the complementary short ad long-distance window should be published together with the total $\ahvp$. So far, only four collaborations have published results in this direction but some preliminary results have been presented recently. The results for the light quark contribution in the isospin limit and with $(\Delta,t_0,t_1) = (0.15, 0.4, 1.0)$ are depicted in \Fig{fig:amu_win}. In this comparison, the R-ratio estimate is extracted from~\cite{Borsanyi:2020mff} by subtracting all lattice contributions, except the light-quark-connected one, to the phenomenological estimate based on the R-ratio. 
We observe some tension between different estimates. More importantly, a signifiant tension appears between the R-ratio estimate and the published lattice values, with the exception of the RBC/UKQCD estimate. 
As stated above, FSE are small in this time region. The main difficulty thus lies in the scale-setting determination and the continuum extrapolation. It is of major importance to have better control on this observables and more lattice determinations would be valuable.

For the strange-quark contribution, the results published in~\cite{Lehner:2020crt,Borsanyi:2020mff,Blum:2018mom} are in good agreement. It suggests that the scale-setting determination is not the cause of the discrepancy observed in the light-quark contribution. 
Concerning the charm quark contribution, a small tension is observed between BMW~\cite{Borsanyi:2020mff} and the RBC/UKQCD~\cite{Blum:2018mom} collaboration who found  $a_{\mu}^{\rm win,c} = 2.7(0.1) \times 10^{-10}$ and $a_{\mu}^{\rm win,c} = 3.0(0.1) \times 10^{-10}$, respectively. In addition to the light-quark contribution, results for the quark disconnected contribution might be valuable to understand the tension in the quark-disconnected contribution discussed in \Section{sec:dischvp}.

To conclude, this observable is in principle much easier to access than the full HVP contribution and a good statistical precision is easily obtained. Many sources of systematic errors are suppressed and the main difficulty originates from the scale-setting and the continuum extrapolation. A comparison between different lattice collaborations provides a strong consistency check, especially for the light and quark disconnected contributions. In particular, it is important to understand the current discrepancy between some lattice results and the R-ratio estimate.

\subsection{Running of alpha}

\begin{figure*}
\begin{center}
\resizebox{0.75\textwidth}{!}{  \includegraphics{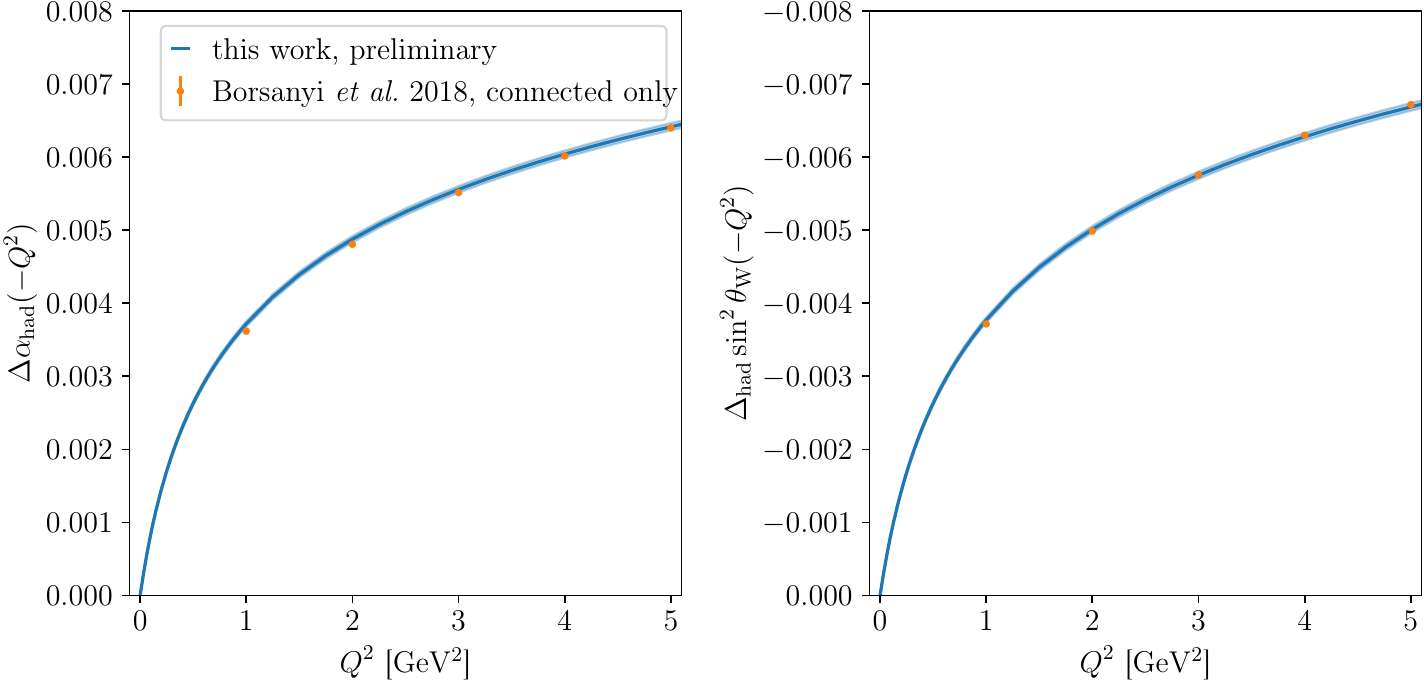}  }
\end{center}
\caption{Preliminary results for the running of the electromagnetic coupling $\alpha$ (right panel) and the Weinberg angle (right panel) as a function of the space-like momentum transfert $Q^2$. The blue line corresponds to the preliminary Mainz result presented in Ref.~\cite{Ce:2019imp}. The red dots are obtained from the results published in Ref.~\cite{Borsanyi:2017zdw}. Figure extracted from~~\cite{Ce:2019imp}. }
\label{fig:alpha}      
\end{figure*}

The fine structure constant is a fundamental parameter of the Standard Model of particle physics that characterizes the strength of the electromagnetic interaction. In the Thomson limit ($Q^2=0$), it is known with an impressive precision of 0.08 ppb~\cite{alphanature,Tanabashi:2018oca}. However, at the Z-pole mass, the value of the coupling increases by approximately 7\% where half of the correction is due to hadronic effects. As a consequence, almost 5 order of magnitude are lost in precision:
\begin{align}
\alpha &= 1/137.035 999 206(11) \,, \\
\alpha(M^2_Z) &= 1/127.955(10) \,.
\end{align}
In this running, the major source of uncertainty comes from low-energy hadronic contributions. 
This important parameter enters precision tests of the Standard Model and the relative precision of $5\times 10^{-5}$ would be required by future colliders~\cite{Baak:2014ora}. It corresponds to a reduction of the error by a factor 2-3.

The effective running coupling at a scale $q^2$ is conventionally written as
\begin{equation}
\alpha(q^2) = \frac{ \alpha }{1 - \Delta \alpha(q^2) } \,,
\end{equation}
where $\alpha = \alpha(0)$ is fine-structure constant and $\Delta \alpha(-Q^2) = \Delta \alpha_{\rm lep}(-Q^2) + \Delta \alpha^{(5)}_{\rm had}(-Q^2) + \Delta \alpha_{\rm top}(-Q^2)$ contains the contribution from leptons and the top quark, that can be estimated using perturbation theory, and the non perturbative hadronic vacuum polarization from the five light quarks, $\Delta \alpha^{(5)}_{\rm had}(Q^2)$. As for $\ahvp$, it can be estimated in a data-driven approach from $e^+ e^- \to$ hadron cross sections using dispersion relations~\cite{Davier:2019can,Keshavarzi:2019abf}. As compared to \Eq{eq:disp}, only the weight function is different and, for the running coupling, the contribution from high energy is much more pronounced such that 70\% of $\Delta \alpha^{(5)}_{\rm had}(Q^2)$ comes from pQCD and contributes to a large part of the final error.

The lattice calculation of the hadronic contribution to the running starts with
\begin{equation}
\Delta \alpha^{(5)}_{\rm had}(Q^2) = 4\pi \alpha \widehat{\Pi}(Q^2) \,,
\end{equation}
where the right-hand side is evaluated using \Eq{eq:tmrPi} in the time-momentum representation. The exact same correlation functions as for $\ahvp$ are required. Here, the main challenge lies in the control of discretization effects that become large as $Q^2$ increases. Ideally, very fine lattice spacings are needed. Then, the question is which $Q^2$ is high enough such that a matching with perturbation theory is under control, within the lattice estimate uncertainty. One possibility is to use Euclidean split technique~\cite{Jegerlehner:2008rs} where
\begin{multline}
\Delta \alpha^{(5)}_{\rm had}(M_Z^2) = \Delta \alpha^{(5)}_{\rm had}(-Q_0^2) +  \\ \left[ \Delta \alpha^{(5)}_{\rm had}(-M_Z^2) - \Delta \alpha^{(5)}_{\rm had}(-Q_0^2) \right] \\ + \left[ \Delta \alpha^{(5)}_{\rm had}(M_Z^2) - \Delta \alpha^{(5)}_{\rm had}(-M_Z^2) \right] \,.
\end{multline}
For $Q_0^2$ high enough, the last two term can be computed using perturbation theory and the first term is the lattice input.

The running coupling has been studied on the lattice for the first time in~\cite{Renner:2012fa,Feng:2012gh}, using twisted mass fermions. Later, the ETM collaboration~\cite{Burger:2015lqa} published results for the running, in the range $Q^2 \in [0-10]~\GeV^2$, that includes a chiral and continuum extrapolation to the physical point. The Mainz group has also presented preliminary results in~\cite{Francis:2014yga,Francis:2015grz} with two dynamical Wilson quarks. 
During last years Lattice conference, an analysis based on the CLS ensembles with $2+1$ dynamical quarks including one ensemble at the physical pion mass has been presented~\cite{Ce:2019imp}. In ~\cite{Borsanyi:2017zdw} the BMW collaboration published results for $\hat{\Pi}(Q^2)$ using staggered quarks at the physical pion mass. A comparison between the most recent BMW and the preliminary Mainz results is shown in the left panel of \Fig{fig:alpha}. 

Another observable of phenomenological interest is the electroweak mixing (Weinberg) angle $\Theta_W$ that parametrizes the mixing between the electromagnetic and the weak interactions. It defined through
\begin{equation}
\sin \Theta_W = \frac{ g^{\prime 2} }{ g^2 + g^{\prime 2} } \,,
\end{equation}
where $g$ and $g^{\prime}$ are the $SU(2)_L$ and $U(1)_Y$ couplings. Contrary to $\alpha$, the Weinberg angle is measured with a sub-permil precision at the Z-pole mass and the running at low energy is affected by hadronic uncertainties. In this case, the hadronic part of the running is defined by~\cite{Jegerlehner:1985gq}
\begin{equation}
\Delta_{\rm had} \sin^2 \Theta_W(-Q^2) = - \frac{e^2}{ \sin^2 \Theta_W^2 } \hat{\Pi}^{\gamma Z}(Q^2)
\end{equation}
where $\hat{\Pi}^{\gamma Z}$ is the hadronic vacuum polarization mixing between the electromagnetic current defined below \Eq{eq:vactensor} and the vector part of the neutral weak current
\begin{equation}
j_{\mu}^Z = j_{\mu} - \sin^2 \Theta_W \left( \frac{1}{4} \ubar \gamma_{\mu} u - \frac{1}{4} \dbar \gamma_{\mu} d - \frac{1}{4} \sbar \gamma_{\mu} s + \frac{1}{4} \cbar \gamma_{\mu} c \right) \,.
\end{equation}
Compared to the phenomenological estimate, it is much easier to perform the required flavor decomposition in lattice QCD simulations.

The electroweak mixing angle has ben studied on the lattice in~\cite{Burger:2015lqa,Francis:2015grz,Guelpers:2015nfb,Ce:2018ziv,Ce:2019imp} and the results of the running for the Mainz group is shown in the right panel of \Fig{fig:alpha}. 

\subsection{Electron $g-2$}

\begin{figure}[t!]
\center
\resizebox{0.45\textwidth}{!}{  \includegraphics{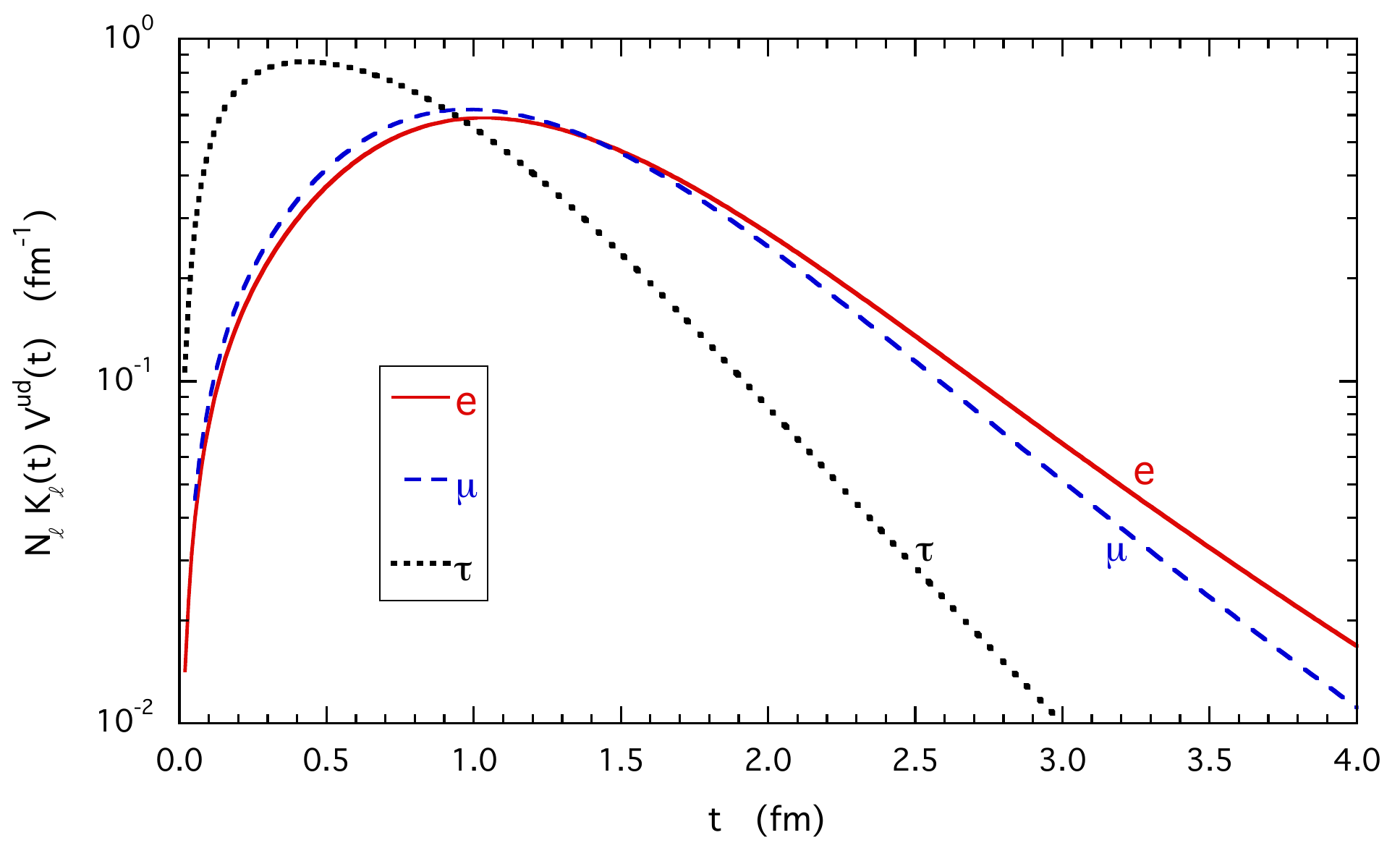}  } 
\caption{Integrand for the lepton ($\ell=e,\mu\tau$) anomaly in the time-momentum representation in \Eq{eq:lohvp} by the ETM collaboration. The integrand is normalized for convenience. Extracted from~\cite{Giusti:2020efo}. }
\label{fig:int_al}      
\end{figure}

The anomalous magnetic moment of the electron has been measured by the Harvard group with an accuracy of 0.24 ppb~\cite{Hanneke:2010au} and present a slight tension with the SM~\cite{Keshavarzi:2019abf} 
\begin{equation}
a_e^{\rm SM} - a_e^{\rm exp} = -89(23)_{\rm th}(28)_{\rm exp} \times 10^{-14}
\end{equation}
corresponding to about $-2.5~\sigma$. Very recently, a new measurement of the fine structure constant was published~\cite{alphanature} with a relative accuracy of 81 parts per trillion. This new value differs by by more than 5 standard deviations with the previous best result~\cite{Parker:2018vye} and leads to 
\begin{equation}
a_e^{\rm SM} - a_e^{\rm exp} = +48(30) \times 10^{-14} \,,
\end{equation}
corresponding to a tension of $+1.6~\sigma$. 
Interestingly, this new result leads to a deviation with the same sign as compared to the muon anomaly. In the case of the electron, the experimental determination based on atomic interferometry, is completely different from the one used for the muon anomaly. The phenomenology determination is, however, based on the same data set and can be computed from the $e^+e^-\to$~hadrons cross sections via dispersive methods. On the lattice, only the QED weight function differs through the lepton mass and $a_e$ can be easily obtained as a side results of the $(g-2)_{\mu}$, using the same set of correlation functions. The integrand, in the time momentum representation, and for lepton$=e,\mu,\tau$ is depicted in \Fig{fig:int_al}. Results for individual leptons have been presented in Refs~\cite{Burger:2015oya,Chakraborty:2016mwy,Borsanyi:2017zdw}.

Recently, the ETM collaboration proposed to study the following ratio~\cite{Giusti:2020efo}
\begin{equation}
R_{e/\mu} = \left( \frac{ m_{\mu} }{ m_e } \right)^2 \frac{ a_{e}^{\rm hvp} }{ a_{\mu}^{\rm hvp} }\,,
\label{eq:Rem}
\end{equation}
constructed from the anomalous magnetic moment of the electron and the muon. 
Because the integrand of the two lightest leptons are similar, this ratio can be used to reduce the statistical and systematic errors. In particular, it has been pointed out that the error coming from the scale setting (see the discussion in \Section{sec:scaleset}) is strongly reduced as compared to $\ahvp$. Finite-size effects also largely compensate as well as renormalization factors. Finally, the chiral extrapolation is also much milder. In~\cite{Giusti:2020efo}, the authors observed a reduction of the error by a factor close to 4 for individual ensembles.

In Ref.~\cite{Giusti:2020efo}, the ETM collaboration has published a lattice determination of the ratio~(\ref{eq:Rem}), including the dominant isospin-breaking corrections. Their result at the physical point reads
\begin{equation}
R_{e/\mu} = 1.1456(63)_{\rm stat}(54)_{\rm stat} \,.
\end{equation}
This result is in good agreement with the estimate obtained from the dispersive analyses based on $e^+e^- \to$ hadron cross sections~\cite{Keshavarzi:2019abf} and presents a tension of about 2.7 standard deviation with the experimental result~\cite{Hanneke:2008tm,Hanneke:2010au,Bennett:2006fi}. However, on the phenomenological side, this tension is mostly due to QED uncertainties and not to hadronic ones.

From a purely lattice point of view, this observable might be interesting to compare the results obtained by different lattice collaborations.

\subsection{Time moments}

The time moment of order $n$ has been defined in \Eq{eq:timemoment} and can be used to determine the hadronic vacuum polarization contribution using Pad\'e approximants. This is actually the strategy followed by the Fermilab-HQPCD-MILC collaboration, where the authors use two series of approximants $[n,n]$ and $[n,n-1]$~\cite{Chakraborty:2014mwa}. It has been proven that the true result lies between them, such that only a few terms of the series are required at a given precision. 

The time moments can be computed using the exact same set of lattice data and results that have been presented by various collaborations. As such, they played an important role in the past years when comparing results from different lattice collaborations. 

\begin{figure}[b]
\center
\resizebox{0.45\textwidth}{!}{  \includegraphics{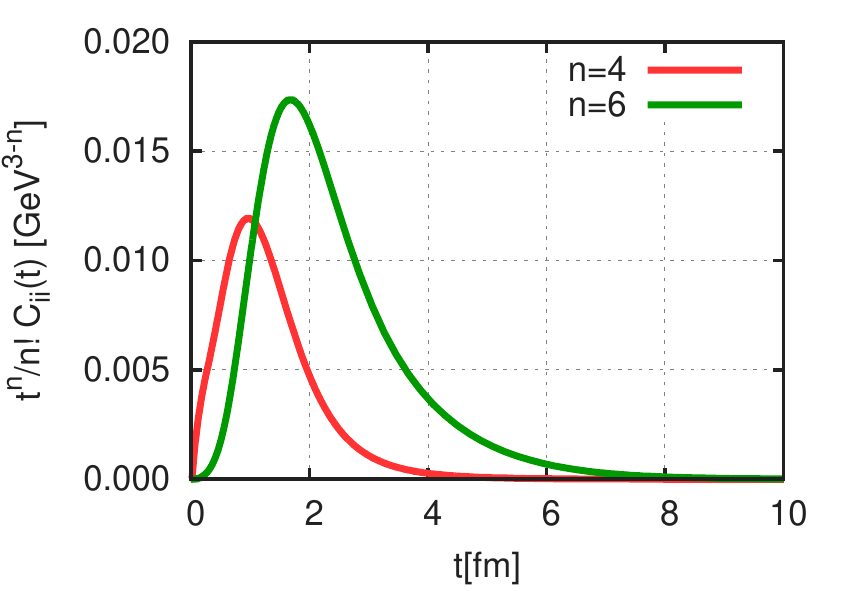}  } 
\caption{Integrand for the first two time-moment as a function of the Euclidean time. The electromagnetic current correlator is obtained using a phenomenological description of the $e^+e^-$ data and dispersion relations, as explained in~\cite{Borsanyi:2016lpl}. The figure is extracted from Ref.~\cite{Borsanyi:2016lpl}. }
\label{fig:moments}      
\end{figure}

As for the time momentum representation, one can use the flavor or isospin decompositions of the vector correlator. The statistical and systematic errors of the time moments can be treated in a very similar fashion as for the HVP. The integrand for the first two moment ($n=1,2$) is depicted in \Fig{fig:moments} and they probe larger Euclidean times as $n$ increases. It is interesting to note that the kernel function $\tilde{K}$, used in lattice calculations of the HVP in \Eq{eq:lohvp}, behaves as $\tilde{K} \sim \frac{\pi^2}{9} m_{\mu}^2 t^4$ at short distances $t \ll m^{-1}_{\mu}$ and as $\tilde{K} \sim 2 \pi^2 t^2$ at long distances $t \gg m^{-1}_{\mu}$~\cite{Gerardin:2019rua}. Thus, the second moment probes much longer distances and is even more subject to the noise problem discussed in \Section{sec:light}. It is also more affected by FSEs or by the finite time-extent of the lattice. On the other hand, the integrand of the first moment is closer to the one for the HVP. It indeed behaves as the HVP at short distances and is also peaked at $t \approx 1~\fm$. Finally, one notices that, contrary to $\amu$, the time moments are dimensionful quantities. 

Because of those difficulties, the time moments are, to some extent, even more challenging to evaluate that the HVP itself and makes them less useful for a direct comparison between lattice results. 

\begin{figure}[h!]
\begin{center}
\resizebox{0.4\textwidth}{!}{  \includegraphics{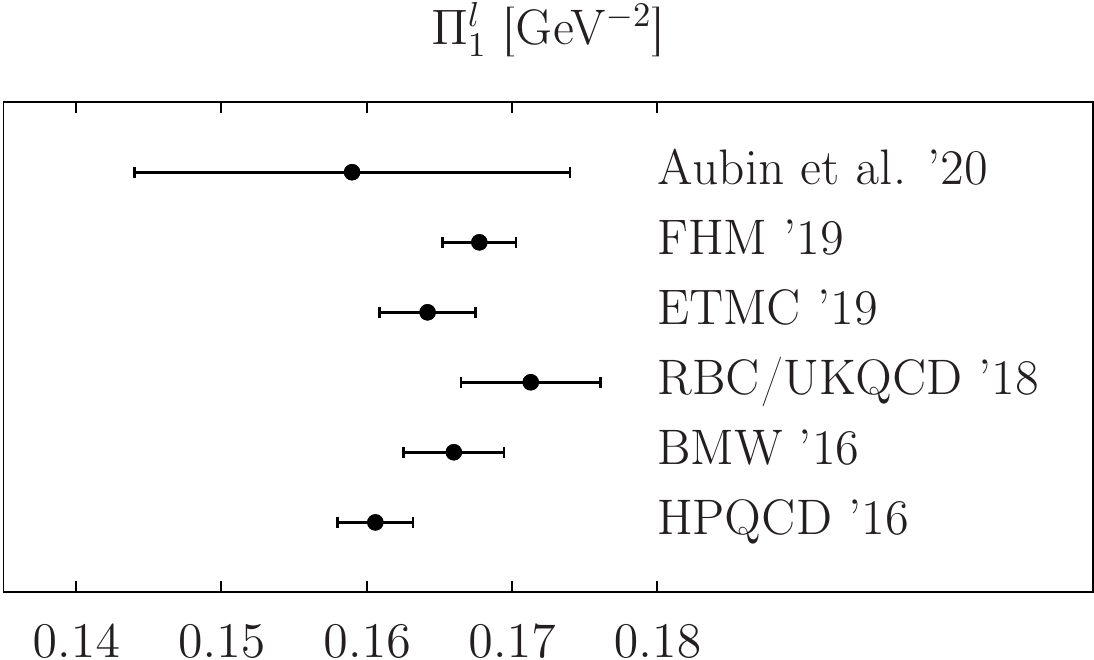}  } \\[3mm]
\resizebox{0.4\textwidth}{!}{  \includegraphics{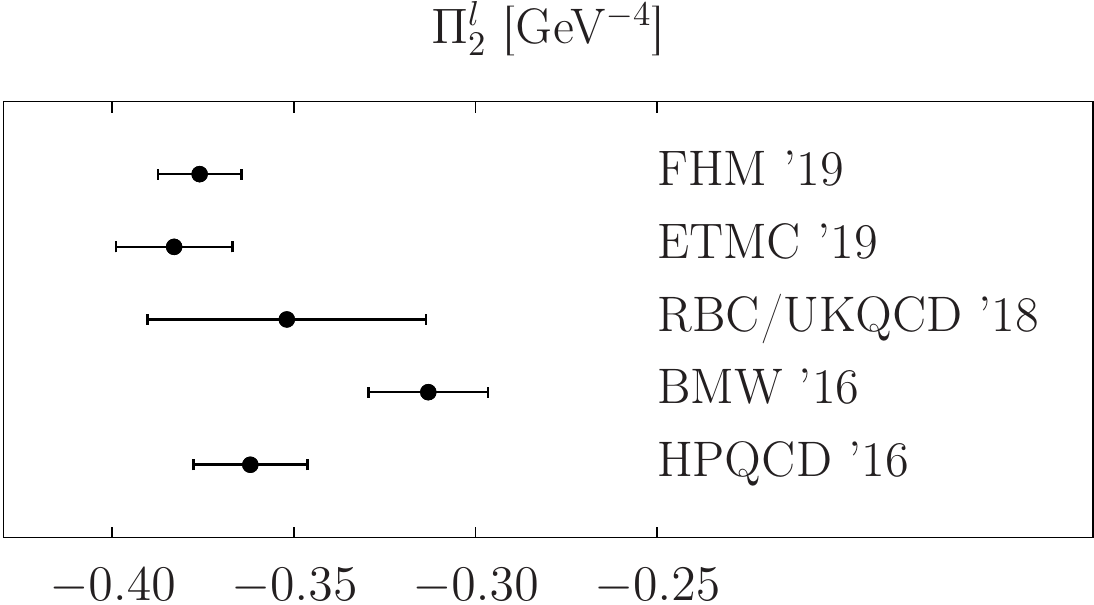}  } 
\end{center}
\caption{Comparison of lattice results for the light-quark contribution (in the isospin limit) to the first two time moments defined in Eq.~(\ref{eq:timemoment}). The results are extracted from~\cite{Aubin:2019usy,Davies:2019efs,Giusti:2019xct,Blum:2018mom,Borsanyi:2016lpl,Chakraborty:2016mwy}.} 
\label{fig:cmp_disc}      
\end{figure}

Results for the first moment, restricted to the light quark contribution in the isospin limit, have been published by several collaborations~\cite{Aubin:2019usy,Davies:2019efs,Giusti:2018mdh,Borsanyi:2016lpl,Chakraborty:2016mwy} and, as can be seen in \Fig{fig:cmp_disc}, turn out to be in good agreement. Results for the second moment have also been presented by various collaborations~\cite{Davies:2019efs,Giusti:2018mdh,Borsanyi:2016lpl}. In this case, there is however a slight tension  between the FHM and BMW collaborations. It might underline the difficulty to get a good description of the correlator at large Euclidean times, in the treatment of FSEs or in the scale setting determination that affect this dimensionful observable.

\section{The hadronic light-by-light contribution}
\label{sec:hlbl}

The hadronic light-by-light contribution enters at order $\alpha^3$ in the perturbative expansion and is expected to be much smaller than the LO-HVP contribution. The relevant diagram is depicted in the right panel of \Fig{fig:diag} and the current best-estimate, extracted from~\cite{Aoyama:2020ynm}, is given in Table~\ref{tab:status}. If this contribution is about 70 times smaller than the LO-HVP, its evaluation is also more challenging. Fortunately, an overall precision of  10\% would already suffice to match the future experimental precision on $\amu$.  

Interestingly, this contribution is two times larger than the current precision on $\amu$ and it should be about 3 times larger to explain the discrepancy in the muon $(g-2)$ by itself. The recent lattice results by the RBC/UKQCD and Mainz collaborations, as well as the dispersive result, suggest that this is very unlikely. 

In the past, this contribution was estimated using model estimates~\cite{Bijnens:1995cc,Bijnens:1995xf} where systematic errors are difficult to estimate and the Glasgow consensus~\cite{Prades:2009tw} reads $\ahlbl = (102 \pm 39) \times 10^{-11}$. One of the main messages is that pseudoscalar exchanges are by far the dominant contribution. There are currently two independent approaches that aim to provide a model-independent determination of this contribution. 

The first method consists in calculating the full HLbL diagram using ab-initio lattice calculations with all errors under control. Recently, two collaborations have published results in this direction: the RBC/UKQCD collaboration and the Mainz group. Although they use slightly different strategies, they both rely on the numerical evaluation of a four-point correlation function in position space
\begin{multline}
\Pi_{\mu\nu\lambda\sigma}(q_1,q_2,q_3) = \int \mathrm{d}^4x_1 \mathrm{d}^4x_2 \mathrm{d}^4x_3 e^{-i(q_1 x_1 + q_2 x_2 + q_3 x_3)} \\ \langle J_{\mu}(x_1) J_{\nu}(x_2) J_{\lambda}(x_3) J_{\sigma}(0) \rangle
\label{eq:4pt}
\end{multline}
where $J_{\mu}$ is the hadronic part of the electromagnetic current defined below \Eq{eq:vactensor}. This method will be discussed in \Section{sec:direct_hlbl}.

The second method to tackle this challenge is to consider contributions from individual intermediate states. This is done in a systematic and rigorous way in the dispersive approach~\cite{Colangelo:2014dfa,Colangelo:2014pva,Pauk:2014jza,Colangelo:2015ama}. This data-driven method has been successfully applied to the vector 2-point correlation function to estimate the LO-HVP contribution but the situation is much more difficult for the HLbL due to the complicated analytic structure of the 4pt correlation function. Fast progress has been achieved in recent years and an overview of this approach can be found in~\cite{Aoyama:2020ynm}. 
Fortunately, it is expected that, among all possible intermediate states, the largest contributions are given by a handful of states: the light pseudoscalar mesons $\pi^0$, $\eta$ and $\eta^{\prime}$. All the non-perturbative information is encoded in the (space-like) transition form factors that describe the interaction of the meson with two virtual photons. However, the experimental data, needed for this approach, are often missing. Lattice QCD can provide valuable information, especially for the dominant pseudoscalar pole contributions. This will be discussed in Sections~\ref{sec:hlblpi0} and \ref{sec:hlbl_amps}.

\subsection{Direct lattice calculation}
\label{sec:direct_hlbl}

The first lattice calculation of the HLbL diagram was performed by the RBC/UKQCD collaboration in~\cite{Blum:2014oka}, where the authors focused on the connected contribution only. The form factor $F_2$ was computed at different values of $Q^2$ using a non-perturbative treatment of QED and a pion mass of $329~\MeV$. The large statistical error and the extrapolation to $Q^2=0$ prevented them from a comparison with phenomenology. Later, in~\cite{Blum:2015gfa}, several major improvements in methodology were presented, first to reduce the statistical error and the numerical cost and, second, to avoid the need for an extrapolation to $Q^2=0$ by introducing the moment method. For the first time, a very clear signal was obtained for the connected contribution with a pion mass of $171~\MeV$, close to the physical one.

All recent calculations consist in evaluating a four-point correlation function in QCD and in position space. The latter is summed with a QED weight function that represents the muon and quark lines in \Fig{fig:diag}, such that the Pauli form factor is directly obtained at vanishing momentum, $F_2(0) = a_{\mu}$. No extrapolation $q\to 0$ is required. So far, only two collaborations have presented results and the difference in their setups mostly lies in the determination of the QED weight function: either in finite volume for the RBC/UKQCD collaboration or in the continuum and infinite volume for the Mainz Group. The RBC/UKQCD collaboration also presented preliminary results for the second case~\cite{Blum:2017cer}. 

\begin{figure}[t]
\begin{center}
\resizebox{0.12\textwidth}{!}{  \includegraphics{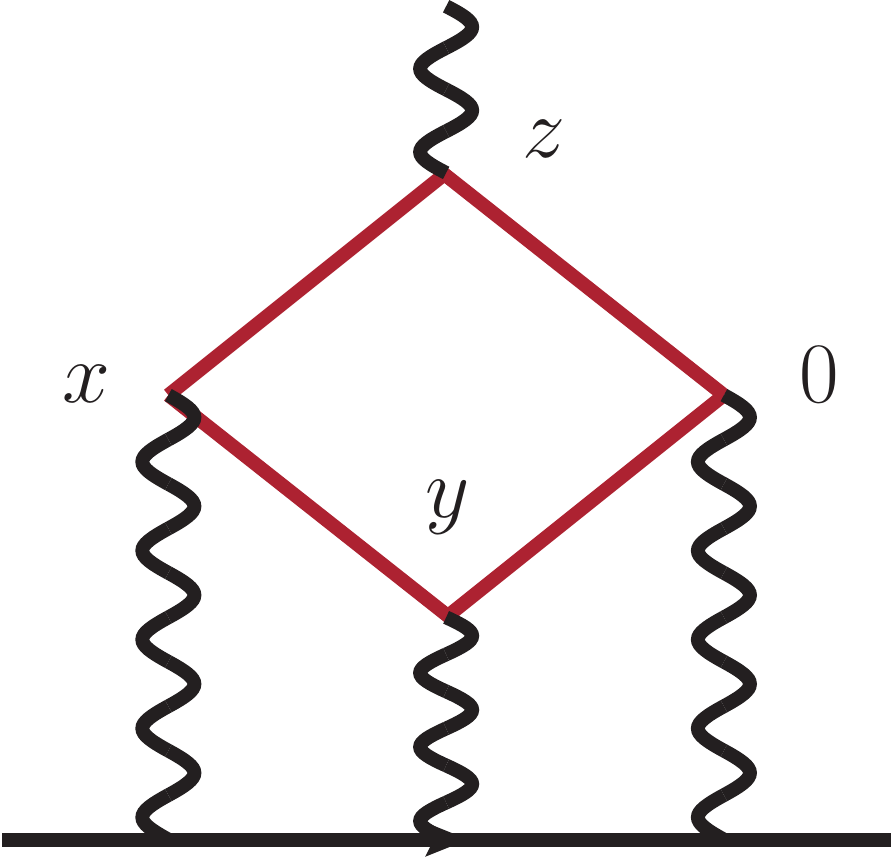}  }  \qquad
\resizebox{0.12\textwidth}{!}{  \includegraphics{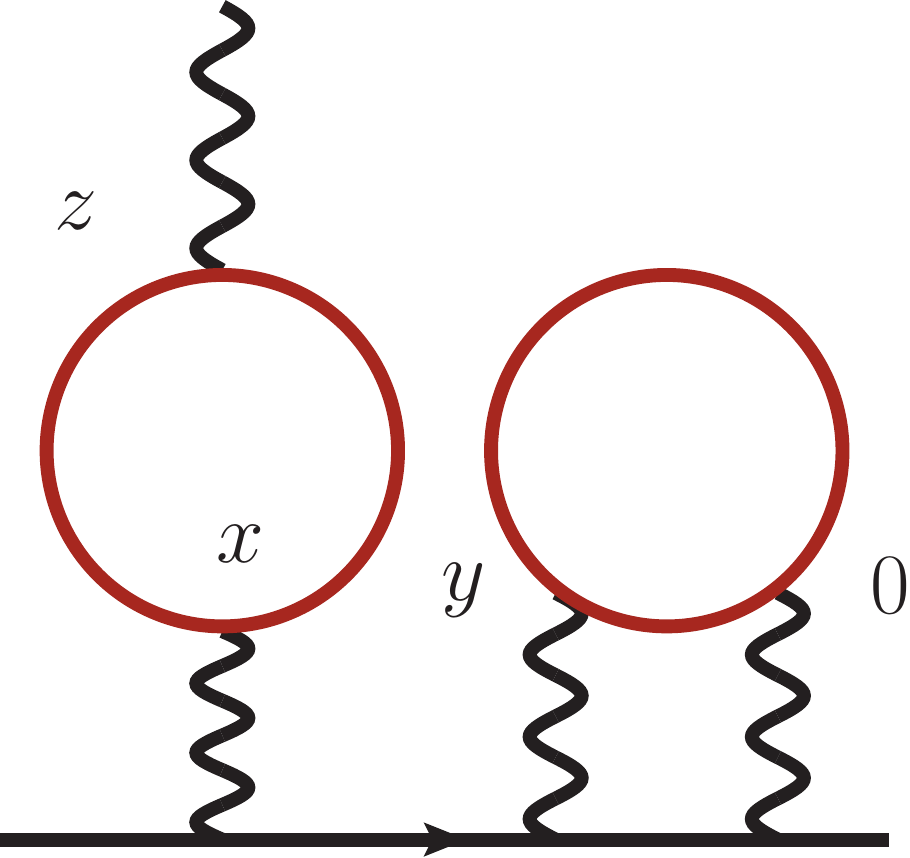}  } \\[7mm]

\resizebox{0.12\textwidth}{!}{  \includegraphics{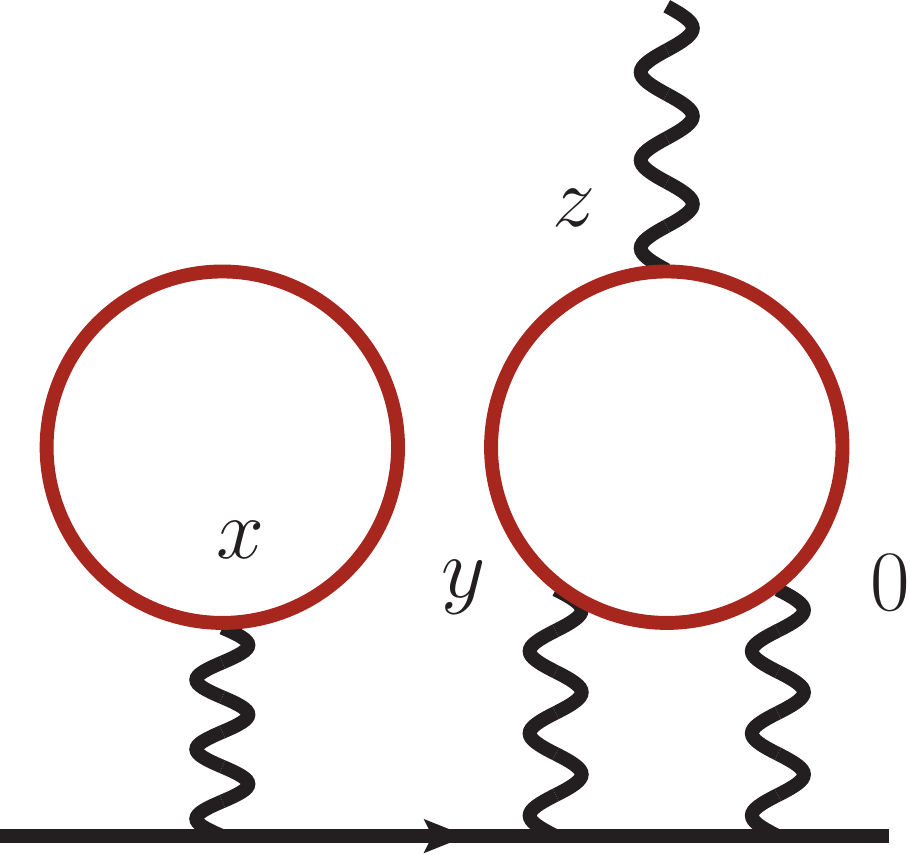}  } \qquad
\resizebox{0.12\textwidth}{!}{  \includegraphics{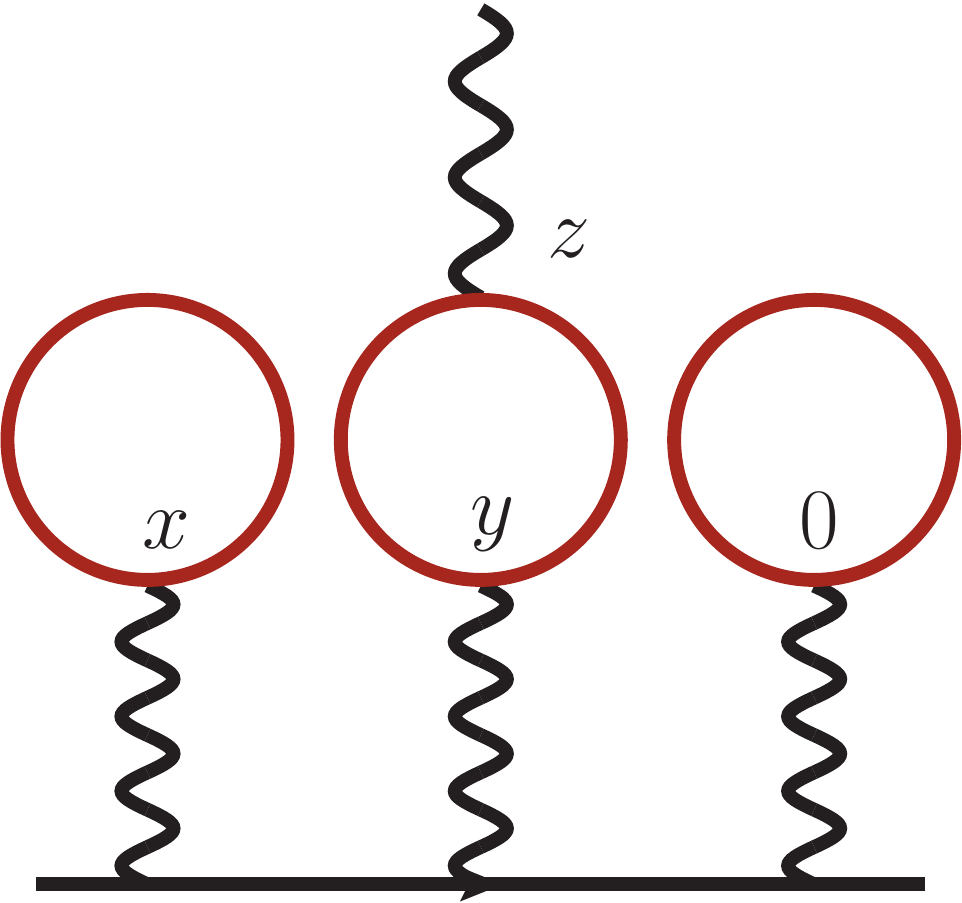}  } \qquad
\resizebox{0.12\textwidth}{!}{  \includegraphics{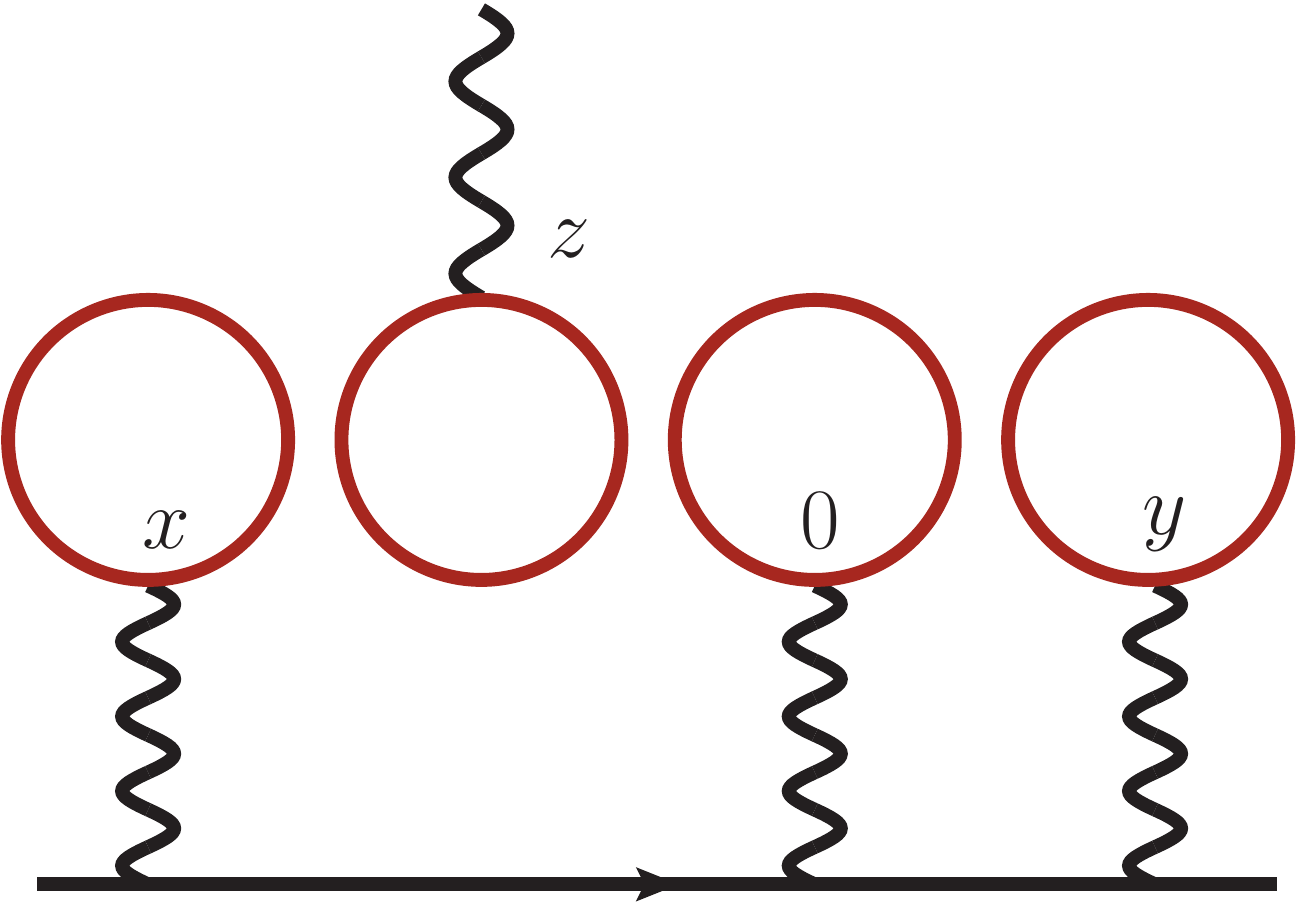}  } \qquad
\end{center}
\caption{The five classes of diagrams for the HLbL calculation. The first two diagrams are dominant one. The other three diagrams vanish in the SU(3)$_f$ limit.}
\label{fig:hlbl}      
\end{figure}

The four-point correlation function in \Eq{eq:4pt} involves five different classes of Wick contractions that are depicted in \Fig{fig:hlbl}. The quark disconnected contributions are conveniently called $(2+2)$, $(3+1)$, $(2+1+1)$ and $(1+1+1+1)$, depending on the number of vertices in each quark loop. 
The fully connected and 2+2 disconnected diagrams, that does not vanish in the SU(3)$_f$ limit (upper diagrams of \Fig{fig:hlbl}) are expected to be dominant. 
In~\cite{Bijnens:2015jqa}, based on $\pi^0$ exchange, the authors anticipated that the disconnected contribution might be large to cancel a part of the connected contribution. This has been confirmed numerically by both the RBC/UKQCD and the Mainz collaborations. 
The other, sub-dominant diagrams vanish exactly in the SU(3)$_f$ symmetry limit and are suppressed by a factor $1/N_c^n$ where $n$ is the number of single closed loops. Although their contribution is expected to be smaller than the target precision, an upper bound on their size would be valuable to assess systematic errors. 

To reach the target precision of 10\% on the hadronic light-by-light contribution, the main challenges consist in reducing the statistical error as well as controlling the main sources of systematic errors, namely the finite-volume effects and the extrapolation to the physical point. The light-quark contribution is by far the dominant one, the strange quark contribution being further suppressed by its electric charge. The charm quark can be safely neglected. 
Reaching a high statistical precision is made difficult by the large cancellation between the connected and disconnected contributions. The numerical implementation used so far leads to contributions that are statistically uncorrelated and little is gained when taking the difference such that the relative precision gets worse. Finally, finite-size effects are found to be large even on big lattices. In this case, the situation differs significantly when one uses QED$_L$ or QED$_\infty$.
Finally, the evaluation of the QED weight function is by itself difficult, as no fully-analytical expression is known and the cost associated to its evaluation is not necessarily negligible. 

\subsubsection{The Mainz approach}
\label{sec:hlbl_qedInf}

The Mainz group starts with the master formula
\begin{equation}
\ahlbl = \frac{m_\mu e^6}{3}\int d^4y \int d^4x \; \kernel_{[\rho,\sigma];\mu\nu\lambda}(x,y)\;i\widehat\PI_{\rho;\mu\nu\lambda\sigma}(x,y),
\label{eq:Mainzmaster}
\end{equation}
with $m_{\mu}$ the muon mass, $\kernel$ a QED weight function that describes the muon and photon lines in \Fig{fig:diag}, and $i\widehat\PI$ is a spatial moment of the Euclidean four-point function in QCD,
\begin{eqnarray}
i\widehat \PI_{\rho;\mu\nu\lambda\sigma}( x, y)  &=& -\int d^4z\,  z_\rho\, \widetilde\PI_{\mu\nu\sigma\lambda}(x,y,z), \label{eq:pihat} \\ 
\widetilde\PI_{\mu\nu\sigma\lambda}(x,y,z)&\equiv& \langle \,j_\mu(x)\,j_\nu(y)\,j_\sigma(z)\, j_\lambda(0) \rangle_{\rm QCD}\,.
\end{eqnarray}
The QED weight function $\kernel$ has been computed semi-analytically in the continuum and infinite volume, in position space. It is expressed in terms of a few scalar weight functions that can be easily evaluated numerically, on the fly, during the contractions of the quark propagators. First results and checks of the kernel have been presented in previous conferences and workshops~\cite{Asmussen:2016lse,Asmussen:2017bup,Asmussen:2018lcw}. A theoretical advantage of treating QED in infinite volume is that no power-law finite-volume effect appears. Such terms are expected to arise in the QED$_L$ formulation (see the next section), due to the massless photon propagators. Instead, finite-volume errors are exponentially suppressed with the lattice size. 

\begin{figure}[b!]
\begin{center}
\resizebox{0.43\textwidth}{!}{  \includegraphics{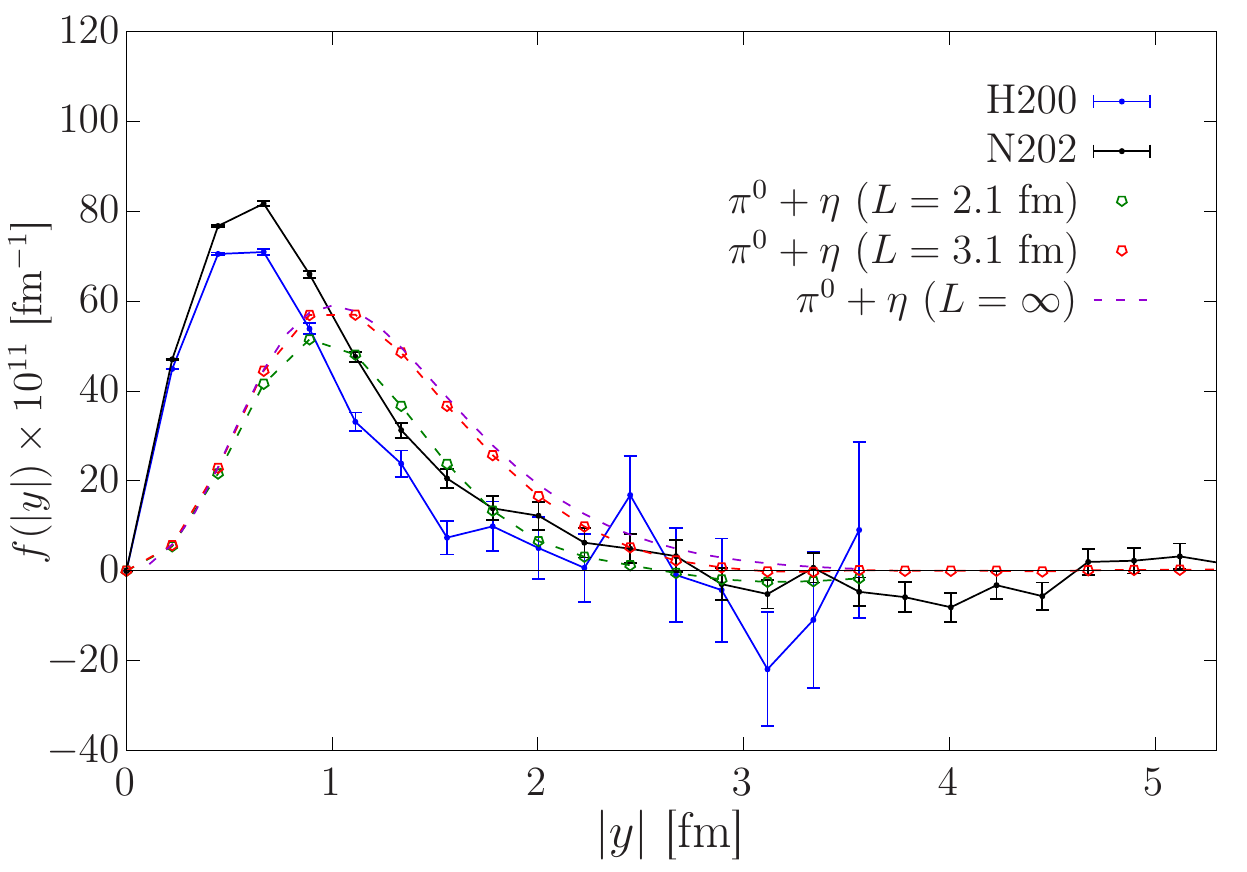}  } 
\end{center}
\caption{Integrand for the quark-connected contribution for two ensembles at the SU(3)$_f$ symmetric point with $m_{\pi} = m_{K} \approx 420~\MeV$ (H200 with $m_{\pi}L = 4.4$ and N202 with $m_{\pi}L = 6.4$) for the Mainz group~\cite{Chao:2020kwq}.}
\label{fig:int_mainz}      
\end{figure}

\Eq{eq:Mainzmaster} requires three sums over the whole lattice volume and an exact summation would be prohibitively expensive. In the numerical setup presented by the Mainz group, only two sums over $x$ and $z$ are performed explicitly on the lattice. This is possible using propagators with sources at the origin and $y$ since the contractions involve only O($V$) operations, because the weight factor factorizes as a function of $(x,y)$ (the QED kernel) times $z_{\rho}$ (see~\Eq{eq:pihat}). After contracting all the Lorentz indices, the result depends only the norm of the vector $y$ (up to discretization effect at finite lattice spacing) and the integrand is sampled reliably by choosing a few values of $|y|$. The shape of the integrand as a function of $|y|$ and for the connected contribution is shown in \Fig{fig:int_mainz}. 

The two main challenges associated with this calculation are the control of statistical errors (especially for the disconnected contribution) and the correction for finite-size effects that are large even for lattices with $m_{\pi} L >4$. Various techniques have been proposed to deal with those challenges.

Concerning statistical errors, as first noted in~\cite{Blum:2017cer}, the weight function in \Eq{eq:Mainzmaster} is not unique. This is a consequence of the conservation of the vector current in the continuum limit (in particular any function that depends only on the variable $x$ or $y$ can be added to the kernel). Such subtractions are equivalent, in the continuum and infinite-volume limits, but provide different lattice estimators that can be affected by different statistical and systematic errors. The following subtraction
\begin{equation}\label{eq:lamsub}
\begin{aligned}
\kernel^{(\Lambda)}_{[\rho,\sigma];\mu\nu\lambda}(x,y) &= \kernel_{[\rho,\sigma];\mu\nu\lambda}(x,y)\\ 
	&-\partial_\mu^{(x)} (x_\alpha e^{-\Lambda m_\mu^2 x^2/2}) \kernel_{[\rho,\sigma];\alpha\nu\lambda}(0,y) \\&- \partial_\nu^{(y)} (y_\alpha e^{-\Lambda m_\mu^2 y^2/2})\kernel_{[\rho,\sigma];\mu\alpha\lambda}(x,0),
\end{aligned}
\end{equation}
where $\Lambda$ is an arbitrary dimensionless parameter reduces the long-distance contribution and the discretization effects that arise with the standard kernel. Using the un-subtracted kernel, the signal is lost at very short distance, below 1~fm.

As a second strategy to address the noise problem at long distance, the Mainz group has considered two different estimators for the connected-quark contribution. A direct implementation of the master formula in \Eq{eq:Mainzmaster} is rather expensive and requires the use of many sequential propagators. Fortunately, the integral can be re-arranged in such a way that only two inversions are required for each value of $|y|$. Keeping all propagators in memory, it allows for a significant reduction of the noise~\cite{Chao:2020kwq}. 

\begin{figure}[h!]
\begin{center}
\resizebox{0.42\textwidth}{!}{  \includegraphics{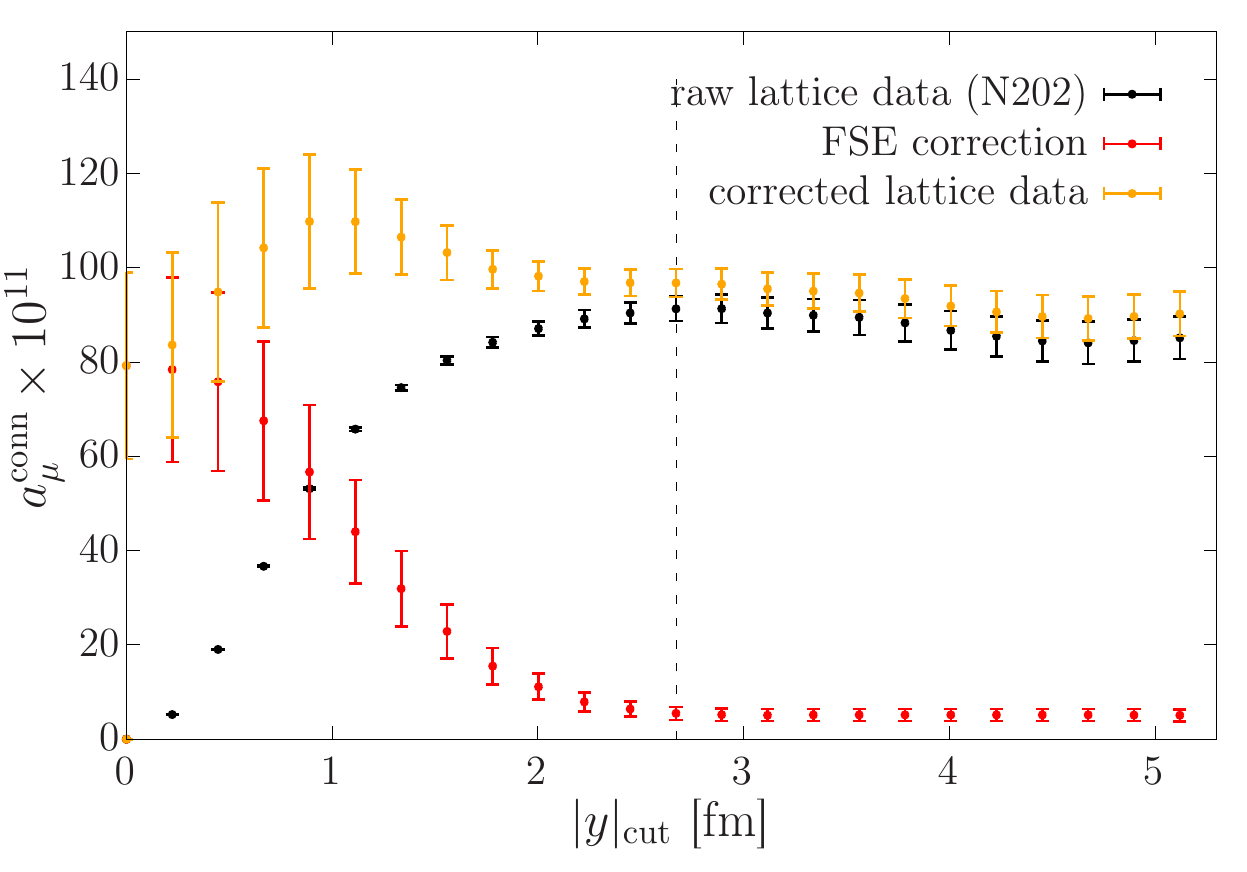}  } 
\end{center}
\caption{Value of the connected contribution using the long-distance correction prescription described in the text: below $|y|_{\rm cut}$ the lattice data are used. Above the cut, the prediction from the pion-pole contribution is used. Extracted from~\cite{Chao:2020kwq}.}
\label{fig:fse_mainz}      
\end{figure}

Finally, assuming a Vector Meson Dominance (VMD) model, with parameters obtained from a dedicated lattice calculation of the pion transition form factor (see \Section{sec:hlblpi0}), the pion-pole contribution has been evaluated both in finite and infinite volume~\cite{workhopMainz:Harvey,Asmussen:2019act}. First, the light pseudoscalar is expected to give the dominant contribution at long distance and the noisy lattice data can eventually be replaced by the pion-pole contribution only. 
In \Fig{fig:fse_mainz} the lattice data are used below the cut and, above the cut, the pion-pole calculation is used. As expected, one observes a plateau if the cut is such that $|y|_{\rm cut} > 2.5~\fm$. Second, this method allows to correct for the dominant source of finite-size effect that arises from the pion and a comparison in different volumes is shown on \Fig{fig:fse_mainz}. 

\subsubsection{RBC/UKQCD : QED in finite volume}
\label{sec:hlbl_qedL}

\begin{figure}
\center
\resizebox{0.32\textwidth}{!}{  \includegraphics{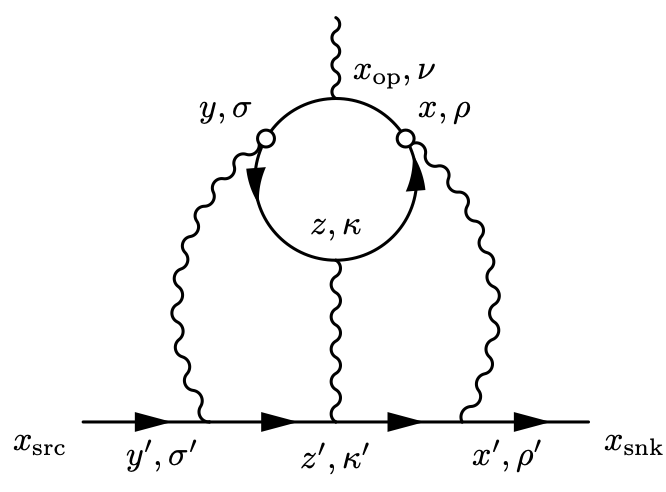}  } 
\caption{Notations used by the RBC/UKQCD collaboration for the quark-connected contribution. The origin is set at $w=\frac{x+y}{2}$ and $r=x-y$. Figure extracted from~\cite{Blum:2016lnc}.}
\label{fig:rbcsetup}      
\end{figure}

In the moment method~\cite{Blum:2015gfa,Blum:2016lnc,Blum:2019ugy}, the RBC/UKQCD collaboration starts with the following expression
\begin{multline}
\frac{\ahlbl}{2 m_\mu}\overline{u}(\vec 0,s')\,\vec\Sigma\,  u(\vec 0,s) = \\
 \frac{1}{2}\sum_{r,z,x_{\rm op}} \vec{x}_{\rm op} \times i\overline{u}(\vec 0,s')\, \vec{\mathcal{F}}^C \left(\frac{r}{2},-\frac{r}{2},z,x_\text{op}\right) u(\vec 0,s),
\label{eq:moment_Xprod}
\end{multline}
where $\Sigma_i=\frac{1}{4i}\epsilon_{ijk}[\gamma_j, \gamma_k]$. The notations are explained in \Fig{fig:rbcsetup} and $r = x-y$. The amplitude $\mathcal{F}^C_{\nu} \left( x, y, z, x_{\text{op}} \right)$ is obtained from the average of the function $\mathcal{F}_{\nu} \left( x, y, z, x_{\text{op}} \right)$ over the three cyclic permutations of the positions $x$, $y$ and $z$ with
\begin{multline}
\mathcal{F}_{\nu} \left( x, y, z, x_{\text{op}} \right) = (-ie)^6 \times \\ \mathcal{G}_{\rho, \sigma, \kappa} \left( x, y, z; x_{\text{snk}}, x_{\text{src}} \right) \mathcal{H}^C_{\rho,\sigma,\kappa,\nu}(x,y,z,x_{\rm op}) \,.
\end{multline} 
where $i^4 \mathcal{H}^C_{\rho,\sigma,\kappa,\nu}(x,y,z,x_{\rm op})$ represents the hadronic part of the diagram that is expressed in terms of four quark propagators and $i^3 \mathcal{G}_{\rho, \sigma, \kappa} \left( x, y, z; x_{\text{snk}}, x_{\text{src}} \right)$ is the QED weight function expressed in terms of muon and photon propagators. The later includes a sum over the variables $x'$, $y'$ and $z'$. 
The QED part is evaluated on the same finite lattice and the spatial zero modes of the photons are explicitly removed. This procedure is known as QED$_L$~\cite{Hayakawa:2008an}. As compared to the Mainz approach, one expects power-law, finite-volume effects and an extrapolation $L \to \infty$ is eventually required. In this setup, the numerical cost associated with the evaluation of the QED weight function is also not negligible.
 
\begin{figure}
	\center
	\resizebox{0.4\textwidth}{!}{  \includegraphics{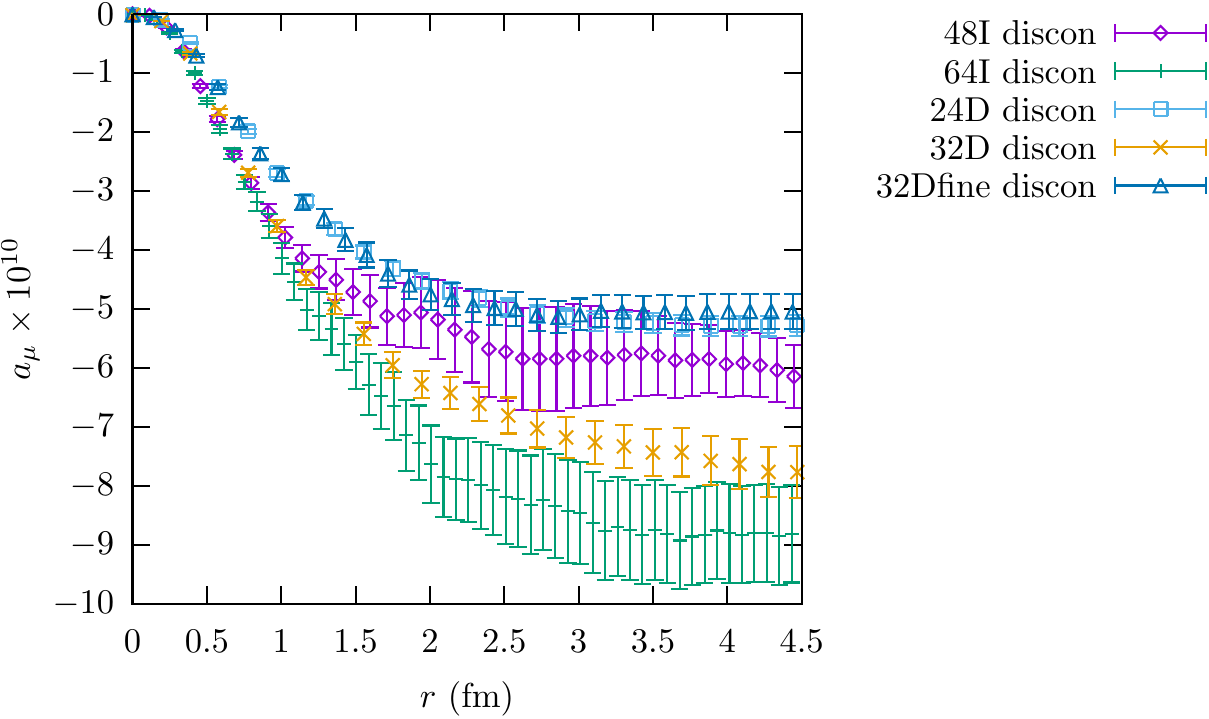}  } \\[3mm]
	\resizebox{0.4\textwidth}{!}{  \includegraphics{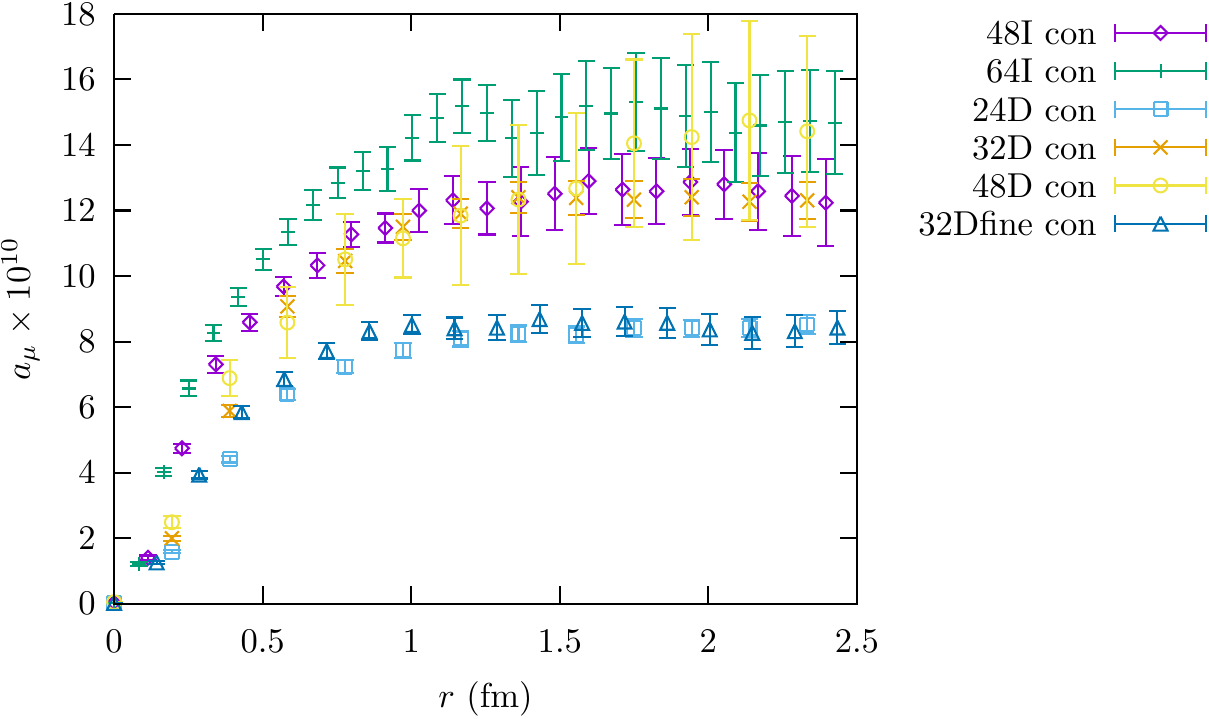}  } 
\caption{Cumulative contributions to the muon anomaly for the connected (upper) and leading disconnected (lower) diagrams in the QED$_F$ formalism. Here $r$ is the distance between the two sampled currents in the hadronic loop (the other two currents are summed exactly).}
\label{fig:qedL_int}    
\end{figure}

Similarly to the Mainz approach, due to the huge computational cost, all spacetime summations cannot be performed exactly over the whole lattice. The summation over $r = x-y$ is done stochastically with a distribution chosen such that points that contribute most to the signal are more frequent. As can be seen in Fig.~\ref{fig:qedL_int}, the dominant contribution indeed comes from distances less than 1 fm. The two propagators, with sources at $x$ and $y$, are contracted for all sink values such that the sums over $x_{\rm op}$ and $z$ are performed exactly over the whole lattice. 

For the connected contribution, the hadronic part is explicitly given by
\begin{multline}
\mathcal{H}^C_{\rho,\sigma,\kappa,\nu}(x,y,z,x_{\rm op}) = - \sum_{q = u, d, s} \left(\frac{e_q}{e}\right)^4 \times 
\Bigl\langle \mathrm{tr} \bigl[ \gamma_{\rho} S_q (x, z) \\ \cdot \\ \gamma_{\kappa} S_q (z, y) \gamma_{\sigma} S_q \left( y, x_{\text{op}} \right) \gamma_{\nu} S_q \left( x_{\text{op}}, x \right) \bigr] \Bigr\rangle_{\text{QCD}} \,.
\label{eq:lbl-amp}
\end{multline}
where $S_q$ is the propagator for a quark of flavor $q$ and electric charge $e_q$. To reduce the noise, the Ward identity on the current insertion $x_{\rm op}$ is enforced exactly, on each gauge configuration. This is done by inserting the external photon at all possible locations on the quark loop on each gauge configuration. This provides a significant increase in the statistical precision.

The moment method can be applied to the disconnected contribution as well, in this case the hadronic part reads
\begin{multline}
\mathcal{H}^D_{\rho,\sigma,\kappa,\nu}(x,y,z,x_{\rm op}) = \Bigl\langle \frac{1}{2} \Pi_{\nu,\kappa}(x_{\rm op}, z) \times \\
 \left[  \Pi_{\rho,\sigma}(x, y) - \Pi^{\rm avg}_{\rho,\sigma}(x, y) \right]  \Bigr\rangle_{\text{QCD}} 
\end{multline}
where
\begin{multline}
\Pi_{\rho,\sigma}(x, y) = -\sum_q \left( \frac{e_q}{e} \right)^2 \mathrm{Tr} \left[ \gamma_{\rho} S_q(x,y) \gamma_{\sigma} S_q(y,x) \right] \,.
\end{multline}
To evaluate this contribution, if $M$ propagators can be kept in memory, with different source positions, then $M(M-1)/2$ combination of points $(x,y)$ can be used to evaluate the integrand. For large values of $M$, it leads to a significant reduction of the noise. 

To study the finite-size effects of the form $1/L^n$ with $n\geq 2$, the RBC/UKQCD collaboration has performed simulations with different volumes: such infinite-volume extrapolations are shown in \Fig{fig:hlbl_rbc_extrap}.

\subsubsection{The RBC/UKQCD approach in infinite volume}

In Ref.~\cite{Blum:2017cer}, the RBC/UKQCD collaboration has presented first results using QED in infinite volume. This approach is more similar to the one followed by the Mainz group. Here, one still starts from \Eq{eq:moment_Xprod}, but the weight function $\mathcal{G}$ is now evaluated in the continuum and infinite-volume limits. The 12-dimensional integration is reduced to a 4-dimensional integration by analytic calculations. The resulting expression is then integrated numerically. To reduce the associated numerical cost, the weight function is pre-computed on a grid of points and an approximation is obtained by interpolating the computed values. A subtraction of the QED weight function has been used to reduce the noise. 

\begin{figure}[t]
	\center
	\resizebox{0.4\textwidth}{!}{  \includegraphics{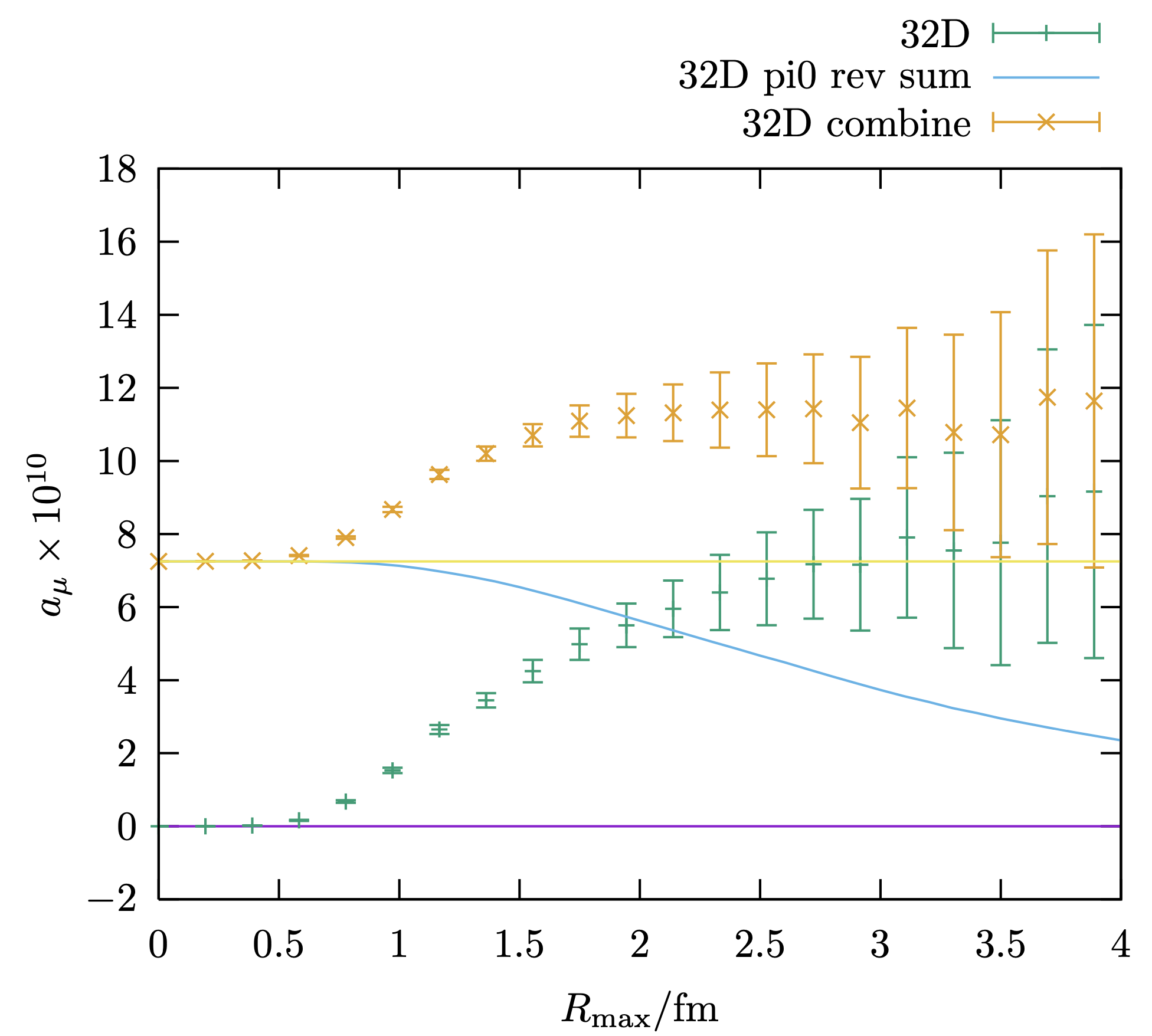}  } 
\caption{Preliminary results for the RBC/UKQCD collaboration in the QED$_{\infty}$ setup. The results, obtained for a pion mass of $m_{\pi} = 142 \MeV$, include both the connected and the dominant disconnected contribution. At short distances, the lattice data are used. At long distance, the pion-pole contribution is used (assuming an LMD model). One expects a plateau a sufficiently large value of $R_{\rm max}$  }
\label{fig:qedinf}    
\end{figure}

Preliminary results have been presented during workshops and conferences~\cite{Jin:2018wln,WorkshopSeatlle:rbc} and are shown in \Fig{fig:qedinf}. A similar strategy to the Mainz group is used: at short distance, the lattice data are used and at long distance, the prediction from the pion-pole contribution is used. Using a lattice with $a=0.2~\fm$ and $L=4.8~$fm, a plateau at $R_{\rm max} \approx 2.5~$fm is observed leading to $\ahlbl = 11.40(1.27)_{\rm stat} \times 10^{-10}$. The error is statistical only and 60\% of the results comes from actual lattice data. A more systematic study of FSE and an extrapolation to the physical point is still required.

\subsubsection{Cross check}

\begin{figure}
	\center
	\resizebox{0.4\textwidth}{!}{  \includegraphics{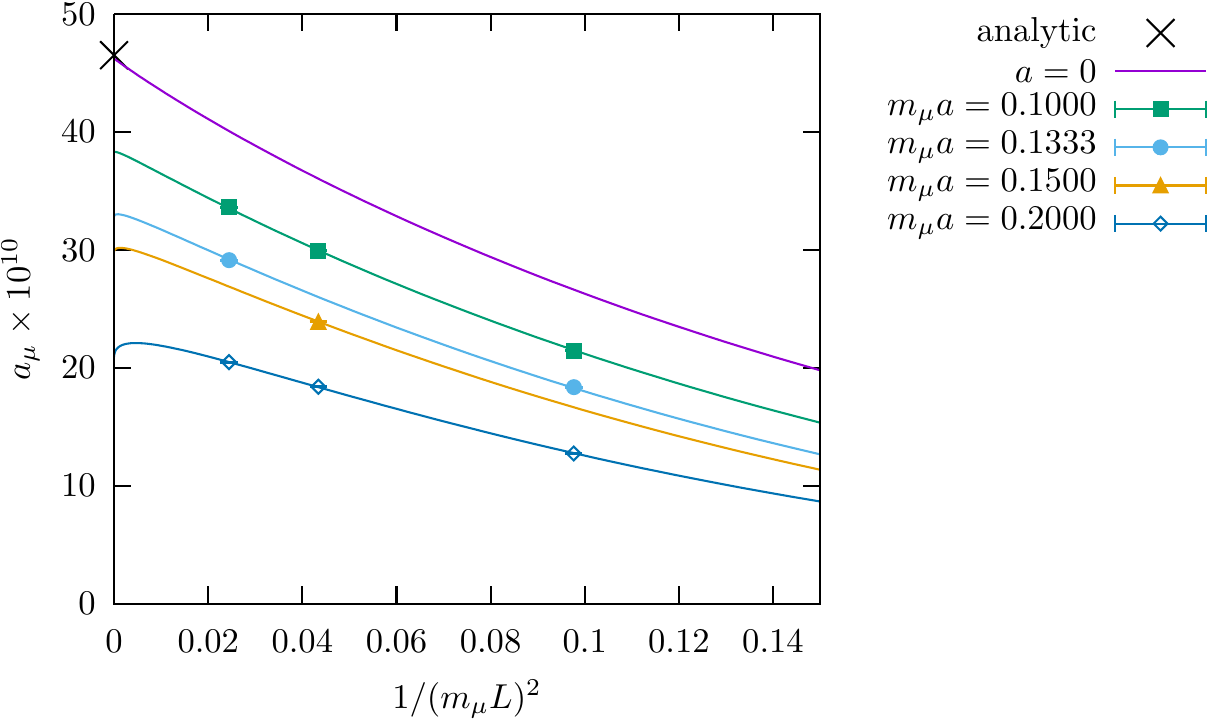}  } \\[3mm]
	\resizebox{0.4\textwidth}{!}{  \includegraphics{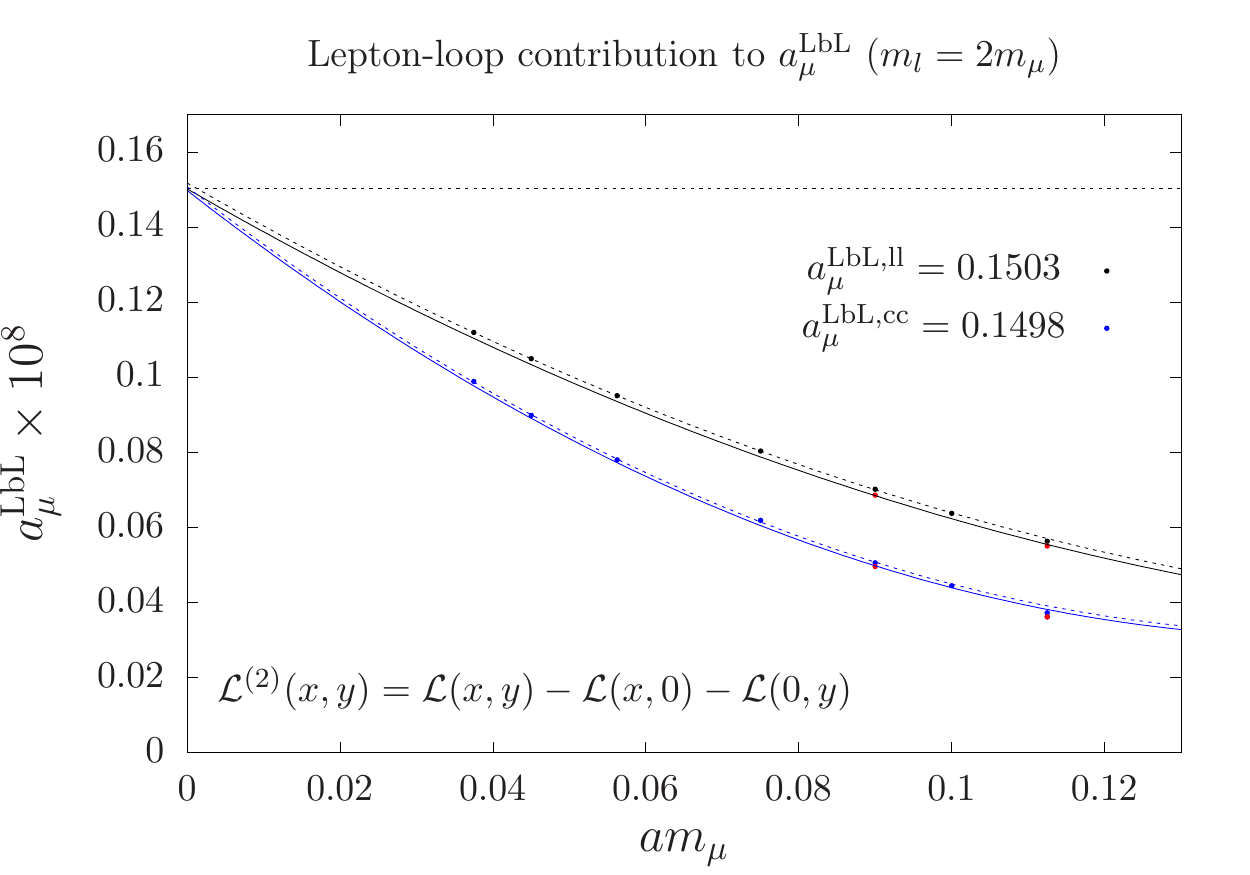}  } 
\caption{QED light-by-light scattering contribution to the muon anomaly. The top panel if the result from the RBC collaboration using the method described in Section~\ref{sec:hlbl_qedL}. Different colors correspond to different lattice spacings and the cross is the extrapolated value. The bottom panel shows the Mainz results using the method described in \Section{sec:hlbl_qedInf}. The two colors correspond to two different discretizations of the vector current, the dashed lines does not include the FSE correction.}
\label{fig:hlblcheck}    
\end{figure}

Both the RBC/UKQCD and the Mainz collaborations have tested their method and lattice setup in pure QED to reproduce the well-known lepton-loop contribution. This contribution enters at order $\alpha^3$ in the QED contribution and is known exactly~\cite{Laporta:1992pa}. The results for both collaboration are shown in \Fig{fig:hlblcheck}. In this extrapolation, finite-size effects are taken into account by both collaborations.

\subsubsection{Results for $\ahlbl$}

This year, the RBC/UKQCD has published the first ab-initio calculation of the the hadronic light-by-light scattering contribution to the anomalous magnetic moment of the muon, at the physical pion mass, with a continuum extrapolation and an estimate of finite-size effects corrections~\cite{Blum:2019ugy}. The study is based on Domain-Wall fermions at the physical point using two sets of ensembles generated with different gauge actions. Simulations include the fully connected contribution for the light quark as well as the leading $2+2$ quark disconnected contribution with both the light and the strange quarks. The latter contributes only at the level of 5\% of the disconnected contribution.

\begin{figure}
	\center
\resizebox{0.43\textwidth}{!}{  \includegraphics{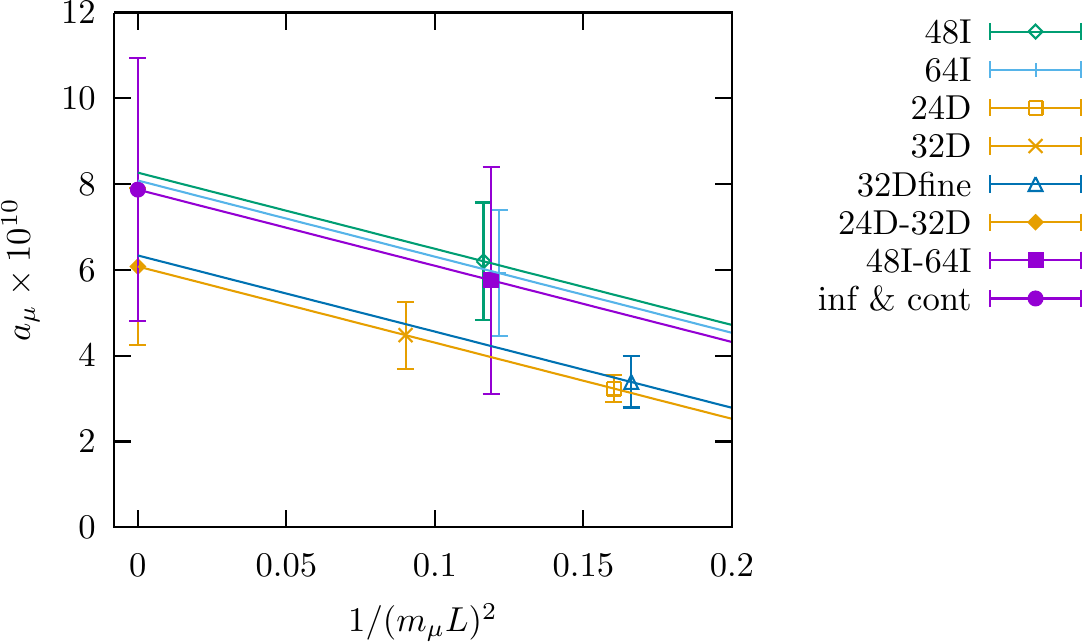}  } 
\resizebox{0.43\textwidth}{!}{  \includegraphics{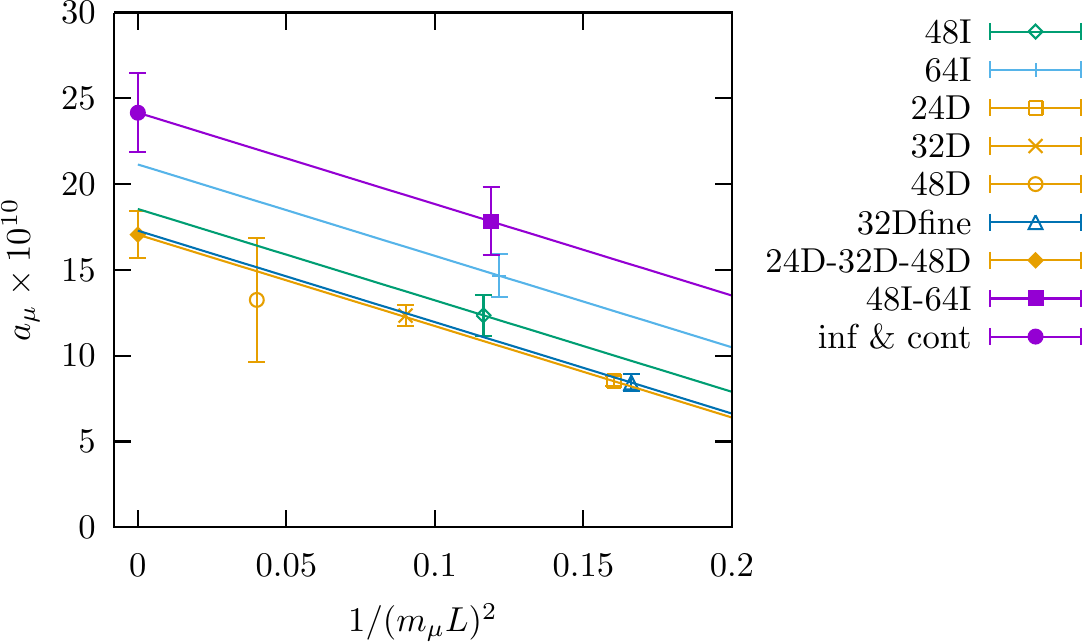}  } 
\resizebox{0.43\textwidth}{!}{  \includegraphics{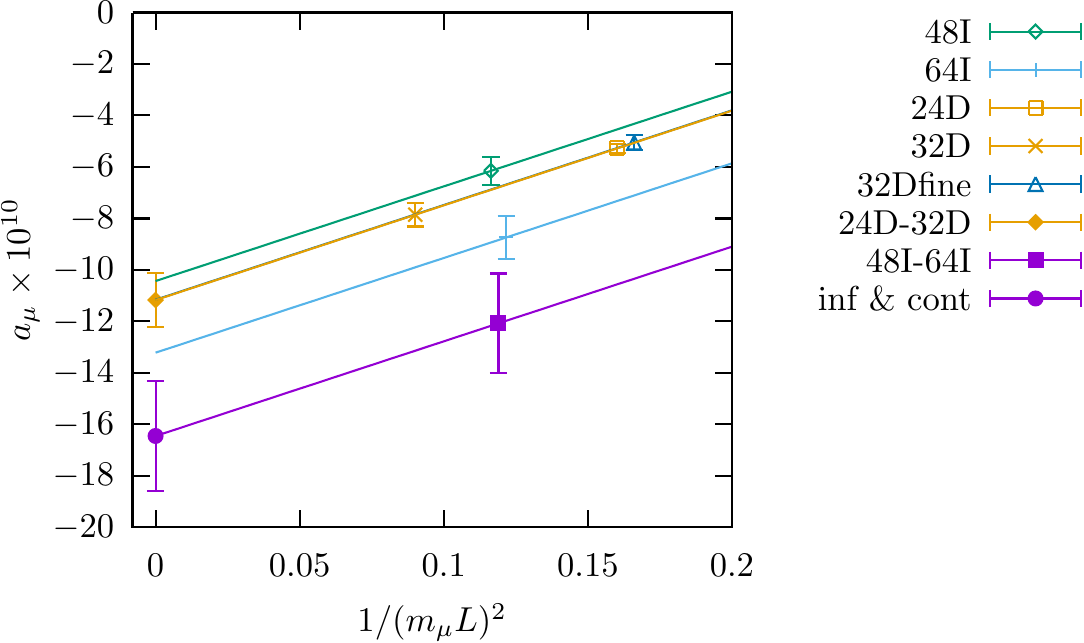}  }
\caption{Infinite volume extrapolation for the RBC/UKQCD collaboration using the QED$_L$ formalism. Connected (top), (2+2) disconnected (middle), and total (bottom). The continuum limit (purple line) is obtained from the two ensembles 48I and 64I sharing the same action. The other ensemble are used to performed the infinite volume extrapolation only. Figure extracted from~\cite{Blum:2019ugy}. }
\label{fig:hlbl_rbc_extrap}       
\end{figure}

The result is obtained using the QED$_L$ formulation described above. A part of the results, on a single ensemble, have been published earlier in~\cite{Blum:2016lnc}. This update includes several lattice spacings, as well as different volumes. Two ensembles using the same action are used to perform the continuum extrapolation and three (two) other ensembles are used to constrain the infinite-volume extrapolation of the quark connected (disconnected) contribution. The value at the physical point reads
\begin{equation}
a_{\mu}^{\rm tot} = 7.87(3.06)_{\rm stat}(1.77)_{\rm syst} \times 10^{-10} \,,
\label{eq:res_hlbl_rbc}
\end{equation}
where the total error is dominated by statistics. It corresponds to a precision of about $45\%$. The systematic error is dominated by finite-size effects and by the continuum extrapolation, see \Fig{fig:hlbl_rbc_extrap}. A large cancellation between the connected and leading disconnected contributions is observed, as expected. In fact the two contributions taken individually are known with much better relative precision
\begin{align}
a_{\mu}^{\rm conn} = 23.76(3.96)_{\rm stat}(4.88)_{\rm syst} \times 10^{-10} \,, \\
a_{\mu}^{\rm disc} = -16.45(2.13)_{\rm stat}(3.99)_{\rm syst} \times 10^{-10}\,.
\end{align}
The result (\ref{eq:res_hlbl_rbc}) is compatible with the most precise determination based on the dispersive approach~\cite{Masjuan:2017tvw,Colangelo:2017fiz,Hoferichter:2018kwz,Bijnens:2019ghy,Colangelo:2019uex,Danilkin:2016hnh,Jegerlehner:2017gek,Knecht:2018sci,Eichmann:2019bqf,Roig:2019reh} (see \Table{tab:status}), although with larger error-bars. It demonstrates the feasibility of a lattice QCD calculation with controlled statistical and systematic errors. Further studies, in particular to better control the extrapolation to the physical point, are needed to reach 10\% level accuracy.\\

\begin{figure}[h!]
	\center
\resizebox{0.45\textwidth}{!}{  \includegraphics{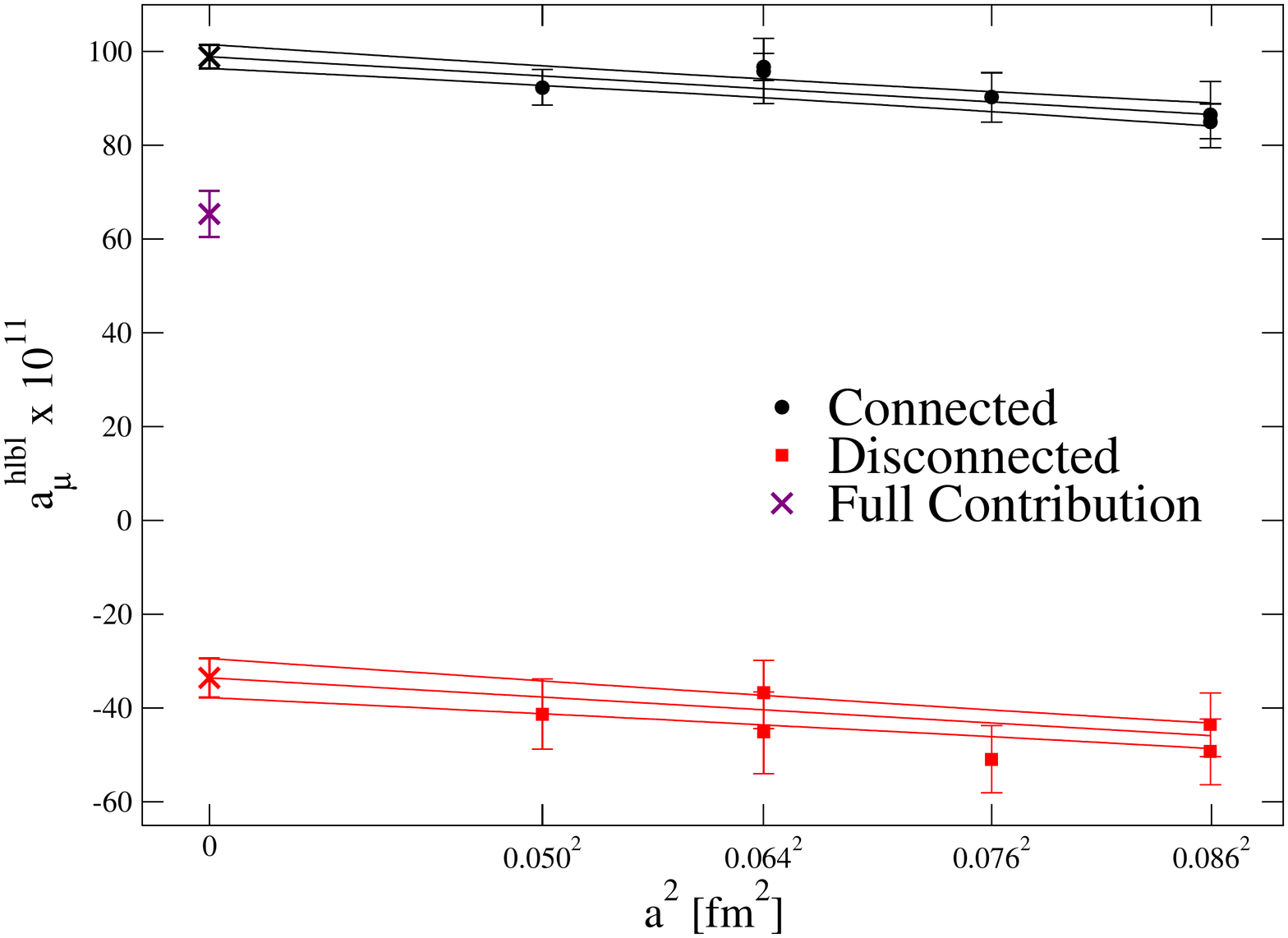}  } 
\caption{Combined continuum-extrapolation analysis for the connected (black) and disconnected (red) data. The purple cross represents the addition of the continuum-extrapolated results for the connected and disconnected contributions. Figure extracted from~\cite{Chao:2020kwq}. }
\label{fig:hlbl_mainz_extrap}       
\end{figure}

This year, the Mainz group published their first result~\cite{Chao:2020kwq} for the hadronic light-by-light contribution to the anomalous magnetic moment of the muon, in the continuum limit, but at the $SU(3)$-flavor symmetric point with $m_{\pi} = m_K \approx 400~\MeV$. The results are obtained with $N_f=2+1$ Wilson fermions and several ensembles have been used to study the main sources of systematic errors : FSEs and discretization effects. The restriction to the $SU(3)_f$ point is motivated by the moderate computational cost as compared to the physical pion mass ensembles and the reduced number of disconnected diagrams that need to be evaluated. In this limit, only the 2+2 disconnected contribution survives since other diagrams with at least one single loop vanish due to the  charge factors. The result in the continuum limit reads
\begin{equation}
a_{\mu}^{\rm tot} = 65.4(4.9)_{\rm stat}(6.6)_{\rm syst} \times 10^{-11} \,.
\label{eq:resMainz}
\end{equation}
and corresponds to a precision better than 15\%. The extrapolation to the continuum limit is shown in \Fig{fig:hlbl_mainz_extrap} and, as for the RBC/UKQCD collaboration, the statistical correlations between the connected and disconnected contributions are found to be very small.

Subtracting from \Eq{eq:resMainz} their lattice result for the pion-pole contribution discussed in \Section{sec:hlblpi0}, and assuming that the remaining part has a mild chiral dependance, the Mainz group obtains a prediction at the physical pion mass
\begin{multline}
a_{\mu}^{ {\rm hlbl}, SU(3)_f} - a_{\mu}^{ {\rm hlbl}, \pi^0, SU(3)_f}  + a_{\mu}^{ {\rm hlbl}, \pi^0, {\rm phys}} = \\ (104.1 \pm 9.1) \times 10^{-11} \,.
\end{multline}
Using a slightly more sophisticated method to correct for the quark-mass effects, that originate from non pion-pole contributions, the authors associate an additional error of 20\% to this result. Additional ensembles are required to perform a more careful chiral extrapolation and first results have been presented during workshops and conferences~\cite{Asmussen:2019act,MainzWorkshop}.
This value is higher than (\ref{eq:res_hlbl_rbc}) but compatible within error bars. It is also in good agreement with the most recent dispersive results~\cite{Masjuan:2017tvw,Colangelo:2017fiz,Hoferichter:2018kwz,Bijnens:2019ghy,Colangelo:2019uex,Danilkin:2016hnh,Jegerlehner:2017gek,Knecht:2018sci,Eichmann:2019bqf,Roig:2019reh}.

\subsubsection{Sub-dominant disconnected contributions}

The RBC/UKQCD collaboration also published results for the first sub-leading diagram~\cite{Blum:2019ugy} (third diagram in \Fig{fig:hlbl}). The latter is evaluated on a single ensemble using the QED in infinite-volume setup. In practice, the whole diagram is not computed and only the expected dominant Wick contraction is included. Above 1~fm the contribution is consistent with zero and this result is used to set a bound of the neglected diagrams : $|a_{\mu}^{\rm hlbl; 3+1}| < 0.5 \times 10^{-10}$.

\begin{figure}[h!]
\begin{center}
\resizebox{0.35\textwidth}{!}{  \includegraphics{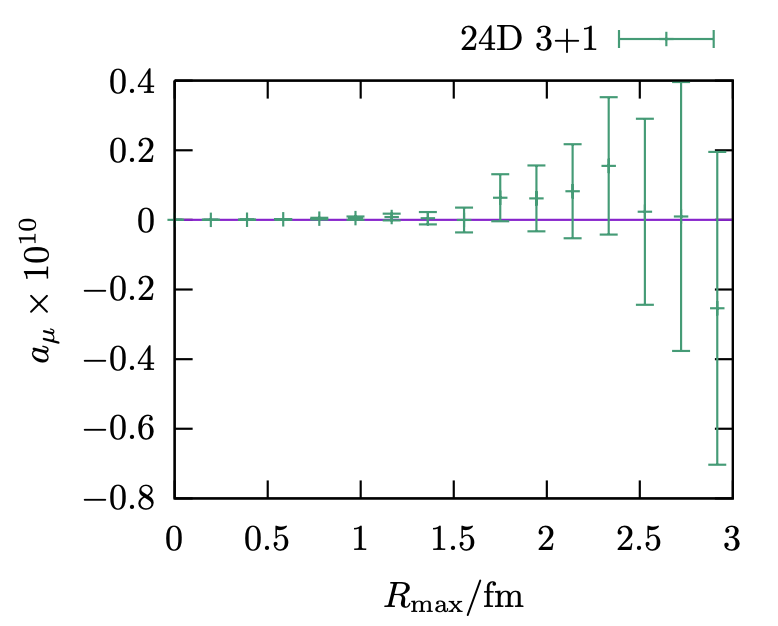}  } 
\resizebox{0.35\textwidth}{!}{  \includegraphics{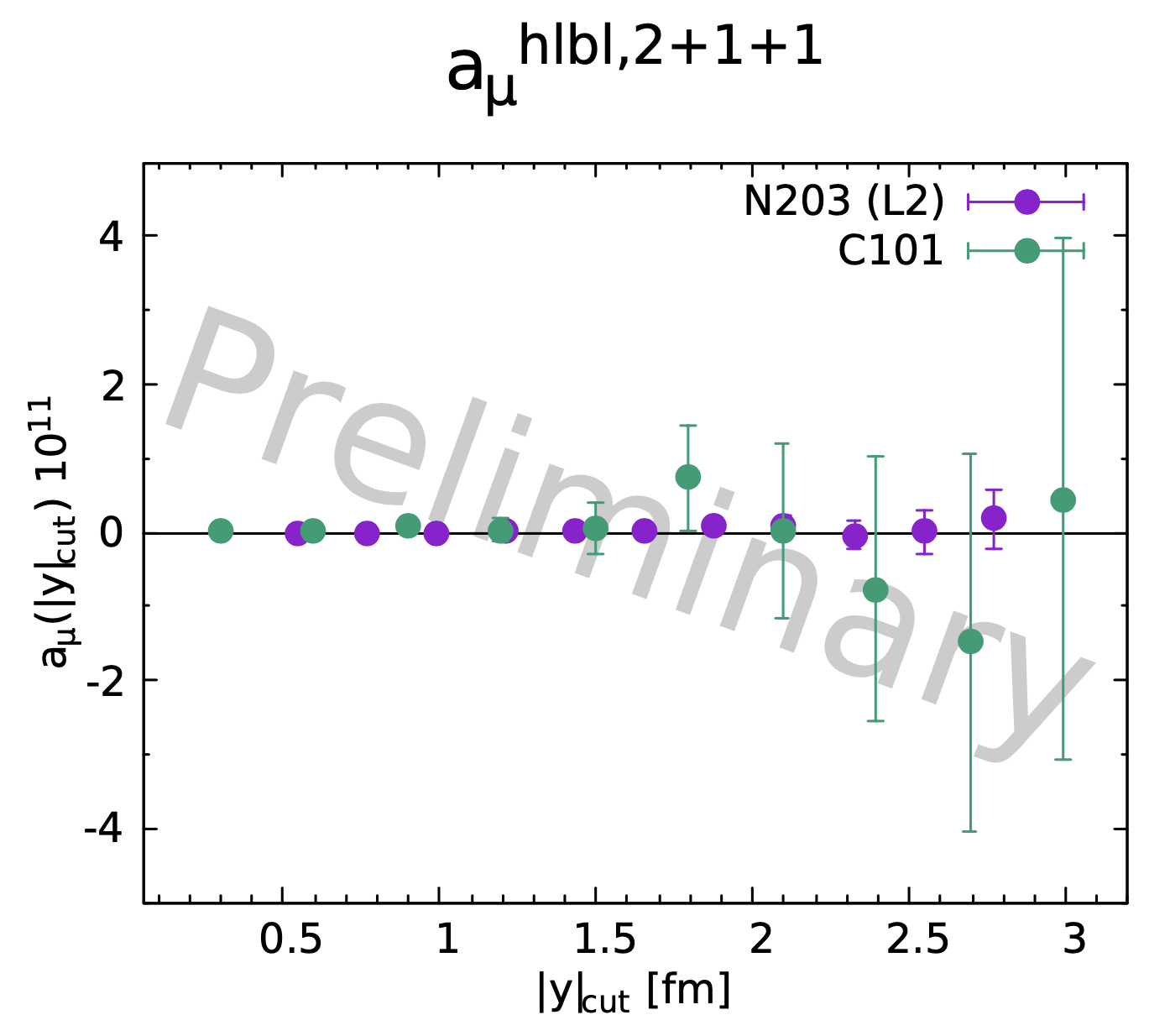}  } 
\resizebox{0.35\textwidth}{!}{  \includegraphics{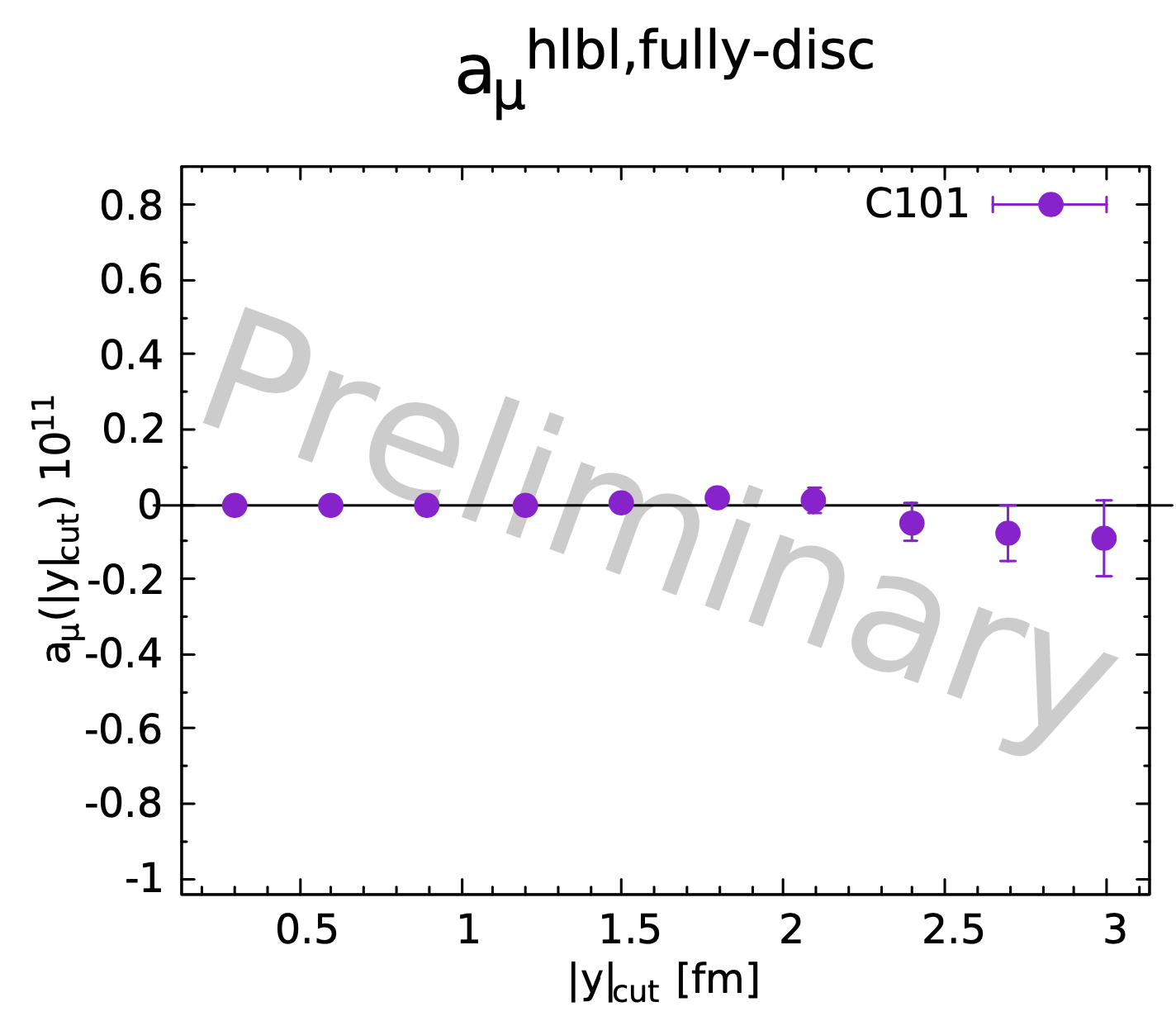}  } 
\end{center}
\caption{Integrated value for the sub-dominant contributions to $\ahlbl$ as a function of the integration range. For the (3+1) topology~\cite{Blum:2019ugy} and the (2+1+1) and (1+1+1+1) topologies~\cite{Main2020}.}
\label{fig:hlbl_sub}      
\end{figure}

The Mainz group has presented preliminary results for all sub-dominant disconnected contributions~\cite{Main2020}. The study is performed on ensembles with pion masses down to 220~MeV (Fig.~\ref{fig:hlbl_sub}). The signal is lost at short distances but it provides important upper bounds on the amplitude of the contributions : they appear to be negligible at the level of precision required by future experiments and they confirm the observation made by the RBC/UKQCD collaboration on the $3+1$ topology. A more systematic study, that includes a proper continuum and chiral extrapolation, as well as an estimate of the FSE correction, would be welcome to confirm this observation.

\subsection{The pseudoscalar-pole contributions}
\label{sec:hlblpi0}

In the dispersive approach to the HLbL contribution to te muon $g-2$, the dominant contribution is given by the light pseudoscalar-pole contributions, namely the pion, the $\eta$ and the $\eta^{\prime}$. According to recent estimates~\cite{Aoyama:2020ynm,Masjuan:2017tvw}, the pion-pole contributes roughly two times more than the $\eta$- and $\eta^{\prime}$-pole together. Furthermore, the two singlet mesons have a comparable contribution in magnitude and are therefore not negligible. In~\cite{Aoyama:2020ynm}, the best estimate for the HLbL contribution to the muon $(g-2)$ is $\ahlbl = 92(19) \times 10^{-11}$ while the pseudoscalar-pole contribution itself is $a_{\mu}^{\rm hlbl, ps} = 93.8(4.0) \times 10^{-11}$ and large cancellations are observed for the remaining, smaller, contributions. Thus, in parallel to the direct lattice calculation of the HLbL contribution, a first-principle determination of the light pseudoscalar-pole contribution is an important challenge for lattice QCD. In addition, this calculation provides valuable information for the direct lattice calculation presented in the previous section. It can be used to reduce significantly the statistical error but also to estimate the dominant source of systematic error. See~\Section{sec:hlbl_qedInf}.

\begin{figure}[t!]
\begin{center}
\resizebox{0.2\textwidth}{!}{  \includegraphics{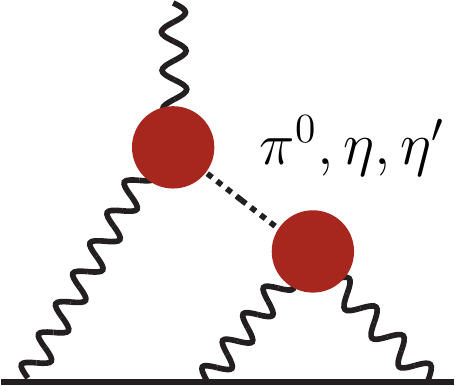}  } 
\end{center}
\caption{Pseudoscalar-pole contribution to hadronic light-by-light scattering in the muon $g-2$. The blobs on the represent the $T \to \gamma^* \gamma^*$
transition form factors with $P=\pi^0, \eta, \eta^{\prime}$}
\label{fig:pionpole}      
\end{figure}

The lattice calculation of the pion-pole contribution is based on the definition of the pseudoscalar-pole contribution of Ref~\cite{Jegerlehner:2009ry} and the result is obtained as a convolution integral between some QED weight functions and a product of two transition form factors (TFF) as shown in \Fig{fig:pionpole}. The formula involves only space-like momenta and the dominant part of the signal come from virtualities below $2~\GeV^2$. Those two points make lattice QCD an excellent candidate for a first-principle determination. 

The pion-pole contribution to the HLbL diagram has been computed in~\cite{Gerardin:2016cqj,Gerardin:2019vio} with a precision of 6\% and previous lattice studies have focused on the normalization of the TFF~\cite{Feng:2012ck}.  With the aim of a 10\% precision on the full HLbL amplitude, a precision of 20\% on the  $\eta$ and $\eta^{\prime}$ might already suffice. In this case, the lattice calculation is more challenging due to the mixing between the two states. There is  currently no first-principle determination of the TFF but first preliminary results have been presented by the ETM collaboration~\cite{APLAT20:etmc}. Since experimental data are very sparse and not available in the kinematical range of interest, in particular for the doubly-virtual form factor, a first, even imprecise, lattice calculation would be extremely valuable.

In the next subsections, I start with a brief description of the methodology used to compute the pseudoscalar transition form factors. Then I present the state of the art of such calculations before concluding on possible future directions.

\subsubsection{Methodology}

In Minkowski space-time, the pion transition form factor is given by the following matrix element
\begin{eqnarray} 
\nonumber M_{\mu\nu}(p,q_1)  &=&
i \int \mathrm{d}^4 x \, e^{i q_1 \cdot x} \, \langle \Omega | T \{ J_{\mu}(x) J_{\nu}(0) \} | \pi^0(p) \rangle \\
&=& \epsilon_{\mu\nu\alpha\beta} \, q_1^{\alpha} \, q_2^{\beta} \, \FF(q_1^2, q_2^2) \,,
\label{eq:M}
\end{eqnarray}
where $p=q_1 + q_2$ and $\epsilon_{\mu\nu\alpha\beta}$ is the fully antisymmetric tensor with $\epsilon^{01234} = +1$. In Ref~\cite{Gerardin:2016cqj,Gerardin:2019vio} the authors have computed the following three-point correlation fonction
\begin{equation}
C^{(3)}_{\mu\nu}(\tau,t_{\pi}) = a^6\sum_{\vec{x}, \vec{z}} \, \big\langle   J_{\mu}(\vec{z}, t_i) J_{\nu}(\vec{0}, t_f)  P^{\dag}(\vec{x},t_0) \big\rangle \, e^{i \vec{p}\, \vec{x}} \, e^{-i \vec{q}_1 \vec{z}}
\label{eq:C3} 
\end{equation}
where $\tau=t_i-t_f$ is the time separation between the two vector currents and  $t_{\pi}={\rm min}(t_f-t_0,t_i-t_0)$ is the minimal time separation between the pion interpolating operator and the two vector currents. To project onto the pion state, one can define the amplitude
\begin{equation}
\widetilde{A}_{\mu\nu}(\tau) \equiv \lim_{t_{\pi} \rightarrow + \infty} e^{E_\pi (t_f-t_0)} C^{(3)}_{\mu\nu}(\tau,t_{\pi}) \,,
\label{eq:Amunu}
\end{equation}
from which the matrix element of interest can be extracted \cite{Ji:2001wha,Ji:2001nf}
\begin{equation}
 M_{\mu\nu}^{\rm E} = \frac{2 E_{\pi}}{ Z_{\pi} }  \int_{-\infty}^{\infty} \, \mathrm{d}\tau \, e^{\omega_1 \tau} \, \widetilde{A}_{\mu\nu}(\tau) \,.
\label{eq:Mlat}
\end{equation}
In this equation, $E_{\pi}$ is the pion energy and $Z_{\pi}=\langle0|P(0)|\pi\rangle$ is the overlap of the pseudoscalar operator with the pion state. The free parameter $\omega_1$ allows to scan different photon virtualities through the relation
\begin{align}
\begin{aligned}
q_1^2 &= \omega_1^2 - \vec{q}_1^{\, 2} \\
q_2^2 &= (E_{\pi} - \omega_1)^2 - (\vec{p}-\vec{q}_1)^2
\end{aligned}
\end{align}
with
\begin{equation}
\vec{q}_1 = \frac{2\pi}{L} \vec{n} \,, \quad \vec{n} \in \mathbb{Z}^3 \,,
\end{equation}
and $L$ the spatial extent of the lattice. Using several values of the pion momenta $\vec{p}$ and photon virtuality $\vec{n}$, it is possible to scan a dense region in the $(Q_1^2, Q_2^2)$ plane, especially for large volumes. It should be noted that \Eq{eq:Mlat} is valid only below the threshold $s_0 = 4 m_{\pi}^2$ (assuming the rho meson is heavier). For the muon $g-2$, this limitation is of little relevance since we are interested in space-like virtualities. 
In practice, when using \Eq{eq:Mlat}, one is confronted to similar problems as for the HVP: first, the integration range is limited by the finite time extent of the lattice and second, the noise over signal increases rapidly at large $\tau$ and a careful treatment of the tail is required. 

\subsubsection{Results for the pion-pole contribution}

A first feasibility study was presented in~\cite{Cohen:2008ue}. The results are based on CP-PACS gauge configurations using $2+1$ dynamical flavors of clover fermions at a single lattice spacing $a^{-1} \approx 2.25$~GeV and with a heavy pion mass $M_{\pi} = 725$~MeV. The authors were able to describe their lattice data for the doubly-virtual transition form factor assuming a simple vector-meson dominance (VMD) model with a pole mass approximately equal to the corresponding rho mass.

In a later study~\cite{Lin:2013im}, the pion transition form factor was computed at a single lattice spacing $a\approx 0.12~\fm$ and for pion masses in the range [830-310]~MeV, using gauge configurations from the Hadron Spectrum Collaboration. The later are based on anisotropic clover lattices with $2+1$ dynamical quarks. In the single-virtual case, the study supports the validity of the vector-meson dominance model. In the doubly-virtual case, the data suggest that the vector-meson dominance model does not work very well.

In Refs~\cite{Gerardin:2016cqj,Gerardin:2019vio}, a first systematic study of the TFF, with a proper continuum and chiral extrapolation, as well as an estimate of the pion-pole contribution to the HLbL diagram, was published. The TFF was computed for several values of $Q_1^2$ and $Q_2^2$ to cover the full Euclidean region, below $4~\GeV^2$ in the $(Q_1^2,Q_2^2)$ plane, relevant for the $g-2$. The authors used non-perturbatively O(a)-improved Wilson fermions with $N_f = 2+1$ dynamical quarks. Four lattice spacings in the range $[0.5-0.9]~\fm$ and two lattice discretizations of the vector curent, that differ by $O(a^2)$ lattice artifacts, were used to perform the continuum extrapolation. Several pion masses, down to $200~\MeV$ allow an extrapolation to the physical pion mass. Finite-size effects have been found to be negligible compared to the statistical precision and both the connected and the disconnected contributions have been included; the later contributes at the level of 1\%, and is a sub-leading source of error in this calculation.

\begin{figure*}[t]
	\center
	\resizebox{0.45\textwidth}{!}{  \includegraphics{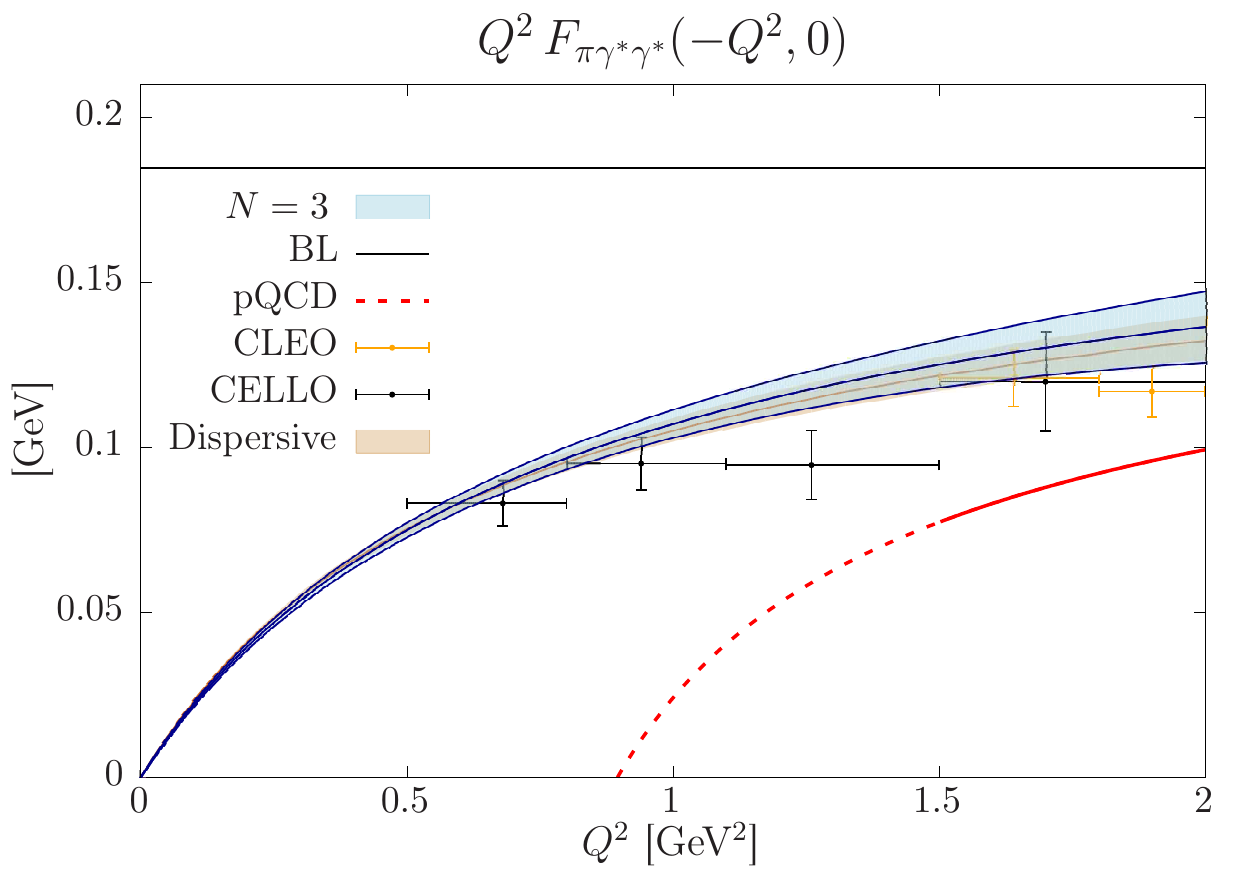}  } 
	\resizebox{0.45\textwidth}{!}{  \includegraphics{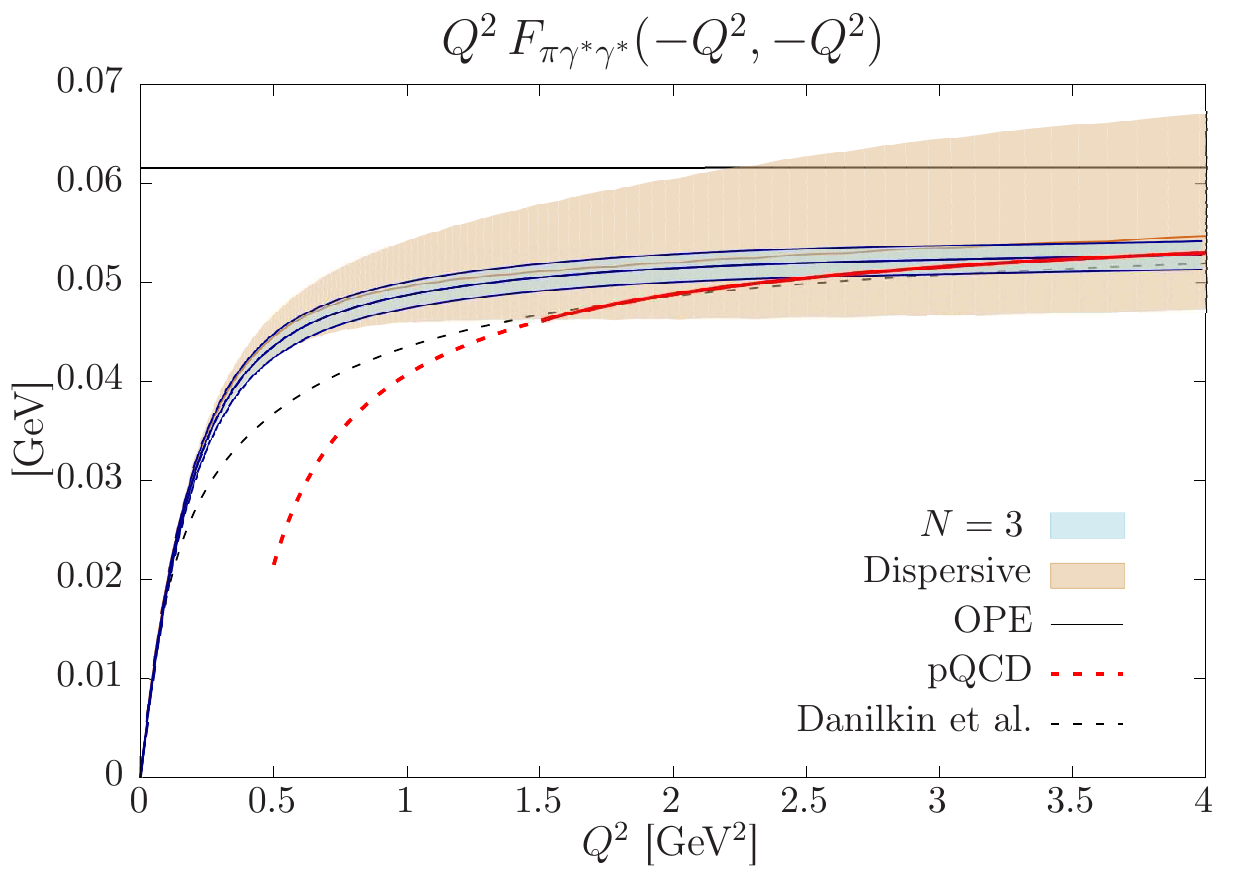}  } 
\caption{Result for the calculation of the pion transition form factor published by the Mainz group~\cite{Gerardin:2019vio}. Left: single virtual transition form factor where a comparison with experimental data is possible. Right: prediction for the double-virtual transition form factor. A confirmation from the dispersive framework is also shown.}
\label{fig:pionTFF}
\end{figure*}

The extrapolation to the physical point was performed using a modified $z$-expansion, which satisfies the kinematical and short-distances contraints,
\begin{multline}
P(Q_1^2,Q_2^2) \ \FF(-Q_1^2, -Q_2^2)  = \\  \sum_{n,m=0}^{N} c_{nm} \, \left( z_1^n - (-1)^{N+n+1} \frac{n}{N+1} \, z_1^{N+1} \right) \, \\ \left( z_2^m -  (-1)^{N+m+1} \frac{m}{N+1} \, z_2^{N+1} \right) \,.
\label{eq:z_exp_mod}
\end{multline}
with 
\begin{equation}
z_k = \frac{ \sqrt{t_c+Q_k^2} - \sqrt{t_c - t_0} }{ \sqrt{t_c+Q_k^2} + \sqrt{t_c - t_0} } \,, \quad k=1,2 \,,
\label{eq:zvar}
\end{equation}
and $t_c = 4m_{\pi}^2$. Here, the free parameter $t_0$ was chosen to reduce the maximum value of $|z_k|$ in the range $[0, Q_{\max}^2]$. The imaginary part of the TFF behaves as $(q^2 - t_c)^{3/2}$ near threshold (P-wave) and this contraint has been implemented in \Eq{eq:z_exp_mod} by imposing~\cite{Bourrely:2008za}
\begin{equation}
\left[ \frac{  \mathrm{d} \FF  }{ \mathrm{d}z_k } \right]_{z_k=-1} = 0 \,, \quad k=1,2 \,.
\end{equation}
The function $P(Q_1^2,Q_2^2)$ is an arbitrary analytical function. The choice
\begin{equation}
P(Q_1^2,Q_2^2) = 1 + \frac{Q_1^2 + Q_2^2}{M_V^2} \,,
\label{eq:P_zexp}
\end{equation}
where $M_V = 775~\MeV$ is the vector meson mass ensures that the TFF has the correct short-distance behavior, as predicted by the Brodsky-Lepage scaling,  in the single-virtual case and by the OPE in the double-virtual case~\cite{Lepage:1979zb,Lepage:1980fj,Brodsky:1981rp,Nesterenko:1982dn,Novikov:1983jt}. At finite value of $N$, the TFF decreases asymptotically as $1/Q^2$ in all directions in the $(Q_1^2,Q_2^2)$ plane. The advantage of this approach is that it is systematically improvable as more precise data become available.

In practice, the value $N=3$ was used and was sufficient to describe the TFF in the whole kinematic range. The final result was obtained through a global fit over 13 ensembles and the parameters $c_{nm}$ were expanded to linear order in $a^2$ and $m_{\pi}^2$, to account for discretization effects. The results for two specific kinematics are shown in \Fig{fig:pionTFF}. The lattice results are in good agreement with the experimental data in the single virtual case. In the doubly-virtual case, no experimental data exist, and the results are in good agreement with other determination based on dispersion relations~\cite{Hoferichter:2014vra,Hoferichter:2018dmo,Hoferichter:2018kwz} and Canterbury approximants~\cite{Masjuan:2017tvw}.

Using the parametrization of the TFF in the continuum limit and at the physical pion mass, one can estimate the pion-pole contribution to the HLbL. The formalism was derived in~\cite{Knecht:2001qf} and the result reads
\begin{equation}
a_{\mu}^{\mathrm{HLbL}; \pi^0} = (59.7 \pm 3.4 \pm 0.9 \pm 0.5) \times 10^{-11} \,, 
\label{eq:amu_zexp}
\end{equation} 
where the first error is statistical, the second is the systematic error associated with the parametrization. This corresponds to a relative precision of 6\%. In fact, this purely lattice result can be further improved if the normalization of the TFF is constrained by the experimental result~\cite{Larin:2010kq}
\begin{equation}
a_{\mu}^{\mathrm{HLbL}; \pi^0} = (62.3 \pm 2.0 \pm 0.9 \pm 0.5) \times 10^{-11} \,, 
\end{equation} 
Both results are in very good agreement with other determinations based on a dispersive analysis $a_{\mu}^{\mathrm{HLbL}; \pi^0} = 63.0^{+2.7}_{-2.1} \times 10^{-11}$~\cite{Hoferichter:2014vra,Hoferichter:2018kwz,Hoferichter:2018dmo} and Canterbury approximants $a_{\mu}^{\mathrm{HLbL}; \pi^0} = 63.6(2.7) \times 10^{-11}$~\cite{Masjuan:2012wy,Sanchez-Puertas:2016mmz}, with a comparable precision.

\subsubsection{The chiral anomaly}

An important cross-check of the lattice calculations is the normalization of the TFF.  In the chiral limit, the normalization of the transition form factor is given by~\cite{Adler:1969gk,Bell:1969ts} $\FF(0,0) = \frac{1}{4\pi^2 F}$ where $F$ the pion decay constant in the chiral limit. Away from the chiral limit, corrections have been computed up to NNLO in ChiPT~\cite{Donoghue:1986wv,Bijnens:1988kx,Kampf:2009tk}. This normalization was measured by the PrimEx-II experiment with a precision of 0.8\% and it translates to $\Gamma(\pi^0 \to \gamma\gamma) = 7.790(56)_{\rm stat}(109)_{\rm syst} \,\textrm{eV}$~\cite{Larin506}.

In~\cite{Feng:2012ck}, the authors have studied the two-photon decay on the lattice by computing the transition form factors at low virtualities using a single lattice spacing. Large FSE corrections were required and their final result reads  $\Gamma_{\pi^0 \to \gamma \gamma} = 7.83(31)(49)~$eV, compatible with the measured value. In~\cite{Lin:2013im}, the authors found $\Gamma_{\pi^0 \to \gamma \gamma} = 8.7(1.4)~$eV, where the error is statistical only. Again, this study is performed at a single lattice spacing and heavy pion masses were used to perform the chiral extrapolation. 
Finally, in~\cite{Gerardin:2019vio}, the authors performed a full continuum and chiral extrapolation to the physical point and they quote $\FF(0,0) = 0.264(8)(4)~\GeV^{-1}$, with a precision of about 3\%. It translates to $\Gamma_{\pi^0 \to \gamma \gamma} = 7.17(50)~$eV. 

Improving this result is also important for future lattice calculations. In~\cite{Gerardin:2019vio}, a significant part of the error in the estimate of the pion-pole contribution comes from very low virtualities.

\subsubsection{Future directions}

Concerning the pion-pole contribution, the current precision is already satisfactory. A more precise determination of the decay width and a comparison with experiment would be valuable but would probably require, in addition to high statistics, the study of isospin breaking effects due to the light quark mass difference. 

\begin{figure}[h!]
	\center
	\resizebox{0.4\textwidth}{!}{  \includegraphics{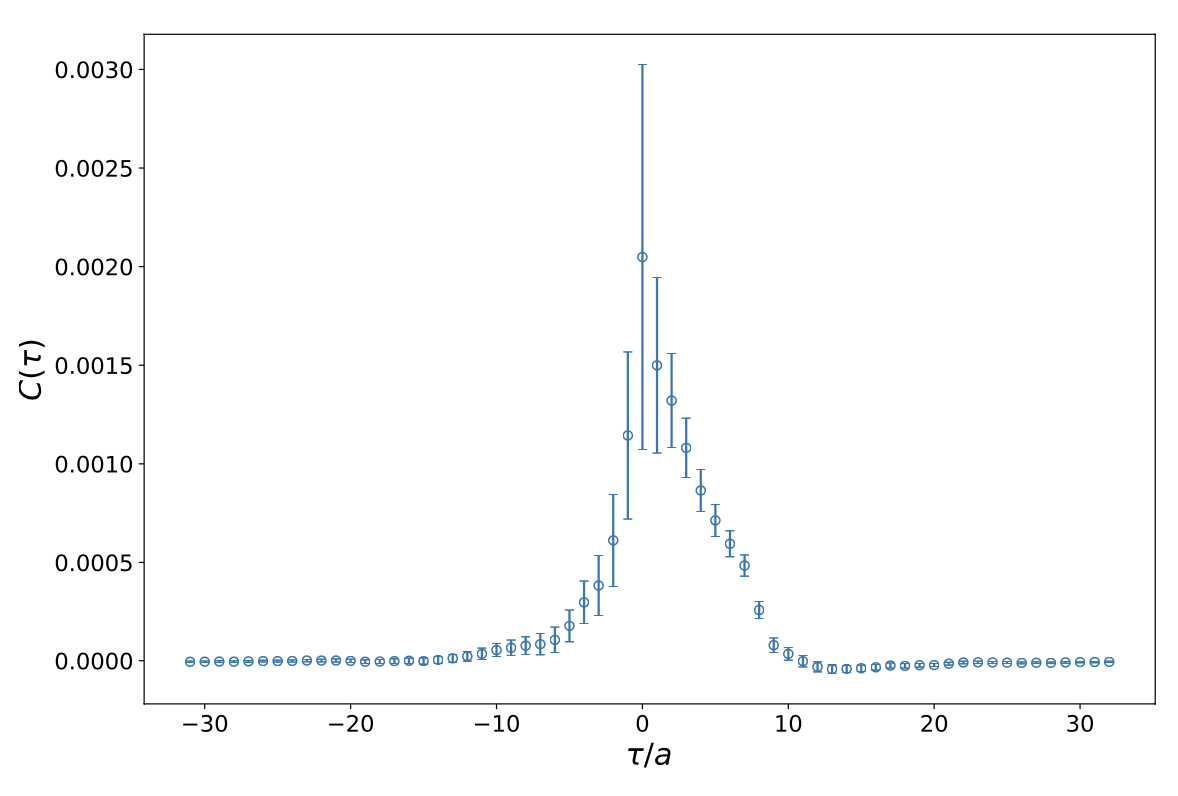}  } 
	\resizebox{0.4\textwidth}{!}{  \includegraphics{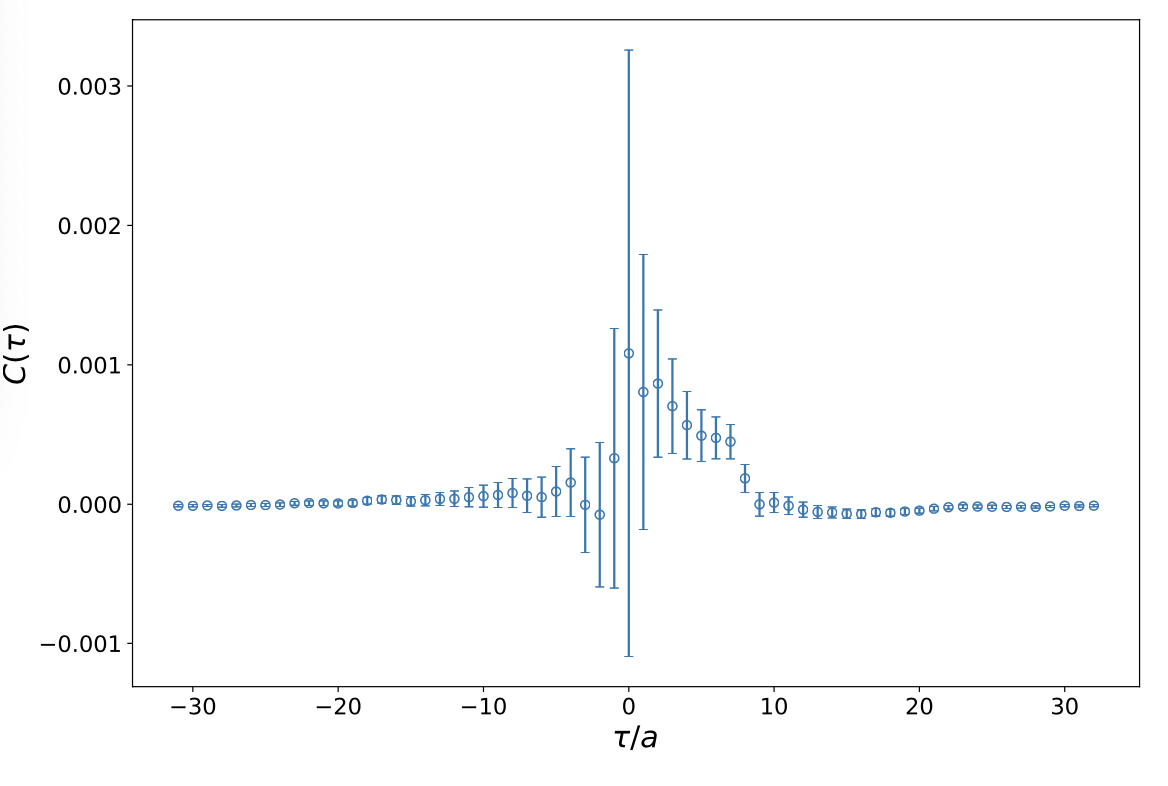}  } 
\caption{Preliminary results for the $\eta$ and $\eta^{\prime}$ transition form factor using a pion mass of 260~MeV. The integrand correspond to \Eq{eq:Amunu} for a photon virtuality $\vec{q}^2 = 1$. Plot extracted from~\cite{APLAT20:etmc}. }
\label{fig:etaTFF}
\end{figure}

In addition to the pion-pole, the $\eta$ and $\eta^{\prime}$ mesons are expected to contribute significantly. The only determination of the corresponding contribution to the HLbL diagram is based on Canterbury approximants~\cite{Masjuan:2017tvw} and a lattice QCD calculation would be extremely valuable. Here one has first to face the problem of the mixing between those two states and the extraction of the TFFs is much more challenging. However, a precision of 20\% would already suffice in view of the forthcoming experimental precision. This year, very first results in this direction have been presented by the ETM collaboration during the APLAT 2020 lattice conference~\cite{APLAT20:etmc} (Fig.~\ref{fig:etaTFF}).

\subsection{Light-by-light forward scattering amplitudes}
\label{sec:hlbl_amps}

\begin{figure*}[t]
	\center
	\resizebox{0.69\textwidth}{!}{  \includegraphics{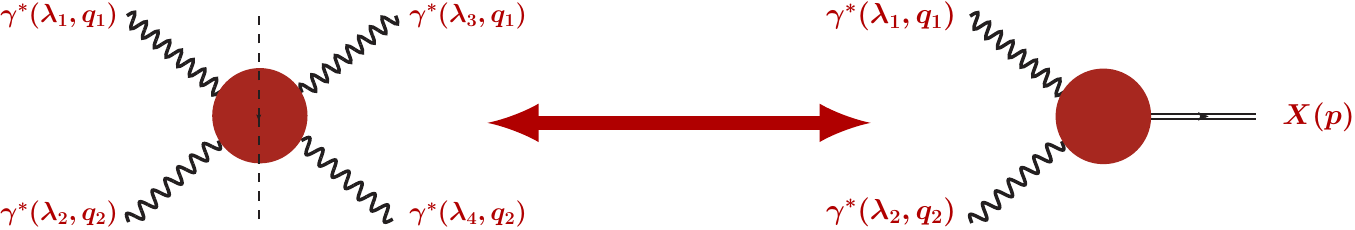}  } 
\caption{The eight light-by-light forward scattering amplitudes (left) that can be computed on the lattice. Using dispersion relation, they are related to two-photon fusion processes (right) that are described in terms of transition form factors.}
\label{fig:hlbl_amps}
\end{figure*}

The hadronic contribution to the scattering of space-like virtual photons, $\gamma^* \gamma^* \to \gamma^* \gamma^*$, can be studied on the lattice for virtualities below the hadronic threshold. From a lattice point of view, this study requires the analysis of the same four-point correlation function as the one used in the direct approach to $\ahlbl$~\Eq{eq:4pt}, but in momentum space. In particular, the knowledge of the QED weight function is not needed here. 
The motivation for such a study is that the forward scattering amplitudes can be related, using dispersive relations, to the $\gamma^* \gamma^* \to M$ fusion cross sections where $M$ stands for any C-parity even final state. 
Thus, the applicability of the hadronic models to $\ahlbl$~\cite{Jegerlehner:2009ry}, where the QCD amplitude has been approximated by the exchange of a few mesonic resonances, can be tested. Furthermore, assuming that only a few states are needed to saturate the sum rules, the lattice calculations can provide constraints on meson transition form factors and thus improve the estimate of $\ahlbl$, especially when experimental measurements are missing. It is complementary to the direct calculation of pseudoscalar transition form factors described in the previous section.

In refs.~\cite{Green:2015sra,Gerardin:2017ryf}, the authors restricted their study to the forward light-by-light scattering amplitudes with real or spacelike photons
\begin{equation}
\gamma^*(\lambda_1,q_1) \gamma^*(\lambda_2,q_2) \to \gamma^*(\lambda_3,q_1) \gamma^*(\lambda_4,q_1) 
\end{equation} 
 where $\lambda = \pm,0$ is the transverse or longitudinal helicity of the virtual photon. 
The relevant process is depicted on the left panel of \Fig{fig:hlbl_amps}. Using parity and time-reversal invariance, the 81 amplitudes associated with this process $\mathcal{M}_{\lambda_1,\lambda_2,\lambda_3,\lambda_4}$ reduce to only 8 independent amplitudes which depend on the 3 kinematical invariants $q_1^2$, $q_2^2$ and $\nu = q_1 \cdot q_2$.

For fixed photon virtualities $Q_1^2$ and $Q_2^2$, the optical theorem can be used to write sum rules that relate each of the eight amplitudes to $\gamma^* \gamma^* \to M$ fusion cross sections where $M$ stands for any C-parity-even states. In the specific case of two transverse photons, one obtains 
\begin{equation}
\overline{\mathcal{M}}_{TT}(q_1^2,q_2^2,\nu)  = 
\frac{4\nu^2}{\pi} \int_{\nu_0}^\infty \!\! d \nu^\prime \, \frac{ \sqrt{\nu^\prime-q_1^2 q_2^2} \, \sigma_{TT}(\nu^\prime) }{ \nu^\prime ( \nu^{\prime \, 2} - \nu^2-i\epsilon )} 
\label{eq:srTT}
\end{equation} 
where $\overline{\mathcal{M}}_{TT}(q_1^2,q_2^2,\nu)  = \mathcal{M}_{TT}(q_1^2,q_2^2,\nu)  - \mathcal{M}_{TT}(q_1^2,q_2^2,0) $ means that a subtraction is done and $\sigma_{TT}(\nu^\prime)$ is the total cross section $\gamma^* \gamma^* \to M$ with total helicities 0 and 2. Similar sum-rules can be written for the other 7 amplitudes (see~\cite{Pascalutsa:2012pr} for explicit formulae). 

The left hand side of \Eq{eq:srTT} is computed on the lattice starting from \Eq{eq:4pt} and using adequate projector of the four-pont correlation function, see Ref~\cite{Green:2015sra} for details. For two transverse photons, $\overline{\mathcal{M}}_{TT}$, one has 
\begin{equation}
\overline{\mathcal{M}}_{TT}(q_1^2,q_2^2,\nu) = e^4 \, P_{\mu \nu \lambda \sigma}(q_1,q_2) \, \Pi_{\mu\nu\sigma\lambda}(q_1,q_2, q_1) \,,
\end{equation} 
where $P_{\mu \nu \lambda \sigma}$ is a projector. As in the direct lattice calculation of $\ahlbl$, there are 5 classes of diagrams: the fully connected and the $(2+2)$ quark-disconnected diagrams are expected to be dominant, whereas the $(3+1)$, $(2+1)$ and $(1+1+1+1)$ disconnected contractions are expected to be subdominant~\cite{Bijnens:2016hgx,Bijnens:2017trn}. 
Assuming SU(2) flavor symmetry, if one neglects those sub-leading diagrams and if one assumes that resonance exchanges indeed dominate the amplitudes, then one can show that only isovector mesons contribute to the connected diagrams, with a weight factor $34/9 > 1$. In the (2+2) disconnected contribution, the isovector mesons contribute with the negative weight factor $-25/9$ that compensate part of the connected contribution, while the iso-scalar mesons contribute with weight factor 1. Those weight factors have been used in the comparison with the lattice data obtained with $N_f = 2$ degenerate quarks. Similar counting rules can be derived assuming SU(3) flavor symmetry~\cite{Gerardin:2017ryf}. 

In a second step, the right hand side of \Eq{eq:srTT} is obtained using a phenomenological description of the two-photon fusion processes for mesons with quantum number $J^{PC} = 0^{-+}$, $0^{++}$, $1^{++}$, $2^{++}$. 
As an example, the pseudoscalar meson contribution to $\sigma_{TT}(\nu^\prime)$ is
\begin{equation}
\sigma_{TT}  = 8 \pi^2 \delta(s-m_P^2)  \frac{ \GammaGG }{ m_P } \frac{ 2\sqrt{X} }{ m_P^2 }  \left[ \frac{  F_{{\cal P} \gamma^\ast \gamma^\ast}(Q_1^2, Q_2^2) }{ F_{{\cal P} \gamma^\ast \gamma^\ast}(0, 0)  } \right]^2 \,, 
\end{equation}
where $X$ is a kinematical factor. In addition to the pseudoscalar meson mass $m_P$, the cross section simply depends on the transition form factor of the $P \to \gamma^* \gamma^*$ process (the two-photon decay width is directly related to the normalization of the transition form factor). Similar expressions are obtained for the other channels. Thus, the only unknowns are the masses of the particles and their transition form factors. 
Hadronic resonances with different quantum numbers appear with different weight in each sum rule. If one assumes that only a few states contribute significantly, then we obtain an over-constrained system that allow us to constrain the form factors. 

\begin{figure}[b]
\vspace{-0.5cm}
\begin{center}
\resizebox{0.45\textwidth}{!}{  \includegraphics{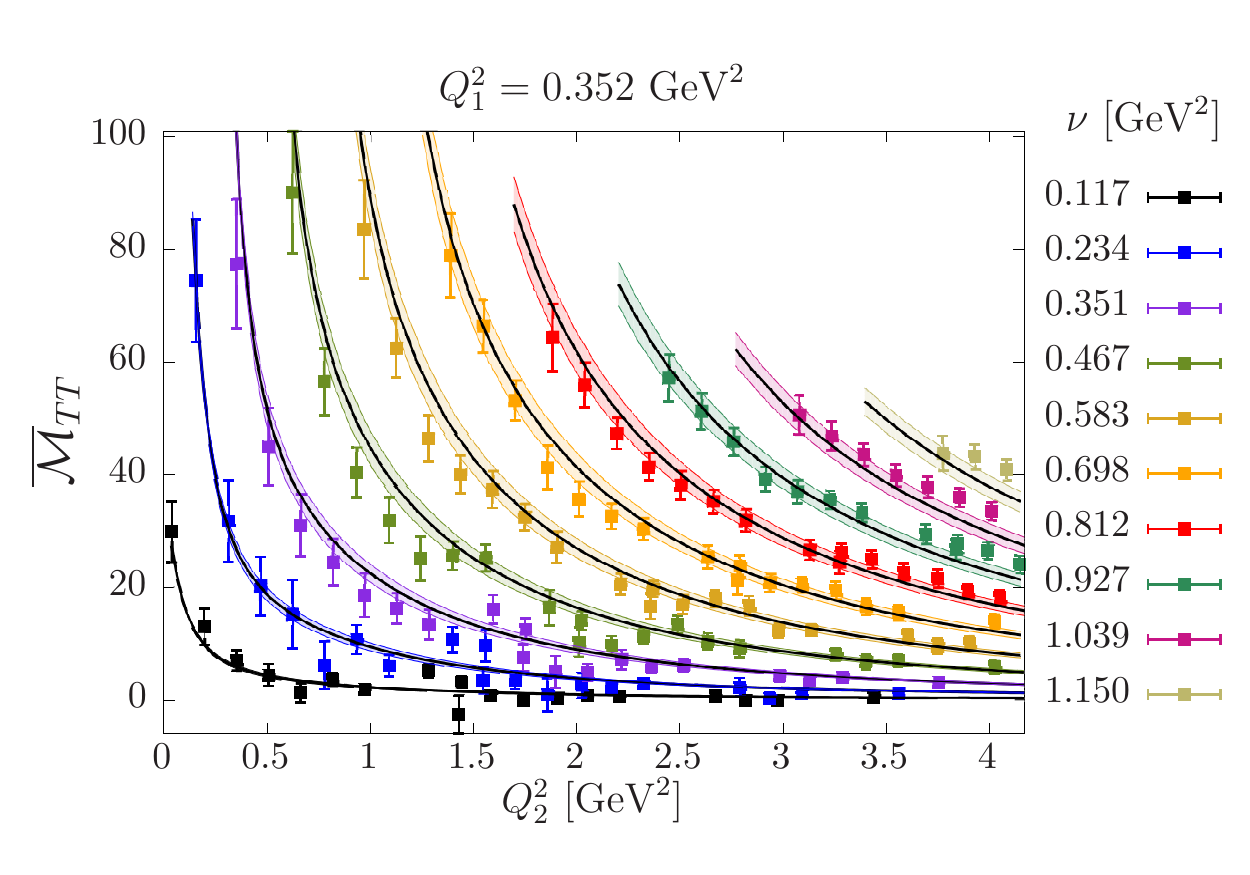}  }
\end{center}
\caption{ Amplitude $\overline{\mathcal{M}}_{TT}$ with two transverse photons. The results correspond to a lattice ensemble with a pion mass of 270 MeV. The curves with error-bands represent the result of a fit using a phenomenological model where the monopole and dipole masses of the TFFs are considered as free fit parameters. Extracted from~\cite{Gerardin:2017ryf}. }
\label{fig:amps_TT}
\end{figure}

The Mainz collaboration has computed the eight light-by-light forward scattering amplitudes~\cite{Green:2015sra,Gerardin:2017ryf} for different virtualities and an example is depicted in \Fig{fig:amps_TT}. They used $N_f=2$ Wilson fermion with two lattice spacings to study discretization effects and four pion masses to perform an extrapolation to the physical pion mass. Only the fully-connected and the dominant quark-disconnected contributions have been computed. To estimate the right hand side of the sum rules, only one resonance was included in each channel. 
Using the pion TFF from a dedicated lattice simulation~\cite{Gerardin:2019vio}, the exploratory study~\cite{Gerardin:2017ryf} found that it is possible to describe the whole set of data, for virtualities in the range [0-4]~$\GeV^2$, with a single resonance in each channel and assuming simple monopole or dipole parameterizations of the virtuality-dependance of the form factors. The results of the global fit to the eight amplitude, in the case of $\overline{\mathcal{M}}_{TT}$, is given by the colored bands in \Fig{fig:amps_TT}. A comparison of the TFFs with phenomenology has been performed. This analysis was mostly limited by the large statistical error in the disconnected diagrams and if a chiral limit has been taken, the continuum limit and the possibly large finite-size effects remain to be studied. 

A more systematic study, which includes both connected and quark-disconnected contributions as well as a study of FSEs might provide valuable information on resonance transition form factors. In a second step, the latter can be used in phenomenological models or as input in the dispersive framework to improve the determination of $\ahlbl$.

\section{Conclusion}
\label{sec:ccl}

This year, the first sub-percent estimate of the leading-order hadronic vacuum contribution to the anomalous magnetic moment of the muon has been published by the BMW collaboration and the RBC/UKQCD collaboration has published the first estimate for $\ahlbl$ at the physical point with controlled systematic errors. On the experimental side, the first results of the Fermilab experiment~\cite{Venanzoni:2014ixa} are expected to be published early next year, before the final results expected within the next few years.

Concerning the LO-HVP contribution, many collaborations have presented results in the last years. Most systematics are now properly addressed and it becomes a mature field of research. A confirmation of the recent BMW result by other groups is highly desired. 
Meanwhile, other cross-checks between lattice collaboration are possible and do not require additional computer ressources : the window method, presented in Section~\ref{sec:window}, might help to understand the spread in the lattice results for the dominant light quark contribution. 

Concerning the hadronic light-by-light contribution, the lattice community is likely to play a major role in the comparison with experiment. The current precision of the RBC/UKQCD collaboration is of $45\%$~\cite{Blum:2019ugy} and a precision of $20\%$ might suffice. The Mainz group has presented preliminary results at the SU(3)$_f$ point where an error of 15\% has been reached~\cite{Chao:2020kwq} and some preliminary results at lower pion masses have been presented. The lattice community can also provide important inputs to the dispersive framework, especially for the light pseudoscalar-pole contributions. The pion-pole contribution is already known with a precision of $6\%$~\cite{Gerardin:2016cqj,Gerardin:2019vio} and preliminary results for all three light pseudoscalars have been presented by the ETM collaboration this year~\cite{APLAT20:etmc}.

\vspace*{0.8cm}
\noindent
\centerline{\textbf{Acknowledgements}}
\vspace*{0.1cm}

I would like to thank the organizers of the 38th International Symposium on Lattice Field Theory for the invitation to write this review.  I would like to thank Laurent Lellouch for interesting discussions and feedback on this manuscript. 

This publication received funding from the Excellence Initiative of Aix-Marseille University - A*MIDEX, a French “Investissements d’Avenir” programme, AMX-18-ACE-005.


\bibliographystyle{epj}
\bibliography{review}{}
\end{document}